\documentclass[11pt]{article}
\pdfoutput=1
\usepackage{jcapmod}

\usepackage[english]{babel}
\usepackage{booktabs}
\usepackage{multirow}
\usepackage{subcaption}
\usepackage[per-mode=reciprocal, mode=math, group-digits=integer]{siunitx}[=v2]
\usepackage[babel=true]{microtype}

\numberwithin{equation}{section}
\allowdisplaybreaks[1]

\newcommand{\Neff}{N_\mathrm{eff}}
\newcommand{\sigmaNeff}{\sigma(\Neff)}
\newcommand{\fsky}{f_\mathrm{sky}}
\newcommand{\thetaFWHM}{\theta_b}
\newcommand{\alphaKnee}{\alpha_\mathrm{knee}}
\newcommand{\ellKnee}{\ell_\mathrm{knee}}
\newcommand{\sfour}{\mbox{CMB-S4}}
\newcommand{\sfourwide}{\mbox{\sfour~Wide}}
\newcommand{\sfourdeep}{\mbox{\sfour~Ultra-deep}}
\newcommand{\Planck}{Planck}
\newcommand{\sfourhybridwide}{\mbox{\sfour~Hybrid Wide}}
\newcommand{\sfourhybriddelensing}{\mbox{\sfour~Hybrid Delensing}}

\DeclareSIUnit{\muKelvin}{\micro\kelvin}
\DeclareSIUnit{\jansky}{Jy}
\DeclareSIUnit{\parsec}{pc}

\begin{document}

\pagenumbering{roman}
\begin{titlepage}
	\baselineskip=15.5pt \thispagestyle{empty}
	
	\bigskip\
	
	\vspace{0.52cm}
	\begin{center}
		{\huge \sffamily \bfseries Sensitivity of Next-Generation CMB~Surveys\\[8pt]to Neutrinos and Other Light Relics}
	\end{center}
	
	\vspace{0.2cm}
	\begin{center}
	{\large Cynthia~Trendafilova,$^{1,2}$ Srinivasan~Raghunathan,$^{3,2}$ Benjamin~Wallisch,$^{4,5,6}$ Joel~Meyers,$^{7}$ Kevork~N.~Abazajian,$^{8}$ Edoardo~Altamura,$^{9}$ Carlo~Baccigalupi,$^{10,11,12,13}$ Kimberly~K.~Boddy,$^{6}$ Thejs~Brinckmann,$^{14}$ Yuji~Chinone,$^{15}$ Gabriele~Coppi,$^{16,17}$ Francis-Yan~Cyr-Racine,$^{18}$ Jacques~Delabrouille,$^{19,20}$ Katherine~Freese,$^{6,4,5,21}$ Helena~García~Escudero,$^{22}$ Martina~Gerbino,$^{23}$ Shamik~Ghosh,$^{20}$ Vera~Gluscevic,$^{22}$ Daniel~Green,$^{24}$ Daniel~Grin,$^{25}$ Kevin~M.~Huffenberger,$^{26}$ Mudit~Jain,$^{27}$ Lloyd~Knox,$^{3}$ Anto~I.~Lonappan,$^{24}$ Marilena~Loverde,$^{28}$ Philip~Lubin,$^{29}$ Gabriele~Montefalcone,$^{6,30}$ Valentine~Novosad,$^{31,32}$ Marco~Raveri,$^{33}$ Christian~L.~Reichardt,$^{34}$ Sayan~Saha,$^{35}$ Murali~M.~Saravanan,$^{28}$ Emmanuel~Schaan,$^{36,37}$ Sara~M.~Simon,$^{38}$\\Julien~Tang,$^{39,19,20}$ and Scott~Watson\hskip1pt$^{40}$\\(\mbox{\sfour}~Collaboration)}
	
	\bigskip
	\textsl{(Author affiliations are provided on the following pages.)}
	\end{center}
	
	\vspace{0.8cm}
	\hrule \vspace{0.3cm}
	\noindent {\sffamily \bfseries Abstract}\\[0.1cm]
	Neutrinos and other light relics leave characteristic imprints in the cosmic microwave background anisotropies, making their observation a sensitive probe of the particle content and thermal history of the early universe. The energy density in these relativistic species is parameterized by their effective number~$\Neff$. Measuring this parameter at the percent level, which is a long-standing science goal of~\sfour\ and other experiments, would test a wide range of well-motivated physics within and beyond the Standard Model of particle physics. In this paper, we present Fisher-matrix forecasts of the projected sensitivity to~$\Neff$ of several \sfour~survey configurations considered during its extensive design phase. The conceptual design reaches $\sigmaNeff < 0.03$ over its seven-year observing period, while the revised configuration achieves the same precision over a longer timescale. We complement these results with a cosmic-variance-limited survey over the same multipole range to quantify the room for improvement accessible with additional instrumental, observational, and theoretical efforts. Finally, we discuss the broad implications of precise $\Neff$~measurements for the radiation sector, big bang nucleosynthesis, light thermal relics, and other early-universe physics. The forecasts presented in this work are performed with the publicly released~DRAFT~(Dark Radiation Anisotropy Flowdown Team)~tool. It provides an end-to-end pipeline from simulated foreground maps and component separation to delensing and projected sensitivities for any cosmological parameter, and it can be directly applied to other cosmic microwave background survey designs.
	\vskip10pt
	\hrule
	\clearpage

	\begin{center}
		\textsl{$^{1}$ CERCA/ISO, Department of Physics, Case Western Reserve University,\\*Cleveland, OH~44106, USA}
		\thispagestyle{empty}
		
		\vskip8pt
		\textsl{$^{2}$ Center for AstroPhysical Surveys, National Center for Supercomputing Applications,\\*University of Illinois Urbana-Champaign, Urbana, IL~61801, USA}
		
		\vskip8pt
		\textsl{$^{3}$ Department of Physics \& Astronomy, University of California, Davis, Davis, CA~95616, USA}
		
		\vskip8pt
		\textsl{$^{4}$ Oskar Klein Centre, Department of Physics, Stockholm University, 10691~Stockholm, SE}
		
		\vskip8pt
		\textsl{$^{5}$ Nordita, KTH~Royal Institute of Technology and Stockholm University, 10691~Stockholm, SE}
		
		\vskip8pt
		\textsl{$^{6}$ Texas Center for Cosmology and Astroparticle Physics, Weinberg Institute for Theoretical\\*Physics, Department of Physics, The University of Texas at Austin, Austin, TX~78712, USA}
		
		\vskip8pt
		\textsl{$^{7}$ Department of Physics, Southern Methodist University, Dallas, TX~75205, USA}
		
		\vskip8pt
		\textsl{$^{8}$ Center for Cosmology, Department of Physics and Astronomy, University of California, Irvine,\\*Irvine, CA~92697, USA}
		
		\vskip8pt
		\textsl{$^{9}$ Jodrell Bank Centre for Astrophysics, School of Physics and Astronomy,\\*The University of Manchester, Manchester~M13~9PL, UK}
		
		\vskip8pt
		\textsl{$^{10}$ The International School for Advanced Studies~(SISSA), 34136~Trieste, Italy}
		
		\vskip8pt
		\textsl{$^{11}$ Istituto Nazionale di Fisica Nucleare~(INFN), Sezione di Trieste, 34127~Trieste, Italy}
		
		\vskip8pt
		\textsl{$^{12}$ Istituto Nazionale di Astrofisica~(INAF), Osservatorio Astronomico di Trieste,\\*34143~Trieste, Italy}
		
		\vskip8pt
		\textsl{$^{13}$ Institute for Fundamental Physics of the Universe~(IFPU), 34151~Trieste, Italy}
		
		\vskip8pt
		\textsl{$^{14}$ CP3-Origins, University of Southern Denmark, 5230 Odense M, Denmark}
		
		\vskip8pt
		\textsl{$^{15}$ QUP~(WPI), KEK, Tsukuba, Ibaraki~\mbox{305-0801}, Japan}
		
		\vskip8pt
		\textsl{$^{16}$ Department of Physics, University of Milano-Bicocca, 20126~Milano, Italy}
		
		\vskip8pt
		\textsl{$^{17}$ Istituto Nazionale di Fisica Nucleare~(INFN), Sezione di Milano-Bicocca, 20126~Milano, Italy}
		
		\vskip8pt
		\textsl{$^{18}$ Department of Physics and Astronomy, University of New Mexico,\\*Albuquerque, NM~87106, USA}
		
		\vskip8pt
		\textsl{$^{19}$ CNRS-UCB International Research Laboratory, Centre Pierre Bin\'etruy, IRL~2007, CPB-IN2P3, Berkeley, CA~94720, USA}
		
		\vskip8pt
		\textsl{$^{20}$ Lawrence Berkeley National Laboratory, Berkeley, CA~94720, USA}
		
		\vskip8pt
		\textsl{$^{21}$ Department of Physics, University of Michigan, Ann Arbor, MI~48109, USA}
		
		\vskip8pt
		\textsl{$^{22}$ Department of Physics and Astronomy, University of Southern California,\\*Los Angeles, CA~90007, USA}
		
		\vskip8pt
		\textsl{$^{23}$ Istituto Nazionale di Fisica Nucleare~(INFN), Sezione di Ferrara, 44122~Ferrara, Italy}
		
		\vskip8pt
		\textsl{$^{24}$ Department of Physics, University of California San Diego, La Jolla, CA~92093, USA}
		
		\vskip8pt
		\textsl{$^{25}$ Department of Physics and Astronomy, Haverford College, Haverford, PA~19041, USA}
		
		\vskip8pt
		\textsl{$^{26}$ Mitchell Institute for Fundamental Physics and Astronomy, Department of Physics\\*and Astronomy, Texas A\&M University, College Station, TX~77843, USA}
		
		\vskip8pt
		\textsl{$^{27}$ Department of Physics and Astronomy, Rice University, Houston, TX~77005, USA}
		
		\vskip8pt
		\textsl{$^{28}$ Department of Physics, University of Washington, Seattle, WA~98195, USA}
		
		\vskip8pt
		\textsl{$^{29}$ Physics Department, University of California, Santa Barbara, Santa Barbara, CA~93106, USA}
		
		\vskip8pt
		\textsl{$^{30}$ Wisconsin IceCube Particle Astrophysics Center~(WIPAC), University of Wisconsin,\\*Madison, WI~53703, USA}
		
		\vskip8pt
		\textsl{$^{31}$ Materials Science Division, Argonne National Laboratory, Lemont, IL~60439, USA}
		
		\vskip8pt
		\textsl{$^{32}$ Institute of Multidisciplinary Research for Advanced Materials, Tohoku University, Sendai~\mbox{980-8577}, Japan}
		
		\vskip8pt
		\textsl{$^{33}$ Department of Physics, INFN and INAF, University of Genova, 16146~Genova, Italy}
		
		\vskip8pt
		\textsl{$^{34}$ School of Physics, The University of Melbourne, Parkville, VIC~3010, Australia}
		
		\vskip8pt
		\textsl{$^{35}$ Department of Physics, Northeastern University, Boston, MA~02115, USA}
		
		\vskip8pt
		\textsl{$^{36}$ SLAC National Accelerator Laboratory, Menlo Park, CA~94025, USA}
		
		\vskip8pt
		\textsl{$^{37}$ Kavli Institute for Particle Astrophysics and Cosmology, Stanford University,\\*Stanford, CA~94305, USA}
		
		\vskip8pt
		\textsl{$^{38}$ Fermi National Accelerator Laboratory, Batavia, IL~60510, USA}
		
		\vskip8pt
		\textsl{$^{39}$ Universit\'e Paris Cit\'e, CNRS, AstroParticule et Cosmologie, 75013~Paris, France}
		
		\vskip8pt
		\textsl{$^{40}$ Department of Physics, Syracuse University, Syracuse, NY~13244, USA}
	\end{center}
	\thispagestyle{empty}
\end{titlepage}

\thispagestyle{empty}
\setcounter{page}{4}
\tableofcontents

\clearpage
\pagenumbering{arabic}
\setcounter{page}{1}
\section{Introduction}
\label{sec:introduction}

The universe provides a unique laboratory for fundamental physics. The long timescales and extreme environments of the cosmos provide ideal conditions to probe physics in and beyond the Standard Model of particle physics~\cite{Green:2022hhj, Chang:2022lrw, Green:2022bre}, making cosmological surveys complementary to terrestrial experiments and astrophysical observations. The high temperatures and number densities in the early universe allow us in particular to probe light and weakly-interacting particles that are highly motivated theoretically, hard to measure in laboratory and collider searches, and among the prime observational targets for the next decade~\cite{Green:2019glg, Gerbino:2022nvz, Dvorkin:2022jyg}.\medskip

Cosmology offers a range of observational probes of the early universe and these light particles, in particular the cosmic microwave background~(CMB), big bang nucleosynthesis~(BBN), and the large-scale structure of the universe. Observations of the CMB~anisotropies have been at the forefront of this endeavor over the last few decades and will continue to be the leading source of new information.\footnote{Measurements of primordial light-element abundances produced during big bang nucleosynthesis have also played an important and complementary role in constraining the radiation content of the early universe at temperatures of order~\SI{1}{MeV}~(see e.g.~\cite{Lague:2019yvs, Pisanti:2020efz, Yeh:2020mgl, Matsumoto:2022tlr, Yeh:2022heq, Yanagisawa:2025mgx, Yeh:2026pil} or the recent reviews~\cite{Pitrou:2018cgg, Grohs:2019cae, Fields:2019pfx, Grohs:2023voo, Cooke:2024nqz}). In addition, large-scale structure observables will dramatically gain in constraining power over the next decade, both in general and in particular for neutrinos and other light relics~\cite{Baumann:2017gkg, Sprenger:2018tdb, CosmicVisions21cm:2018rfq, Baumann:2019keh, PUMA:2019jwd, Sailer:2021yzm, MoradinezhadDizgah:2021upg, Karkare:2022bai, Shi:2022drq, Lee:2023uxu, Euclid:2024imf, MoradinezhadDizgah:2026hrg}, with ongoing surveys such as DESI~\cite{DESI:2016fyo} and Euclid~\cite{Euclid:2021icp}, proposed surveys such as the MUltiplexed Survey Telescope~\cite{Zhao:2024alp} and Spec-S5~\cite{DESI:2022lza}, and more futuristic galaxy and line-intensity-mapping observations.} The~CMB itself consists of the photons that have been freely propagating through the universe since the epoch of recombination, about \num{400000}~years after the big bang, when electrons and protons combined into neutral hydrogen and the primordial photon-baryon plasma first became transparent. After traveling for more than 13~billion years, this light is now measured with increasing precision by a variety of ground- and space-based telescopes. While the~CMB is nearly isotropic, its temperature anisotropies are at the level of one part in~\num{e5} and its polarization is roughly another order of magnitude smaller, both of which have now been measured and characterized in detail. These anisotropies are a snapshot of the sound waves in the primordial plasma taken at the time of recombination when the photons decoupled and started to freely stream through the universe. Their observations have been a treasure trove of information not only about the conditions around the time of recombination and the constituents that make up our cosmos, but also about earlier times back to the first instants of the universe.\medskip

Neutrinos and other light relics leave distinct imprints in the CMB~anisotropies that we can use to distinguish them from other components of our universe, such as baryons and dark matter, and to probe their properties~(see e.g.~\cite{Lesgourgues:2013sjj, Abazajian:2013bxd, Abazajian:2016hbv, Alexander:2016aln, CMB-S4:2016ple, Lattanzi:2017ubx, Wallisch:2018rzj, Green:2019glg, Asadi:2022njl, Gerbino:2022nvz, Dvorkin:2022jyg, Green:2022bre, Antel:2023hkf} for reviews). Relativistic particles affect the~CMB through the following main effects: the damping of power on small angular scales, changes to the sound horizon and matter-radiation equality, and, for free-streaming species, coherent shifts in the acoustic peaks. The first of these drives the sensitivity to~$\Neff$: the damping tail responds to the total radiation energy density through the background expansion rate around recombination, and provides most of the constraining power when the precisely measured angular scale of the sound horizon is held fixed~\cite{Hou:2011ec}. At the level of the perturbations, free-streaming radiation induces characteristic shifts in the amplitude and phase of the acoustic oscillations, which provide additional sensitivity and can be used to distinguish different particle properties, such as free-streaming from fluid-like radiation~\cite{Bashinsky:2003tk, Follin:2015hya, Baumann:2015rya, Montefalcone:2025unv, Montefalcone:2025ibh}. Even the small phase-shift has been detected in \Planck~data~\cite{Follin:2015hya, Baumann:2015rya}, and more recently also in data of the Atacama Cosmology Telescope~(ACT) and the South Pole Telescope~(SPT)~\cite{Montefalcone:2025unv}. Importantly, the damping-tail and acoustic-peak features that drive the~$\Neff$~sensitivity are imprinted in both the CMB~temperature and the $E$-mode polarization power spectra. At the relevant angular scales and frequencies, the foregrounds in polarization are much smaller and the acoustic peaks much sharper than in temperature, making polarization observations a particularly clean probe of the radiation content of the universe. Gravitational lensing of the~CMB by the intervening large-scale structure between the last-scattering surface and our telescopes, however, broadens the acoustic peaks and partially obscures the damping tail. The high-fidelity removal of this lensing effect, referred to as delensing, is therefore essential to extract the full $\Neff$~information from observations of the primary CMB~anisotropies.

These signatures allow us to infer the non-photon radiation energy density, or the effective number of relativistic~(free-streaming) species present in our universe, as conventionally parameterized by~$\Neff$. This parameter is defined through the total radiation density,~$\rho_\mathrm{r}$, according to
\begin{equation}
	\rho_\mathrm{r} = \left[ 1 + \frac{7}{8} \left(\frac{4}{11}\right)^{\!4/3} \Neff \right] \rho_\gamma \,,
\end{equation}
where~$\rho_\gamma$ is the photon energy density. The prefactors arise from the Fermi-Dirac statistics of neutrinos and from the neutrino-to-photon temperature ratio due to electron-positron annihilation in the standard thermal history. If we consider only the known particle species in the Standard Model~(SM) of particle physics, the expected value is $\Neff = \Neff^\mathrm{SM} = 3.044$, which has been computed with an error smaller than the digits given and well below any projected observational sensitivity~\cite{Akita:2020szl, Froustey:2020mcq, Bennett:2020zkv, Cielo:2023bqp, Drewes:2024wbw, Binder:2024vmy, Ihnatenko:2025kew, Escudero:2025kej}. This value differs slightly from~$3.0$ for the three neutrino species due to the non-instantaneous decoupling of neutrinos and corrections from quantum electrodynamics. A measurement of~$\Neff$ that deviates from this value, parameterized as $\Delta\Neff \equiv \Neff - 3.044$, would therefore signal a clear departure from the Standard Model.\medskip

\begin{figure}
	\centering
	\includegraphics{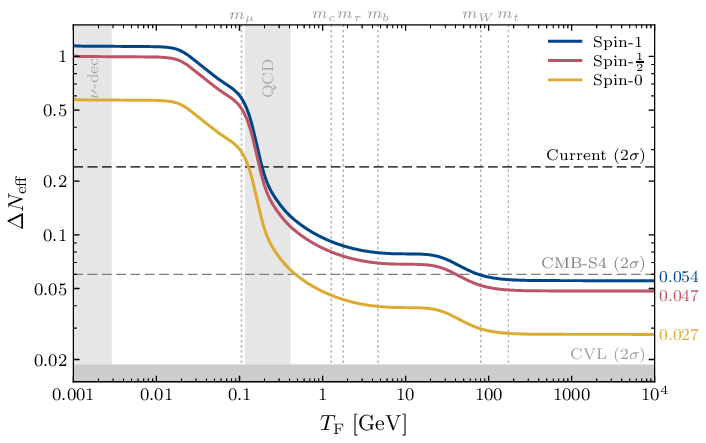}\vspace{-3pt}
	\caption{Contribution to the effective number of relativistic species,~$\Neff$, from a single thermally-produced relic, $\Delta\Neff = \Neff - \Neff^\mathrm{SM}$, with Standard Model value $\Neff^\mathrm{SM} = 3.044$, as a function of its freeze-out~(decoupling) temperature~$T_\mathrm{F}$ for three different spin values~(adapted from~\cite{Wallisch:2018rzj}; cf.~also~\cite{Green:2019glg, Dvorkin:2022jyg} for reviews). As the decoupling temperature increases, more Standard Model species are relativistic and in thermal equilibrium, so that their subsequent annihilation dilutes the relic abundance relative to the photon bath. The vertical dotted lines indicate the masses of some of these Standard Model particles~(muon, charm quark, tau lepton, bottom quark, $W$~boson, and top quark), while the vertical gray bands mark the epochs of neutrino decoupling and the~QCD~phase transition. A new light particle that decoupled above all Standard Model mass thresholds contributes $\Delta\Neff \approx 0.027$ for a real scalar~(spin-$0$), about~$0.047$ for a Weyl fermion~(spin-$1/2$), and roughly~$0.054$ for a massless vector boson~(spin-$1$). Detecting or excluding such contributions is a primary science goal of next-generation CMB~experiments, for which we display the~$2\sigma$~target for \sfour-like observations. We also show the currently most precise constraint of $\sigmaNeff = 0.12$ combining~\Planck, ACT, and SPT~data~\cite{Planck:2018vyg, AtacamaCosmologyTelescope:2025nti, SPT-3G:2025bzu}, but note that we have neglected the shift of the mean value by approximately~$2\sigma$ away from~$\Neff^\mathrm{SM}$ to only compare the level of precision with the projected sensitivities. These forecasts can also be compared to the floor set by a cosmic-variance-limited~(CVL) experiment, with the upper edge of the gray band corresponding to a survey covering~60\% of the sky and multipoles up to $\ell = 5000$. The band illustrates that both a larger sky coverage and higher multipoles push this threshold to even lower values. We finally note that the displayed CMB~forecasts will be complemented by data from large-scale-structure surveys, which will allow us to further tighten the bounds with future observations~(see e.g.~\cite{Baumann:2017gkg, Sprenger:2018tdb, CosmicVisions21cm:2018rfq, Baumann:2019keh, PUMA:2019jwd, Sailer:2021yzm, MoradinezhadDizgah:2021upg, Karkare:2022bai, Shi:2022drq, Lee:2023uxu, Euclid:2024imf, MoradinezhadDizgah:2026hrg}).}\vspace{-1pt}
	\label{fig:neff_freezeout}
\end{figure}

Such departures are well-motivated theoretically. Many extensions of the Standard Model predict additional light particles that would have been produced in the thermal bath of the early universe. A particularly well-motivated class of scenarios contributing to~$\Delta\Neff$ therefore involves light species that were in thermal equilibrium with the SM~plasma in the early universe and decoupled while still being relativistic. The contribution of such a thermal relic to~$\Neff$ is set by the number of relativistic degrees of freedom in the plasma when they decouple: an earlier freeze-out leaves more SM~species available to subsequently annihilate and dilute the relic abundance through the entropy transfer to the photon bath. This is illustrated in Fig.~\ref{fig:neff_freezeout},%
which displays~$\Delta\Neff$ as a function of freeze-out temperature for particles with three different spins. A relic that decoupled above all SM~mass thresholds contributes the minimum amount of $\Delta\Neff \approx 0.027$, $0.047$, and~$0.054$ for a single real scalar, a Weyl fermion, and a massless vector boson, respectively, after having been in thermal equilibrium at any point in the standard thermal history of the universe~(see~\textsection\ref{sec:implications_relics} for additional context and details on this statement). The spin-dependent differences capture the additional degrees of freedom of higher-spin states. These thresholds set natural particle-physics targets for next-generation CMB~surveys.

These contributions to~$\Neff$ from thermal freeze-out are particularly relevant since many well-motivated particles beyond the Standard Model~(BSM), including axions and other pseudo-Nambu-Goldstone bosons, sterile neutrinos, dark photons, and gravitinos, can fall in this regime because they may have been thermally produced in the early universe and decoupled while still relativistic. Since stronger SM~couplings lead to later decoupling, a larger interaction strength of these BSM~particles also induces correspondingly larger~$\Delta\Neff$. Beyond the search for light thermal relics, $\Neff$~measurements probe a much broader range of early-universe physics. They are sensitive to the physical properties of the radiation, including the free-streaming nature of the cosmic neutrino background, and constraints on the non-instantaneous decoupling, self-interactions, lifetime, and chemical potential of neutrinos. Additional contributions can arise away from thermal equilibrium, e.g.\ through particles decaying or annihilating into dark-sector species and freeze-in production sensitive to the reheating temperature. On the other hand, $\Delta\Neff < 0$, which is a regime mildly favored by recent ground-based observations~(see below), can arise from, for instance, entropy injection into the photon bath after neutrino decoupling, low reheating temperatures that prevent the neutrino bath from fully thermalizing, or non-standard neutrino decoupling. The parameter~$\Neff$ also sets an integrated bound on any cosmological stochastic gravitational-wave background, which is complementary to bounds from direct-detection experiments that only access relatively narrow frequency bands. Conversely, tightly constraining~$\Neff$ to its SM~value would have far-reaching implications for many BSM~scenarios, such as placing strong limits on the existence and properties of such particles. In addition, such a measurement provides a stringent consistency test of the standard cosmological history across cosmic epochs in combination with~BBN measurements of the primordial helium abundance~$Y_\mathrm{p}$. We refer to Section~\ref{sec:implications} and the reviews~\cite{Lesgourgues:2013sjj, Abazajian:2013bxd, Abazajian:2016hbv, Alexander:2016aln, CMB-S4:2016ple, Lattanzi:2017ubx, Wallisch:2018rzj, Green:2019glg, Asadi:2022njl, Gerbino:2022nvz, Dvorkin:2022jyg, Green:2022bre, Antel:2023hkf} for references to the literature and a comprehensive discussion of these and further implications.\medskip

Current observations have constrained~$\Neff$ with increasing precision. The \Planck~satellite measured $\Neff = 2.89 \pm 0.18$ from its temperature, polarization, and lensing power spectra~\cite{Planck:2018vyg}.\footnote{The analyses of the temperature and polarization spectra derived from \Planck~PR4 NPIPE~maps inferred $\Neff = 3.00 \pm 0.21$ and $\Neff = 3.08 \pm 0.17$ using the \texttt{CamSpec}~\cite{Rosenberg:2022sdy} and the \texttt{HiLLiPoP}~likelihood~\cite{Tristram:2023haj}, respectively. On the other hand, including baryon-acoustic-oscillation data in the inference from \Planck~PR3~spectra minimally tightens the mentioned value to $\Neff = 2.99 \pm 0.17$~\cite{Planck:2018vyg} since current joint constraints are dominated by information from the~CMB. The addition of broadband galaxy power spectra and especially future large-scale-structure surveys will however be able to also contribute meaningful constraining power in combined analyses~(see e.g.~\cite{Baumann:2017gkg, Sprenger:2018tdb, CosmicVisions21cm:2018rfq, Baumann:2019keh, PUMA:2019jwd, Sailer:2021yzm, MoradinezhadDizgah:2021upg, Karkare:2022bai, Shi:2022drq, Lee:2023uxu, Euclid:2024imf, MoradinezhadDizgah:2026hrg}).} This inferred value is consistent with the SM~prediction and already places meaningful limits on thermal relics that decoupled after the phase transition of quantum chromodynamics~(QCD). Current ground-based CMB~experiments are providing complementary measurements with increasing precision at small angular scales and in polarization. The combination of the sixth data release of the Atacama Cosmology Telescope and the first cosmological results from the \mbox{SPT-3G}~camera of the South Pole Telescope resulted in $\Neff = 2.78 \pm 0.17$~\cite{AtacamaCosmologyTelescope:2025nti, SPT-3G:2025bzu}, which now complements the precision of~\Planck\ from the ground. The addition of \Planck~data to~ACT and~SPT leads to the currently most precise value of $\Neff = 2.82 \pm 0.12$~\cite{Planck:2018vyg, AtacamaCosmologyTelescope:2025nti, SPT-3G:2025bzu}, which presents a high-significance detection of the cosmic neutrino background~\cite{Bauer:2022lri, Scott:2024rwc}. The South Pole Telescope \mbox{Ext-10k}~survey is projected to improve upon this precision, achieving $\sigmaNeff \approx 0.069$, by analyzing combined survey data taken through 2026~\cite{SPT-3G:2024qkd}. The Simons Observatory~(SO), which has started to take data, is projected to improve the precision to $\sigmaNeff = 0.045$ with its fully populated receiver over its now planned lifetime until~2034~\cite{SimonsObservatory:2018koc, SimonsObservatory:2025wwn}, further tightening the constraints on light relics and pushing beyond the QCD~phase transition.\medskip

The \sfour~experiment was designed to be a next-generation, ground-based CMB~survey capable of making ground-breaking discoveries in cosmology and astrophysics with even broader scientific implications. The planned observations would have made transformative measurements across a broad range of science targets, including tests of inflationary theories by measuring the primordial tensor-to-scalar amplitude ratio~$r$, constraining neutrinos and other light relics through a measurement of~$\Neff$ with a precision of $\sigmaNeff = 0.030$ at the 68\%~confidence level, studying galaxy clusters and their formation history, and exploring the transient sky in the millimeter-wave frequency range~\cite{CMB-S4:2016ple, Abazajian:2019eic, CMB-S4:2022ght, CMB-S4:2023cdr}. Such a measurement would in particular reach the $\Delta\Neff = 0.06$ threshold at~$2\sigma$, which would allow us to detect or place strong limits on many types of light particles that could have been in thermal contact with the Standard Model in the early universe~(cf.~Figure~\ref{fig:neff_freezeout}).

Probing neutrinos and other light relics at this exquisite level, with major implications for cosmology and particle physics, is enabled by unprecedented experimental sensitivity in measuring small-scale CMB~temperature anisotropies and especially polarization. More specifically, this sensitivity to the CMB~damping tail and acoustic peaks can only be achieved through low-noise, multi-frequency observations covering a large fraction of the sky with high-resolution large-aperture telescopes. While the $\Neff$~constraints from~\Planck\ are driven by its temperature data, the high signal-to-noise ratio and small foreground contamination achievable in $E$-mode polarization at small angular scales, together with high-fidelity delensing, will primarily drive the $\Neff$~sensitivity of future CMB~surveys, as we explicitly show in this work. Although the \sfour~project as originally envisioned was not carried forward following a programmatic decision by the main funding agencies in~2025, the extensive design optimization and forecasting work performed during its development remains directly applicable to the planning of future CMB~surveys. This includes informing the design and further development of current-generation experiments, such as the Simons Observatory and the South Pole Observatory, and potential next-generation ground- and space-based successors~\cite{Chang:2022tzj}.\bigskip

In this paper, we present forecasts of~$\sigmaNeff$ performed as part of the Maps-to-Power-Spectra analysis working group of the \sfour~collaboration. The purpose of this work is to identify experimental configurations capable of reaching the design-driving science target $\sigmaNeff = 0.030$. This in particular includes understanding how the precision on~$\Neff$ depends on key survey design parameters, such as sky fraction, scan strategy, and observation time, in order to optimize the experiment to achieve the desired threshold in~$\Neff$. We report the statistical results of this iterative and years-long work, and refer to~\cite{Raghunathan:inprep} for an investigation of systematic effects, in particular those affecting small-scale CMB~measurements, and associated mitigation techniques. The paper is organized as follows: In Section~\ref{sec:configurations}, we detail the experimental configurations considered in our forecasts. We then discuss the employed foreground modeling and masking, and describe our forecasting pipeline in Section~\ref{sec:forecasting}, which includes the internal-linear-combination method, the Fisher-matrix formalism, and the DRAFT~tool. We present our forecasting results in Section~\ref{sec:results}, and discuss their cosmological and particle-physics implications in Section~\ref{sec:implications}. \mbox{Finally, we conclude in Section~\ref{sec:conclusions}.}

\section{Survey Design and Configurations}
\label{sec:configurations}

Precise measurements of~$\Neff$ require high-sensitivity observations of the CMB~temperature anisotropies and polarization at small angular scales. This demand shapes the core design choices of any survey targeting light relics: multi-frequency coverage with high-resolution large-aperture telescopes to access the relevant multipoles while controlling foregrounds, combined with a large sky coverage to accumulate as many modes as possible and boost the statistical power. Throughout the development of the \sfour~project, many options were examined for the type, number, and location of the telescopes, and for their scan strategies. In this section, we describe several of the resulting configurations, which were explored at various planning stages and which we analyzed with our forecasting pipeline~(with some of these results presented in Section~\ref{sec:results}). While these are the concrete designs considered for~\sfour, they equally illustrate the general choices facing any future survey targeting~$\Neff$, from the choice of site(s) to the trade-off between depth and sky area. In addition, we include a cosmic-variance-limited survey that establishes the ultimate reach over the same multipole range.\medskip

After detailing the common noise model and forecasting choices that apply to all configurations in~\textsection\ref{sec:configurations_common}, we describe four configurations spanning these choices. We first consider the original two-site conceptual design of~\sfour, with telescopes in both the Atacama Desert of Chile and at the South Pole~(\textsection\ref{sec:configurations_conceptual}). We then turn to two single-site configurations: a South-Pole-only survey, which was studied as part of an in-depth analysis of alternatives~(\textsection\ref{sec:configurations_pole}), and a Chile-only revised configuration, which demonstrates how the original \sfour~science goals could be reached from Chile together with existing observational infrastructure~(\textsection\ref{sec:configurations_revised}). Finally, we introduce a cosmic-variance-limited survey in~\textsection\ref{sec:configurations_cvl} to establish the fundamental sensitivity floor over the same range of observed scales. (Table~\ref{tab:results_summary} summarizes these configurations together with their sky fractions, observation times, and projected sensitivities.) Unless stated otherwise, we present each configuration at its nominal observation time, which is typically seven years for the \sfour~surveys, and discuss the year-by-year evolution of the forecasted sensitivity in Section~\ref{sec:results}. The forecasting choices that apply uniformly across all configurations, including the multipole range, fiducial cosmology, component separation, and Fisher~methodology, are detailed in Section~\ref{sec:forecasting}.

\subsection{Common Specifications}
\label{sec:configurations_common}

The sensitivity of a CMB~survey to~$\Neff$ primarily depends on how well it can measure the damping tail and acoustic peak structure at multipoles $\ell \gtrsim 1000$. Put simply, this is controlled by two competing factors: the total effective noise per mode, which determines how far into the damping tail one can measure, and the observed sky fraction~$\fsky$, which sets the number of available independent modes. For~$\Neff$, the constraining power from additional sky coverage generally exceeds that from deeper observations over a smaller area, since the relevant information is spread across many acoustic peaks rather than concentrated in a narrow multipole range. Moreover, the signal is exponentially damped toward smaller scales, while the instrumental noise power decreases as the inverse of the observing time. At fixed observing effort, more independent modes in the relevant regime are therefore gained by observing a wider patch of sky than by integrating more deeply over a smaller one. We demonstrate this quantitatively in~\textsection\ref{sec:results_optimization} using a simplified noise model, while the configurations described below reflect the resulting design choices in light of various constraints and limited resources.\medskip

Throughout this work, we model the noise power spectra in a given frequency band as a function of multipole~$\ell$ according to
\begin{equation}
	N_\ell^X = \Delta_X^2\ \exp\!\left[ \ell (\ell+1) \frac{\thetaFWHM^2}{8\log2} \right] \left[ 1 + \left( \frac{\ellKnee^X}{\ell} \right)^{\!\alphaKnee^X} \right] ,	\label{eq:noise_model}
\end{equation}
with $X = T\text{ and }P$ for the temperature and polarization anisotropies, respectively. The instrumental white-noise level is given by~$\Delta_X$ and the effect of beam deconvolution at high multipoles is captured by the beam full-width at half-maximum~$\thetaFWHM$. The last factor contains the contribution from low-frequency~($1/f$) noise at low multipoles, which is parameterized by~$\ellKnee^X$ and~$\alphaKnee^X$, with $\alphaKnee^X > 0$ in our notation, and is dominated by atmospheric emission for ground-based observations.\medskip

For all configurations, we combine the \sfour~noise spectra after component separation with the effective instrumental noise of the combined frequency channels from the \Planck~satellite using inverse-variance weighting, on the same masked sky as employed for the~\sfour~analysis~(see~\textsection\ref{sec:forecasting_foregrounds}). \Planck~provides essential complementary information at large angular scales and on a wider range of frequencies than those accessible to ground-based experiments. The employed \Planck~instrumental noise parameters, which were obtained in~\cite{Allison:2015qca} to approximately describe the achieved sensitivity of the satellite in forecasts, are listed in Table~\ref{tab:noise_planck}%
\begin{table}
	\centering
	\begin{tabular}{c S[table-format=4.2] S[table-format=4.2] S[table-format=4.2] S[table-format=4.2] S[table-format=4.2] S[table-format=4.2] S[table-format=4.2]}
				\toprule
			\multirow{2}{*}{Parameter}			& \multicolumn{7}{c}{Observational band~[\si{GHz}]}		\\
												  \cmidrule(lr){2-8}
												& 30 	& 44 	& 70	& 100	& 143	& 217	& 353	\\
				\midrule[0.065em]
			$\thetaFWHM$ [\si{arcmin}]			& 33	& 23	& 14	& 10	& 7		& 5		& 5		\\
			$\Delta_T$ [\si{\muKelvin.arcmin}]	& 145	& 149	& 137	& 65	& 43	& 66	& 200	\\
			$\Delta_P$ [\si{\muKelvin.arcmin}]	& {--}	& {--}	& 450	& 103	& 81	& 134	& 406	\\
				\bottomrule
		\end{tabular}
	\caption{Instrumental noise parameters assumed for the \Planck~satellite~(from~\cite{Allison:2015qca}). We neglect low-frequency noise for Planck so that $\ellKnee^X \to 0$ and the last term in the noise spectra of~\eqref{eq:noise_model} is absent. This implies that the instrumental sensitivity can be described by the beam width~$\thetaFWHM$ and the white noise levels~$\Delta_{T,\hskip1pt P}$ for temperature and polarization, respectively. We add \Planck~data described in this way to all \sfour~configurations using inverse-variance weighting of the computed noise spectra.}
	\label{tab:noise_planck}
\end{table}
for each frequency band. Since~\Planck\ is a space-based experiment without the low-frequency noise sourced by the atmosphere, we approximate the noise spectra as being fully specified by these beam widths and white-noise levels, i.e.\ $\ellKnee^X \to 0$ in~\eqref{eq:noise_model}. These \Planck~noise spectra are combined with the \sfour~spectra for $\ell \geq \ell_\mathrm{min} = 30$, with the \Planck~contribution becoming negligible relative to the~\sfour~sensitivity on small scales. At multipoles $\ell < 30$, we include the large-scale~\Planck~temperature data by adding a Fisher matrix for the \Planck~$TT$~power spectrum over $\ell \in [2, 29]$ and a sky fraction of~$\fsky = 0.8$ to all~\sfour~configurations. We do not include the corresponding large-scale polarization spectra of~\Planck, but impose a Gaussian prior of $\sigma(\tau) = 0.007$~\cite{Planck:2018vyg} to instead capture the information on the Thomson optical depth due to reionization,~$\tau$, contained in these measurements.

\subsection{Two-Site Conceptual Design}
\label{sec:configurations_conceptual}

The initially proposed configuration for the \sfour~experiment, referred to as the conceptual design, is described in detail in~\cite{CMB-S4:2023cdr}. By combining telescopes at the two premier sites for ground-based CMB~observations, the Atacama Desert in Chile and the South Pole, this design was developed to simultaneously serve the two main design-driving science goals of~\sfour: the search for primordial gravitational waves and the percent-level measurement of~$\Neff$. In the following, we summarize its surveys and their observational specifications, focusing on those entering our $\Neff$~forecasts.\medskip

The conceptual design consists of three surveys, each contributing to the combined science goals of the project. Two of these surveys are located at the South Pole: an ultra-deep, low-resolution survey using~18~small-aperture telescopes~(SATs) targeting large-angular-scale polarization and an ultra-deep, high-resolution survey using a single five-meter three-mirror-anastigmat large-aperture telescope~(LAT). The third survey is a wide and deep, high-resolution survey using two six-meter crossed-Dragone~LATs in the Chilean Atacama Desert. This wide survey observes in six frequency bands spanning~$\SIrange{25}{280}{GHz}$, while the ultra-deep survey covers seven bands in the range~$\SIrange{20}{280}{GHz}$. Since the~SATs target degree-scale $B$-mode polarization rather than the small-scale damping tail, they do not contribute to the $\Neff$~forecasts presented here, which is why we focus on the two high-resolution LAT~surveys. The wide survey performed with the two Chilean~LATs contributes most significantly to the $\Neff$~science case, since it provides the large sky fraction that drives the constraining power, as discussed above. In our forecasts, we combine both high-resolution surveys from Chile and Antarctica. This achieves the best possible measurement of~$\Neff$ based on the conceptual design over the envisioned observation time of seven years.

\begin{figure}
	\centering
	\includegraphics{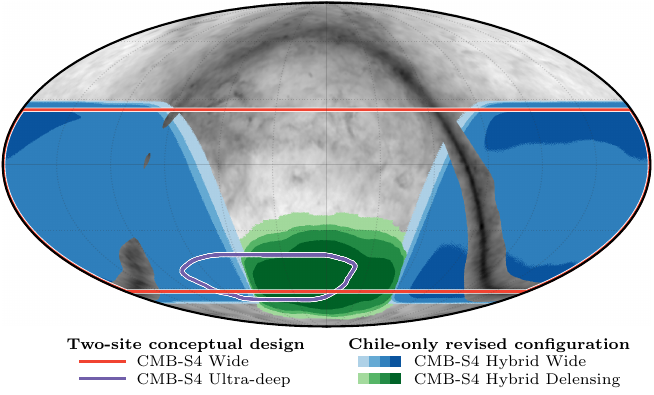}
	\caption{Survey footprints of the \sfour~two-site conceptual design and its Chile-only revised configuration with the \Planck~\texttt{GAL090}~galactic mask on top of the \Planck~galactic dust map. The~\SI{353}{GHz} thermal-dust intensity map from~\Planck~\cite{Planck:2015mvg} is displayed in gray scale with the cleanest parts of the sky in light gray and the regions of highest galactic emission in dark gray. The expected sky coverage of the~\sfourwide\ and \sfourdeep~surveys of the conceptual design~(\textsection\ref{sec:configurations_conceptual}) is indicated by the red and purple lines. This survey employs two large-aperture telescopes covering 68\%~of the sky from the Atacama Desert at an elevation angle of~\SI{40}{\degree} prior to galactic masking, and one large-aperture telescope located at the South Pole, with their frequency-dependent instrumental noise listed in Table~\ref{tab:noise_conceptual}. The footprints of the \sfourhybridwide\ and~Delensing surveys of the revised configuration~(\textsection\ref{sec:configurations_revised}), which uses a single large-aperture telescope located in Chile to observe about 56\%~of the sky, are shown in blue and green, respectively. The color shades indicate the four patches of different depth into which each survey is divided for the forecasts, with darker colors denoting deeper observations and lower noise~(the corresponding noise levels of these subdivisions are provided in Table~\ref{tab:noise_revised}). The Simons Observatory is anticipated to observe the same~68\% of the sky as displayed for~\sfourwide\ of the conceptual design. The \Planck~\texttt{GAL090}~galactic mask~\cite{Planck:2015mrs}, which removes the most contaminated~10\%~of the sky~(the additional apodization with a~\SI{2}{\degree} kernel is not included), is applied to the Hybrid~Wide~footprint to highlight the employed galactic masking for both configurations. Given that the masked region is left uncolored, the underlying galactic dust map remains visible where the mask is applied. The retained sky fractions are $\fsky \approx 0.62$ for the conceptual design and~$0.44$ for the \sfour~Hybrid portion of the revised configuration.}
	\label{fig:footprints}
\end{figure}

We present the instrumental noise in both temperature and polarization for the Chilean \sfourwide\ and South-Pole-based~\sfourdeep~surveys of the conceptual design~\cite{CMB-S4:2023cdr} in Table~\ref{tab:noise_conceptual}.\footnote{We studied alternative allocations of the LAT~detector count across the frequency bands during the \sfour~design process, evaluating each for its influence on foreground mitigation in the context of the damping-tail and Sunyaev-Zel'dovich science cases. Among the considered detector distributions, the baseline allocation, which sets the noise levels reported in Table~\ref{tab:noise_conceptual}, provides the best overall sensitivity for both.} %
\begin{table}
	\centering
	\subfloat[\sfourwide~survey in Chile.]{
		\begin{tabular}{c *{7}{W{c}{3em}}}
				\toprule
			\multirow{2}{*}{Parameter}			& \multicolumn{7}{c}{Observational band~[\si{GHz}]}		\\
												  \cmidrule(lr){2-8}
												& 20 	& 25 	& 40	& 90	& 150	& 230	& 280	\\
				\midrule[0.065em]
			$\thetaFWHM$ [\si{arcmin}]			& {--}	& 7.8	& 5.3	& 2.2	& 1.4	& 1.0	& 0.9	\\
				\midrule[0.03em]
			$\Delta_T$ [\si{\muKelvin.arcmin}]	& {--}	& 27.1	& 11.6	& 2.0	& 2.0	& 6.9	& 16.7	\\
			$\ellKnee^T$						& {--}	& 415	& 391	& 1932	& 3917	& 6740	& 6792	\\
			$\alphaKnee^T$						& {--}	& \multicolumn{6}{c}{3.5}						\\
				\midrule[0.03em]
			$\Delta_P$ [\si{\muKelvin.arcmin}]	& {--}	& 37.6	& 15.5	& 2.7	& 3.0	& 9.8	& 23.9	\\
			$\ellKnee^P$						& {--}	& \multicolumn{6}{c}{700}						\\
			$\alphaKnee^P$						& {--}	& \multicolumn{6}{c}{1.4}						\\
				\bottomrule
		\end{tabular}
	}\\[10pt]
	\subfloat[\sfourdeep~survey at the South Pole.]{
		\begin{tabular}{c *{7}{W{c}{3em}}}
				\toprule
			\multirow{2}{*}{Parameter}			& \multicolumn{7}{c}{Observational band~[\si{GHz}]}		\\
												  \cmidrule(lr){2-8}
												& 20 	& 25 	& 40	& 90	& 150	& 220	& 280	\\
				\midrule[0.065em]
			$\thetaFWHM$ [\si{arcmin}]			& 11.4	& 9.1	& 6.2	& 2.5	& 1.6	& 1.1	& 1.0	\\
				\midrule[0.03em]
			$\Delta_T$ [\si{\muKelvin.arcmin}]	& 11.9	& 6.5	& 3.0	& 0.45	& 0.41	& 1.3	& 3.1	\\
			$\ellKnee^T$						& 1200	& 1200	& 1200	& 1200	& 1900	& 2100	& 2100	\\
			$\alphaKnee^T$						& 4.2	& 4.2	& 4.2	& 4.2	& 4.1	& 4.1	& 3.9	\\
				\midrule[0.03em]
			$\Delta_P$ [\si{\muKelvin.arcmin}]	& 16.7	& 9.2	& 4.2	& 0.63	& 0.59	& 1.8	& 4.3	\\
			$\ellKnee^P$						& 150	& 150	& 150	& 150	& 200	& 200	& 200	\\
			$\alphaKnee^P$						& 2.7	& 2.7	& 2.7	& 2.6	& 2.2	& 2.2	& 2.2	\\
				\bottomrule
		\end{tabular}
	}
	\caption{Noise parameters for the two high-resolution \sfour~surveys of the two-site conceptual design~\cite{CMB-S4:2023cdr}: the wide survey covering approximately~68\%~of the sky from the Atacama Desert in Chile~(\sfourwide), and the ultra-deep survey covering 3\%~of the sky from the South Pole~(\sfourdeep). The beam width~$\thetaFWHM$, white noise levels~$\Delta_{T,\hskip1pt P}$, atmospheric-noise knee~$\ellKnee^{T,\hskip1pt P}$, and atmospheric-noise slope~$\alphaKnee^{T,\hskip1pt P}$ completely parameterize the noise model of~\eqref{eq:noise_model} in temperature~($T$) and polarization~($P$). We additionally take the atmospheric noise to be correlated at~90\% across pairs of adjacent frequency bands~(\SIrange[range-units=single, range-phrase = { and }]{25}{40}{GHz}, \SIrange[range-units=single, range-phrase = { and }]{90}{150}{GHz}, and \SIrange[range-units=single, range-phrase = { and }]{230}{280}{GHz}), which share a common optics tube and, therefore, observe the same sky along a common line of sight. We note that~$\alphaKnee^T$, $\ellKnee^P$, and~$\alphaKnee^P$ for the \sfourwide~survey are the same for all frequency bands.}
	\label{tab:noise_conceptual}
\end{table}
These noise levels were derived from detailed hit-map simulations of the proposed scan strategy~(cf.~\cite{Simon:inprep}). To calculate year-by-year forecasts, we rescale these noise levels by the inverse square root of the observation time. The expected sky coverage of the survey is shown in Fig.~\ref{fig:footprints},%
assuming that the~LATs observe at an elevation angle of~\SI{40}{\degree} from the Atacama Desert. The resulting footprint covers 68\%~of the sky prior to any masking of the galactic plane~(see~\textsection\ref{sec:forecasting_foregrounds}). The ultra-deep~LAT~survey from the South Pole additionally observes 3\%~of the sky in a region that overlaps with the footprint of the wide survey and has minimal galactic foreground contamination, which we therefore neglect in our foreground modeling in this case. The primary purpose of the~\sfourdeep~survey is to provide deep, high-resolution maps for internal delensing of the~SAT $B$-mode measurement in support of the search for primordial gravitational waves. Despite its small area, the low noise levels achieved over this patch still make a non-negligible contribution to the $\Neff$~analysis when combined with the wide survey, as we quantify in~\textsection\ref{sec:results_conceptual}. Together, the~\sfourwide\ and \sfourdeep~surveys located in the Atacama Desert and at the South Pole constitute the portion of the conceptual design entering our $\Neff$~forecasts.

\subsection{South-Pole-Only Alternative}
\label{sec:configurations_pole}

As part of an in-depth analysis of alternative experimental configurations performed during the \sfour~design phase~\cite{CMB-S4:2023aoa}, we considered a single-site option in which the entire telescope deployment is located at the South Pole. The motivation was to simplify logistics and consolidate operations at the single premier CMB~observing site in Antarctica. This option however comes at the cost of substantially reduced sky coverage.\footnote{While a given sky region rises and sets in the Atacama Desert due to the rotation of the Earth, the same patch of sky is perpetually visible from the South Pole.} In the conceptual design of~\textsection\ref{sec:configurations_conceptual}, the South-Pole~LAT is primarily intended for delensing and observes only~$\fsky = 0.03$ over its nominal seven-year survey. Since the $\Neff$~science case is driven by sky coverage at fixed observational effort, as anticipated in~\textsection\ref{sec:configurations_common} and quantitatively shown in~\textsection\ref{sec:results_optimization}, we substantially expand the sky coverage in the alternative survey to illustrate the prospects of $\Neff$~science from a single Antarctic~site.\medskip

For this South-Pole-only option, we assume that the entire \sfour~detector count of the conceptual design~(\textsection\ref{sec:configurations_conceptual}) is deployed to Antarctica, corresponding to a total of~\num{403208}~detectors, with~\num{128448} originally planned for the South-Pole~LAT~\cite{CMB-S4:2023cdr}. All detectors are taken to observe a larger portion of the sky over the nominal seven-year observation period with the same per-detector observing efficiency as that of the single \sfourdeep~LAT. The observed sky coverage is assumed to be about~25\%, which corresponds to a minimum elevation angle of~\SI{30}{\degree}. Given the latitude cut described in~\textsection\ref{sec:forecasting_foregrounds}, we neglect the galactic foregrounds in this South-Pole-based survey for simplicity, as for the \sfourdeep~survey, despite the larger sky area.

The temperature and polarization noise levels for this alternative are obtained by rescaling the \sfourdeep~noise of Table~\ref{tab:noise_conceptual} according to $\Delta_{T,\hskip1pt P} \propto \sqrt{\fsky / (N_\mathrm{det}\hskip1pt t)}$ at fixed observing efficiency, with the number of detectors~$N_\mathrm{det}$ and observational time~$t$. This scaling implies that the white-noise levels in every frequency band degrade by about~63\% relative to those of the ultra-deep survey over the same observation time of seven years, because the increase in the analyzed sky area by a factor of roughly~$8.3$ outweighs the increase in the deployed detector count by a factor of approximately~$3.1$. The beam and atmospheric-noise properties are taken to be identical to those of the conceptual South-Pole~LAT in Table~\ref{tab:noise_conceptual}. While a detailed study has not been performed, we note that observing the wider patch requires a different scan strategy than that assumed for the \sfourdeep~survey~(cf.\ the hybrid scan strategy of the revised configuration in~\textsection\ref{sec:configurations_revised}), which may affect the achievable noise levels and other observational specifications, as well as the constraining power for the $B$-mode science, which favors deep observations on a small patch of sky~\cite{CMB-S4:2020lpa}. Our forecasts based on the simplified scaling-based noise estimate should therefore be interpreted as an indicative benchmark for the $\Neff$~sensitivity achievable from a single-site survey at the South Pole.

\subsection{Chile-Only Revised Configuration}
\label{sec:configurations_revised}

The~\sfour~experiment was redesigned in~2024 to reach its science targets without new telescopes in Antarctica and together with other observational infrastructure. The resulting revised configuration~\cite{CMB-S4:2025rpp} relies on a coordinated effort between~\sfour, the Simons Observatory, and the South Pole Observatory, with \sfour~contributing six~SATs and one~LAT to the observing site in Cerro Toco, Chile. Since the $\Neff$~science case is driven by sensitivity to the CMB~damping tail at small angular scales, we focus on the telescopes in the Atacama Desert, in particular the proposed~\sfour~LAT and its observing strategy.\medskip

Our forecasts for this configuration combine three survey components, which we describe in turn. First, a single \sfour~LAT located in Chile performs a hybrid scan strategy to simultaneously serve the science goals of the~\sfourwide\ and \sfourdeep~surveys of the conceptual design described in~\textsection\ref{sec:configurations_conceptual}. This strategy prioritizes the delensing field for the $B$-mode science of the~SATs when it is accessible and spends its remaining time on a wider and shallower survey. This hybrid scan utilizes a minimum elevation angle of~\SI{35}{\degree} for the~LAT, which is lower than the angle of~\SI{40}{\degree} employed in the conceptual design, and therefore additionally accesses sky areas at slightly lower and higher latitudes. This wide survey covers approximately~47\% of the sky in a non-uniform fashion due to the dual-purpose use of the telescope and the time spent on the sky area of the delensing field. In particular, the hybrid survey does not cover the entire sky that is visible from the observing site, as reflected by the uncovered region at the center of the footprint projection in Fig.~\ref{fig:footprints}.\footnote{This hybrid allocation prioritizes the delensing observations and is therefore conservative for the~$\Neff$~science since the wide survey uses only the~LAT~time that remains after maximizing the delensing survey~\cite{Simon:inprep}. Had further study of the $B$-mode science concluded that slightly larger noise in the delensing field was sufficient, part of this time could instead be allocated to the wide survey, increasing the sky coverage and correspondingly improving the projected~$\Neff$~sensitivity. We refer to~\cite{Simon:inprep} for an example of a hybrid observing strategy, and how the available observing time can be divided between the wide and delensing surveys.} The details of this scan strategy and the relevant methodology are discussed~in~\cite{Simon:inprep}.

Second, the~LAT of the Simons Observatory~(or a similar telescope) is an integral part of the \sfour~revised configuration, with the observations of the new \sfour~LAT and the then existing SO~LAT being combined and jointly analyzed. This is possible because~SO~is planned to perform a uniform wide-area survey covering approximately~61\% of the sky from the same Chilean location. Throughout this work, we refer to this setup as~\mbox{`SO-like'} to indicate that any wide-area survey from Chile with noise levels similar to those of~SO is suitable for such a collaborative effort. We take the SO~LAT~noise at the completion of its planned six-year survey to be the expected goal depth~(rather than the more conservative baseline target) in each frequency band as given in Table~1 of~\cite{SimonsObservatory:2025wwn}. To produce year-by-year forecasts, we rescale these depths according to observation time, both during and after the nominal~SO~observation period.

Third, the \sfourhybriddelensing~survey, whose primary purpose is to enable the crucial delensing for the $B$-mode science as noted above, is also included in the analysis for~$\Neff$ and other damping-tail science. With the proposed hybrid scan strategy for the \sfour~LAT, these ultra-deep observations are largely complementary to the wide survey. Although it covers a small sky area of about~9\%, the ultra-deep survey reaches noise levels much lower than those expected for the SO-like~LAT~(similar to the \sfourdeep~survey at the South Pole in the conceptual design of~\textsection\ref{sec:configurations_conceptual}), providing a non-trivial additional contribution to the $\Neff$~analysis.\medskip

Our forecasts assume that the SO-like~LAT begins observing in~2028\hskip1pt\footnote{The detectors for the upgrade of the~SO~LAT have in fact already been installed in~2026~\cite{Dunkley:private}.\label{fn:so}} and is supported past its nominally planned observation time as part of the Simons Observatory. As proposed in the revised configuration~\cite{CMB-S4:2025rpp}, we extend the SO-like~LAT~observations into the early~2040s to facilitate joint collection and analysis of data from both telescopes for the duration of~\sfour. In this timeline, the \sfour~LAT begins taking data in~2033, with its observations combined with SO-like~LAT~data accumulated up to that point and then jointly analyzed on a yearly basis thereafter. On the sky area observed with the SO-like~LAT but not covered by either \sfour~survey, we use those data alone. Since the revised \sfour~configuration assumes a coordinated joint data analysis of the two experiments, we apply the same galactic masking to the SO-like~and \sfour~data~(the masking procedure and the resulting analyzed sky fractions are described in~\textsection\ref{sec:forecasting_foregrounds}).

Due to the non-uniformity of the hybrid scan strategy for the \sfour~LAT, we subdivide both the wide and delensing portions of the \sfour~survey footprint based on noise level into four patches that we display in Fig.~\ref{fig:footprints}. The sky fraction covered by each patch is~$\fsky = 0.081$, $0.34$, $0.017$, and~$0.029$ for patches~1 to~4 of the wide survey for a total of~47\% of the sky. This is significantly less than the sky coverage of the conceptual design, which is one of the main reasons for including the~SO-like~LAT to provide data especially on its additional sky region. For the delensing survey, the observed sky coverage of the four patches is~$\fsky = 0.043$, $0.021$, $0.010$, and~$0.018$, for a total of~9.2\% of the sky, which is unaffected by the galactic mask given its location. The corresponding noise levels for these patches are given in Table~\ref{tab:noise_revised}, %
\begin{table}
	\centering
	\subfloat[\sfourhybridwide~survey in Chile.]{
		\begin{tabular}{c *{7}{W{c}{3em}}}
				\toprule
			\multirow{2}{*}{$\Delta_T$ [\si{\muKelvin.arcmin}]}	& \multicolumn{7}{c}{Observational band~[\si{GHz}]}		\\
																  \cmidrule(lr){2-8}
																& 20 	& 30 	& 40	& 90	& 150	& 220	& 280	\\
				\midrule[0.065em]
			Patch 1												& 41.3	& 21.1	& 11.4	& 1.9	& 1.8	& 6.8	& 19.9	\\
			Patch 2												& 67.7	& 34.2	& 18.5	& 3.1	& 2.9	& 11.3	& 33.1	\\
			Patch 3												& 151.2	& 74.3	& 40.3	& 6.6	& 6.2	& 26.9	& 78.9	\\
			Patch 4												& 205.8	& 183.8	& 100.0	& 31.1	& 29.4	& 44.9	& 132.0	\\
				\bottomrule
		\end{tabular}
	}\\[10pt]
	\subfloat[\sfourhybriddelensing~survey in Chile.]{
		\begin{tabular}{c *{7}{W{c}{3em}}}
				\toprule
			\multirow{2}{*}{$\Delta_T$ [\si{\muKelvin.arcmin}]}	& \multicolumn{7}{c}{Observational band~[\si{GHz}]}		\\
																  \cmidrule(lr){2-8}
																& 20 	& 30 	& 40	& 90	& 150	& 220	& 280	\\
				\midrule[0.065em]
			Patch 1												& 14.0	& 7.1	& 3.8	& 0.6	& 0.6	& 2.4	& 6.9	\\
			Patch 2												& 39.0	& 17.6	& 9.6	& 1.7	& 1.6	& 8.7	& 25.5	\\
			Patch 3												& 67.2	& 40.3	& 21.9	& 3.7	& 3.5	& 17.4	& 51.2	\\
			Patch 4												& 165.6	& 116.1	& 63.2	& 23.2	& 21.9	& 35.3	& 104.2	\\
				\bottomrule
		\end{tabular}
	}
	\caption{Temperature white-noise levels for the four patches of the~\sfourhybridwide\ and~Delensing~surveys of the Chile-only revised configuration illustrated in Fig.~\ref{fig:footprints}, with only the darkest patches~1 and~2 employed in our forecasts. The sky fraction of each patch is provided in the main text. The beam and atmospheric-noise properties are the same as those of the Chilean large-aperture telescope of the conceptual design, as listed in Table~\ref{tab:noise_conceptual}. Unlike the noise specifications in that table, where the polarization noise is listed independently at each frequency, the polarization noise is here taken to be a factor of~$\sqrt{2}$ larger than the temperature noise as a simplifying assumption for fully polarized detectors.}
	\label{tab:noise_revised}
\end{table}
but we only use patches~1 and~2 of each survey in our forecasts since patches~3 and~4 have significantly higher noise levels, and are also largely observed by the SO-like~LAT. Because the~\sfour~LAT of this configuration uses the same design as the conceptual design, the beam and atmospheric-noise properties are identical to those in Table~\ref{tab:noise_conceptual}.

\subsection{Cosmic-Variance-Limited Survey}
\label{sec:configurations_cvl}

In addition to the specific experimental configurations described above, we consider a cosmic-variance-limited~(CVL) survey in order to establish the ultimate precision on~$\Neff$ achievable from the CMB~primary anisotropies. This idealized case sets a fundamental benchmark against which realistic experiments can be compared and quantifies how much room remains for improvement beyond a \sfour-like survey. More specifically, it allows us to assess what fraction of the CMB~primary information remains untapped at \sfour~precision. It also enables us to identify the regimes in which further improvements are limited by the survey design or our understanding and modeling of astrophysics, rather than by cosmic variance, the fundamental limit set by the single realization of the universe available to us.

For the CVL~configuration, we set the instrumental noise to zero, $\Delta_T = \Delta_P \equiv 0$, so that the measurement uncertainty is just given by the cosmic variance of the CMB~signal itself. In addition, we assume perfect component separation by neglecting any foreground emission, which therefore facilitates an ideal survey and analysis pipeline. This is in contrast to the realistic configurations, which are limited by a combination of instrumental noise and foregrounds, with a balance that depends on angular scale and on whether temperature or polarization dominates~(cf.\ Fig.~\ref{fig:ilc_residuals}). We explicitly consider two representative sky fractions:\footnote{Any other sky fraction can simply be deduced by a relative rescaling with~$\sqrt{\fsky}$, given the use of the Gaussian Fisher-matrix covariance~\eqref{eq:covariance} in our forecasts.} $\fsky = 0.62$, which is representative of the usable sky from the Chilean site after galactic masking as employed in the realistic configurations~(and is strictly speaking limited not by cosmic variance but rather by sample variance, which we denote by~SVL, since we only access part of the sky), and $\fsky = 1.0$ as an absolute upper bound. In both cases, we otherwise use the same forecasting framework as for the other configurations.

\section{From Maps to Constraints}
\label{sec:forecasting}

Translating the survey configurations of the previous section into projected constraints on cosmological parameters, such as~$\Neff$, requires a forecasting pipeline that captures the key elements of a realistic CMB~analysis. These include modeling and mitigating the contamination of the observations with astrophysical foregrounds, optimally extracting the CMB~signal from multi-frequency observations, and projecting the resulting sensitivity onto the cosmological parameters of interest. Our pipeline proceeds in several steps that we describe in this section. We first model the relevant galactic and extragalactic foreground emission, and we mask the most contaminated regions of the sky~(\textsection\ref{sec:forecasting_foregrounds}). We then apply an internal linear combination~(ILC) to optimally extract the CMB~signal from the multi-frequency data on the remaining sky~(\textsection\ref{sec:forecasting_ilc}). The resulting noise and residual-foreground spectra are then employed, together with delensed theoretical spectra, in a Fisher-matrix forecast to obtain projected constraints on~$\Neff$ and other cosmological parameters~(\textsection\ref{sec:forecasting_fisher}). We integrated these steps into the publicly available \texttt{DRAFT}~(Dark Radiation Anisotropy Flowdown Team)~tool, which we describe in~\textsection\ref{sec:forecasting_draft}. The results of applying this pipeline to the configurations of Section~\ref{sec:configurations} are presented in Section~\ref{sec:results}.

\subsection{Foreground Modeling and Masking}
\label{sec:forecasting_foregrounds}

Precise measurements of the CMB~damping tail at multipoles $\ell \gtrsim 1000$ are essential for constraining~$\Neff$, but this is also the regime where astrophysical foreground emission becomes increasingly significant relative to the primary CMB~signal. Extragalactic sources dominate at small angular scales, while galactic emission is the primary contaminant at large angular scales and near the galactic plane. The foreground contamination in polarization is comparatively smaller than in temperature at the angular scales and frequencies relevant for~$\Neff$. Modeling and mitigating these foregrounds through a combination of multi-frequency coverage, sky masking, and component separation is therefore a critical element of the forecasting pipeline since their residuals directly affect the projected~$\Neff$ sensitivity. In the following, we describe the adopted foreground models, which are generally decomposed into their galactic and extragalactic origins, and the sky mask to entirely remove the most contaminated regions from the forecasts.

\paragraph{Galactic Foregrounds}~\\
At large angular scales, the dominant galactic foreground signals at the frequencies relevant for~\sfour\ are thermal dust emission and synchrotron radiation. Thermal dust emission arises from interstellar dust grains heated by the interstellar radiation field and is well described by a modified blackbody spectrum. It rises toward higher frequencies and dominates above approximately~$\SI{100}{GHz}$, and is brightest toward the galactic plane where the dust column density is largest. Synchrotron radiation is emitted by cosmic-ray electrons spiraling in the galactic magnetic field and follows a power-law spectrum that falls steeply toward higher frequencies so that it dominates the lowest \sfour~bands. (We neglect the subdominant free-free and anomalous-microwave-emission components, which are small compared to dust and synchrotron at the frequencies and small angular scales relevant for~$\Neff$.) We model these two components with map-based simulations generated for the \sfour~LATs~\cite{CMB-S4:sims} using the publicly available SO~extensions of the Python Sky Model~3~(\texttt{PySM\,3})~\cite{Thorne:2016ifb, Zonca:2021row}, which extend the original simulations to higher angular resolution~\cite{PySM:docs}. These simulations use a spatially uniform spectral energy distribution~(``model~0'') for each component, i.e.\ the dust and synchrotron amplitude templates are those of the \texttt{PySM}~\texttt{d1} and~\texttt{s1} models, but their spectral parameters are held constant across the	sky, with dust temperature $T_d = \SI{19.6}{\kelvin}$, dust emissivity index $\beta_d = 1.53$, and synchrotron spectral index $\beta_s = -3.1$ based on \Planck~observations.

\begin{figure}
	\centering
	\includegraphics{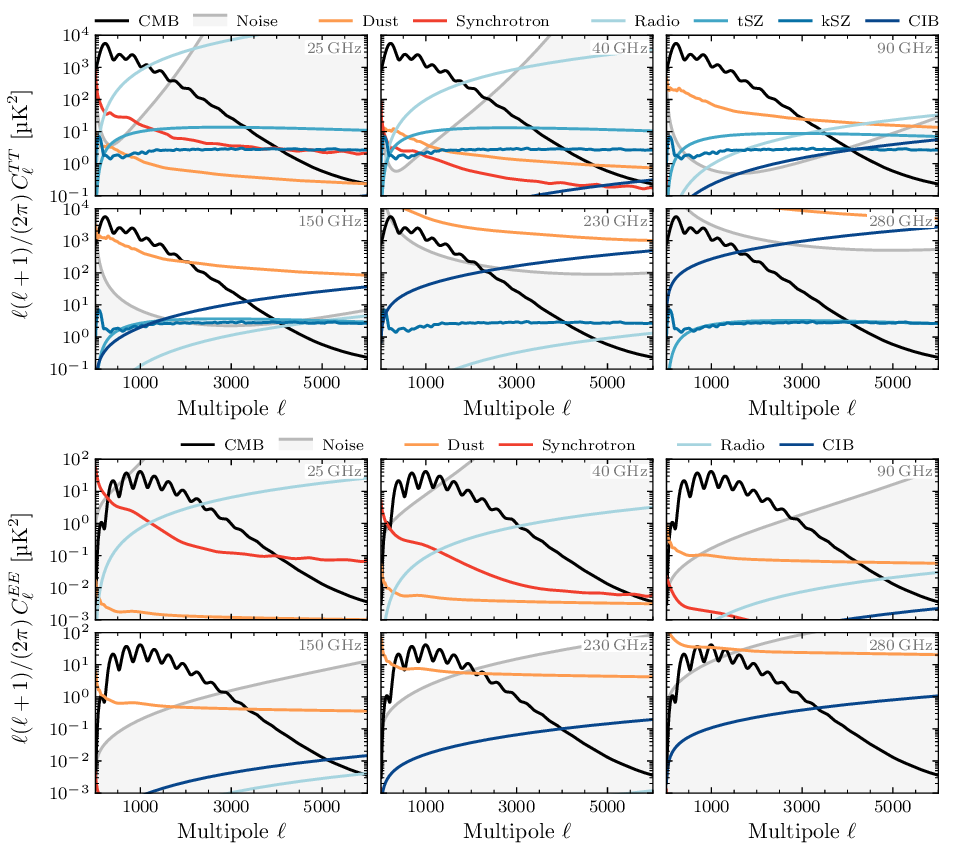}\vspace{-4pt}
	\caption{Auto-frequency power spectra of temperature anisotropies~(\textit{top}) and $E$-mode polarization~(\textit{bottom}) for the~CMB and the foreground components included in our forecasts, shown over the multipole range $\ell \in [10, 6000]$ for six \sfour~frequency bands. The CMB~signal is shown in black. The galactic foregrounds of thermal dust emission and synchrotron radiation are displayed in shades of red and derived from map-based \texttt{PySM\,3}~simulations. The extragalactic foregrounds, i.e.\ radio sources, thermal and kinematic Sunyaev-Zel'dovich emission, and the cosmic infrared background are included in shades of blue and follow the template-based parameterizations described in the main text. The~tSZ and kSZ~signals are taken to be unpolarized so that only the polarized contributions of radio and dusty galaxies appear in the $EE$~spectrum. The displayed foreground spectra are computed on the masked sky region observed with the \sfourwide~survey of the conceptual design~(see Fig.~\ref{fig:footprints}), for which \mbox{we also include the observational noise according to~\eqref{eq:noise_model} for comparison.}}\vspace{-2pt}
	\label{fig:foregrounds}
\end{figure}

From these simulations, we compute the galactic-dust and -synchrotron power spectra in the \sfour~frequency bands on the masked sky regions described below. Since these simulations do not include $TE$~contributions from galactic emission, we compute the correlation as $C_\ell^{TE} = \rho_{TE}\hskip1pt \sqrt{C_\ell^{TT} C_\ell^{EE}}$, with the correlation coefficient set to $\rho_{TE} = 0.35\text{ and }0$ for galactic dust emission and synchrotron radiation, respectively~\cite{Planck:2018gnk}. The resulting power spectra enter the multi-frequency covariance matrix used in the internal linear combination of~\textsection\ref{sec:forecasting_ilc}. We have tested this approach by further subdividing the survey area into multiple sub-regions and verified that the galactic foreground power within the analyzed sky region does not vary significantly between sub-regions~\cite{Raghunathan:2021zfi}. This confirms that a uniform foreground treatment within the analyzed footprint of each configuration is adequate for our $\Neff$~forecasts, which however assume the adopted model to be accurate and do not account for residual uncertainties in the foreground modeling itself. The resulting auto-frequency power spectra are shown in Fig.~\ref{fig:foregrounds}%
for both temperature and $E$-mode polarization, alongside the extragalactic contributions described next. (While we only display the auto-frequency spectra, the cross-frequency spectra are also employed in our pipeline.)

\paragraph{Extragalactic Foregrounds}~\\
At small angular scales, the dominant extragalactic foregrounds contributing to the observed power spectra are the thermal and kinematic Sunyaev-Zel'dovich effects~(tSZ and~kSZ), the cosmic infrared background~(CIB), and emission from radio galaxies. We model the auto- and cross-power spectra of these components across the instrument frequency bands using the parameterization adopted by~SPT~\cite{George:2014oba, SPT:2020psp}. Specifically, we use the template of~\cite{Shaw:2010mn} to describe both Sunyaev-Zel'dovich effects, with the tSZ~signal being sourced by inverse Compton scattering of CMB~photons off the hot electrons in the intracluster medium and the kSZ~effect by the Doppler shift due to the peculiar motion of galaxy clusters. The CIB~emission from dusty star-forming galaxies is decomposed into a Poisson contribution from individually unresolved sources and a spatially clustered contribution, with the frequency dependence of both described by a modified blackbody spectrum. The emission from radio galaxies, which dominates at lower frequencies, is modeled as a Poisson power spectrum whose power-law frequency scaling falls towards higher frequencies. We neglect a potential clustering contribution of these galaxies since there is currently no observational evidence for such a component in the frequency bands considered here~\cite{Hall:2009rv, SPT-3G:2026mcr}. Since we assume that the brightest individual point sources and galaxy clusters are detected and masked, the foreground power spectra that we model are those of the unresolved sources that remain. Rather than injecting these sources into simulated maps, we capture the impact of this masking directly at the power-spectrum level.

We adopt the best-fit foreground parameters inferred from SPT~measurements~\cite{George:2014oba}, which are broadly consistent with those from~ACT~\cite{Dunkley:2013vu, ACT:2020frw}. These parameters are obtained assuming a point-source flux threshold at~\SI{150}{\giga\hertz} of $S_{150} = \SI{6}{\milli\jansky}$, which is highly conservative for~\sfour, roughly corresponding to a detection significance of~$20\sigma$~\cite{Abazajian:2019eic}, and a galaxy-cluster detection threshold of~$5\sigma$, corresponding to a mass limit of $M_{500c} \approx \num{e14}\,M_{\odot}$~\cite{Raghunathan:2021tdc}. The amplitudes and frequency scalings of the various components are summarized in Table~\ref{tab:foregrounds_extragalactic}. %
\begin{table}
	\centering
	\setlength{\tabcolsep}{5.5pt}
	\begin{tabular}{l S[table-format=1.1] l l}
			\toprule
		Component		& {$D_{\ell = 3000}^{TT, 150}$~[\si{\muKelvin\squared}]}	& Multipole dependence			& Frequency scaling 											\\
			\midrule
		tSZ				& 3.4														& Template~\cite{Shaw:2010mn}	& Non-relativistic tSZ~spectrum									\\
		kSZ				& 3.0														& Template~\cite{Shaw:2010mn}	& CMB~blackbody spectrum										\\
		CIB (Poisson)	& 9.1														& $D_\ell \propto \ell^2$		& Modified blackbody: $\beta_\mathrm{CIB}^\mathrm{P} = 1.505$	\\
		CIB (clustered)	& 3.4														& $D_\ell \propto \ell^{0.8}$	& Modified blackbody: $\beta_\mathrm{CIB}^\mathrm{c} = 2.51$	\\
		Radio			& 1.0														& $D_\ell \propto \ell^2$		& Power law with spectral index~$-0.9$							\\
			\bottomrule
	\end{tabular}
	\caption{Characterization of the templates describing the extragalactic foregrounds of the thermal and kinematic Sunyaev-Zel'dovich effects, the Poisson and clustered components of the cosmic infrared background, and the emission from radio galaxies. The power spectrum of each component is characterized by the normalization of its \SI{150}{\giga\hertz}~auto-spectrum at $\ell = 3000$, $D_{\ell = 3000}^{TT, 150} \equiv \left.\ell(\ell+1)/(2\pi)\,C_\ell^{TT, 150}\right|_{\ell=3000}$, its multipole dependence, and its scaling to other frequency bands. The two CIB~components share the same temperature $T_\mathrm{CIB} = \SI{20}{\kelvin}$ but have different modified-blackbody emissivity indices~$\beta_\mathrm{CIB}^\mathrm{P}$ and~$\beta_\mathrm{CIB}^\mathrm{c}$ for the Poisson and clustered components, respectively. These amplitudes and scalings follow the parameterization based on measurements by the South Pole Telescope~\cite{George:2014oba, SPT:2020psp}.}
	\label{tab:foregrounds_extragalactic}
\end{table}
We ignore here the~\mbox{tSZ-CIB}~correlation for simplicity, even though it has been measured at the level of $3\text{ to }4\sigma$~\cite{Dunkley:2013vu, George:2014oba, Planck:2015emq, SPT:2020psp, SPTpol:2026xmb}, since the additional variance that it would introduce is negligible compared to that from the other foreground components.\footnote{We refer to~\cite{Raghunathan:inprep} for a more detailed consideration of this assumption and of the potential systematic uncertainties associated with the adopted foreground templates, which our forecasts otherwise take to be accurate.} Since these extragalactic foregrounds are only weakly polarized, their contribution is subdominant compared to the CMB~$E$-mode polarization signal. We treat the~tSZ~and kSZ~signals as unpolarized, and assign radio and dusty galaxies polarization fractions of~3\% and~2\%, respectively~\cite{Datta:2018oae, SPT:2019wyt}. The resulting template spectra are shown for all non-zero components of our extragalactic foreground model in Fig.~\ref{fig:foregrounds} and enter the multi-frequency covariance matrix used in the internal linear combination described in~\textsection\ref{sec:forecasting_ilc}.

\paragraph{Galaxy Masking}~\\
Since the \sfour~surveys from Chile observe up to about~68\% of the sky, portions of the footprint near the galactic plane are subject to significant galactic-foreground contamination. The multi-frequency ILC~method described in~\textsection\ref{sec:forecasting_ilc} can only partially clean these regions, since foreground removal at the small angular scales that drive the $\Neff$~constraints is limited by our knowledge of the foreground spectral and spatial properties. We therefore additionally mask the most contaminated regions of the sky to reduce the systematic risk from foreground residuals.\footnote{A quantitative investigation of foreground-induced biases and associated mitigation strategies is presented in~\cite{Raghunathan:inprep} based on the pipeline described here.}

We adopt a galactic-foreground mask derived by~\Planck\ for the two-site conceptual design and the Chile-only revised configuration. Specifically, we use the unapodized version of the \Planck~\texttt{GAL090}~mask~\cite{Planck:2015mrs}, which is inferred from the \SI{353}{GHz}~dust map, removes~10\% of the sky closest to the galactic plane, and is shown in Fig.~\ref{fig:footprints} cutting out the wide-survey footprint of the revised configuration. We apodize this mask with a Gaussian kernel of $\thetaFWHM = \SI{2}{\degree}$ to suppress ringing artifacts in the power-spectrum estimation. For the wide survey of the conceptual design of~\textsection\ref{sec:configurations_conceptual}, combining this mask with the survey footprint yields an effective sky fraction of $\fsky = 0.62$. We apply the same mask to the revised configuration of~\textsection\ref{sec:configurations_revised}, which combines the \sfour~Hybrid~surveys with the SO-like survey. The four observed patches of the \sfourhybridwide~survey~(Fig.~\ref{fig:footprints}) cover a total of~47\% of the sky before masking, which the \texttt{GAL090}~mask reduces to~41\%~($\fsky = 0.064$, $0.31$, $0.016$, and~$0.027$ for patches~1 to~4, respectively). Retaining only the lower-noise patches~1 and~2 of the wide and delensing footprints leaves a combined~44\% for the analyzed \sfour~Hybrid~sky. We identically mask the SO-like~observations since we assume a coordinated data analysis of the surveys. This results in a total usable sky fraction of $\fsky \approx 0.62$ for the analysis of the SO-like~LAT~data~(the official Simons Observatory~forecasts retain~40\% after masking~\cite{SimonsObservatory:2025wwn}) and the entire revised configuration, both before~\sfour\ begins observing and during the subsequent joint observations. In contrast, for the South-Pole-only alternative of~\textsection\ref{sec:configurations_pole}, we employ a simple latitude mask of~$\pm\SI{10}{\degree}$ rather than the \texttt{GAL090}~mask. This removes the most contaminated portion of the galactic plane and serves as a simplified footprint, retaining $\fsky = 0.20$ for the analysis.\footnote{The galactic foreground level in the observed patch is set by the combination of the minimum elevation angle and the latitude mask, with a cut of~$\pm\SI{10}{\degree}$ being somewhat more conservative than the \Planck~\texttt{GAL090} mask employed for the Chilean configurations. The same sky fraction $\fsky = 0.20$ could equivalently be reached with a different balance of these two choices, such as a larger minimum elevation and a correspondingly smaller galactic cut, which would in turn increase the galactic-dust power in the analyzed patch.}

\subsection{Internal Linear Combination}
\label{sec:forecasting_ilc}

The multi-frequency observations described in Section~\ref{sec:configurations} are combined into a single estimate of the~CMB signal using a harmonic-space internal linear combination~\cite{Tegmark:1995pn, Cardoso:2008com, Planck:2013fzg} following the implementation of~\cite{Raghunathan:2023yfe}. The~ILC exploits the known frequency independence of the~CMB~signal~(in thermodynamic temperature units), together with the distinct frequency dependence of the foregrounds, to construct an optimal estimate by linearly combining the maps from all available frequency channels, while down-weighting modes that are dominated by foreground emission or instrumental noise. The framework also supports three variants referred to as constrained~ILC~\cite{Remazeilles:2010hq}, partial~ILC~\cite{SPT-SZ:2021gsa}, and cross-ILC~\cite{Raghunathan:2023yfe}. In the constrained~ILC, the weights are additionally tuned to null the frequency response of one or more sky components. Partial~ILC suppresses the targeted component by artificially rescaling the covariance of a specific component and, therefore, reduces the contribution of that component without imposing an explicit nulling constraint. This is particularly useful when the frequency response of an unwanted component is not known a priori or is complicated enough that it cannot be captured by a single spectral energy distribution, as is the case for the cosmic infrared background. By contrast, cross-ILC constructs two separate constrained-ILC maps that null different sky components and estimates the CMB~power spectrum from their cross-correlation, simultaneously suppressing the residuals of all targeted components in the resulting cross-spectrum. The forecasts presented in this paper use the standard minimum-variance~(MV) form of~ILC, which we now describe alongside the generalizations, since all variants are supported by the \texttt{DRAFT}~tool~(cf.~\textsection\ref{sec:forecasting_draft}).\medskip

In harmonic space, the ILC~estimate of the CMB~signal at multipole~$\ell$ is given by
\begin{equation}
	S_\ell = \sum_{i=1}^{N_f} w_\ell^i\, M_\ell^i\, ,
\end{equation}
where~$M_\ell^i$ is the harmonic-space representation of the observed map in frequency channel~$i$ and the sum runs over all~$N_f$ frequency channels. The multipole-dependent weights~$w_\ell^i$ are determined by minimizing the total variance of~$S_\ell$ subject to the constraint that the~CMB signal is preserved, with the option to additionally null or suppress the frequency response of one or more other sky components. We collect the spectral energy distributions of all components included in the solution into a matrix $\mathcal{F} = [A_s, B_{f_1}, C_{f_2}, \ldots]_{N_f \times N_c}$, where~$N_c$~is the number of components. The first column~$A_s$ is the target signal, and~$B_{f_1}, C_{f_2}, \ldots$ are the $N_c - 1$~undesired (foreground)~components. For the purposes of this work, the target signal is the~CMB, and we therefore set $A_s = A_\mathrm{CMB} = [1, 1, \ldots, 1]$ in the following for all frequency bands. The constrained-ILC weights then take the form
\begin{equation}
	w_\ell = \mathbf{C}_\ell^{-1} \mathcal{F} \left( \mathcal{F}^\mathrm{T} \mathbf{C}_\ell^{-1} \mathcal{F} \right)^{\!-1} U\, ,	\label{eq:ilc_weights}
\end{equation}
where the constraint vector $U = [1, 0, \ldots, 0]$ has length~$N_c$ and selects the CMB~signal~$A_\mathrm{CMB}$ as the preserved component. The $N_f \times N_f$~covariance matrix of the noise and foreground power at multipole~$\ell$ is denoted by~$\mathbf{C}_\ell$. It is constructed from the auto- and cross-frequency power spectra of the instrumental noise~\eqref{eq:noise_model}, and of the galactic and extragalactic foreground components described in~\textsection\ref{sec:forecasting_foregrounds}, after computing them on the masked sky.

In the limit where no additional components are nulled, $\mathcal{F}$~reduces to the $N_f \times 1$~frequency response vector~$A_\mathrm{CMB}$ of the CMB~signal, and the expression~\eqref{eq:ilc_weights} simplifies to the standard~MV-ILC~weights,
\begin{equation}
	w_\ell^\mathrm{MV} = \frac{\mathbf{C}_\ell^{-1}\hskip-1pt A_\mathrm{CMB}}{A_\mathrm{CMB}^\mathrm{T}\hskip1pt \mathbf{C}_\ell^{-1}\hskip-1pt A_\mathrm{CMB}}\, .	\label{eq:ilc_weights_mv}
\end{equation}
This~MV~form is the one used for the forecasts presented in this work. The corresponding residual power spectrum after component separation is given by $C_{\ell}^\mathrm{res} = (A_\mathrm{CMB}^\mathrm{T}\hskip1pt \mathbf{C}_\ell^{-1}\hskip-1pt A_\mathrm{CMB})^{-1}$, which represents the effective noise that includes residuals from both the experimental noise and the foreground signals in the component-separated CMB~map. We compute~$C_\ell^\mathrm{res}$ for each survey configuration of Section~\ref{sec:configurations} to get the \sfour~noise spectra, which we then sum with the effective~(frequency-combined) instrumental noise from the \Planck~satellite using inverse-variance weighting as described in~\textsection\ref{sec:configurations_common}, i.e.\ we use the inverse-variance combination of the individual~\Planck~channels for simplicity, rather than including the~\Planck~bands directly in the~ILC.\footnote{This is a very good approximation for our purposes: \Planck~has substantially higher noise and coarser angular resolution than the ground-based surveys at the small scales that drive the~$\Neff$~sensitivity, which means that a joint~ILC would leave the high-$\ell$ residuals essentially unchanged, while the inverse-variance combination captures the complementary information from~\Planck\ at the large scales.} These combined spectra are those that enter the covariance~\eqref{eq:covariance} of the Fisher-matrix forecast described in~\textsection\ref{sec:forecasting_fisher} as the post-ILC noise term~$C_\ell^{\mathrm{res}, XY}$ for $XY = TT, TE, EE$.\medskip

\begin{figure}
	\centering
	\includegraphics{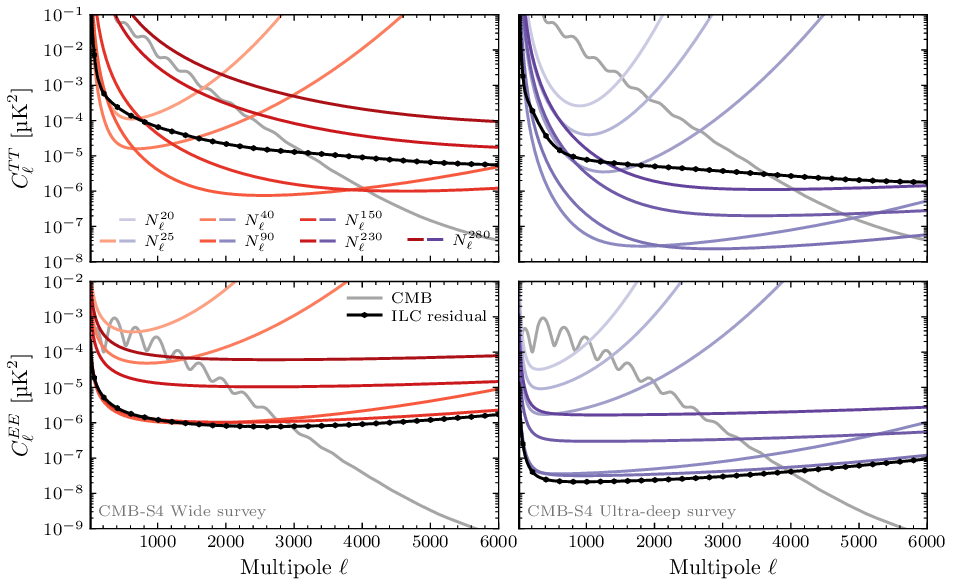}
	\caption{Effective noise and foreground residuals~(black curves) in the minimum-variance internal-linear-combination~map for the \sfourwide~(\textit{left}, shades of red) and \sfourdeep~(\textit{right}, shades of purple) surveys of the conceptual design observing from the Atacama Desert and the South Pole, respectively. The top and bottom panels are for the temperature~($TT$) and polarization~($EE$) auto-spectra, with their primary CMB~signal displayed in gray. The colored curves show the noise spectra of each individual frequency band, shaded from light to dark with increasing frequency and with the \SI{20}{GHz}~channel being only available for the ultra-deep survey. In the absence of foregrounds, the ILC~residuals reduce to the inverse-variance combination of the individual-channel noise curves, which is approximately realized in polarization. In temperature, the impact of the foregrounds is clearly visible and somewhat greater for the wide than the deep survey due to the different exposure to galactic foregrounds. The ILC~residuals are used as the input noise spectra in our Fisher~forecasts for the conceptual design. The left and right panels are therefore generally representative of the other surveys located in Chile and Antarctica, respectively.}
	\label{fig:ilc_residuals}
\end{figure}

We show the minimum-variance ILC~residuals for the temperature and polarization spectra in Fig.~\ref{fig:ilc_residuals}%
for the two high-resolution surveys of the two-site conceptual design, with the Chilean \sfourwide~survey in the left panels and the South-Pole-based \sfourdeep~survey in the right panels. These results are therefore illustrative of the expected foreground residuals for the configurations observing either a wide and partially galaxy-contaminated sky area from Chile or a smaller and cleaner patch from the South Pole. For reference, the beam-deconvolved noise curves of the individual frequency channels and the primary-CMB~signal are also displayed. In temperature, the foreground residuals are clearly visible at high multipoles, in particular for the wide survey where the galactic contamination is more significant. In polarization, the ILC~residuals closely track the inverse-variance-weighted combination of the individual-channel noise curves, which reflects the comparatively small foreground contamination in polarization at these angular scales and frequencies. As expected, we can clearly see the lower noise levels of the deep survey compared to the wide survey, which however observes over a much larger sky area. This difference in noise levels is most visible in polarization, where the comparative lack of foreground contamination leaves the spectra signal-dominated up to $\ell \approx 3000\text{ and }4000$ for the wide and deep surveys, respectively. In temperature, the signal-to-noise ratio exceeds unity only up to $\ell \approx 3200\text{ and }3600$, despite the much larger primary signal.

\subsection{Fisher Information}
\label{sec:forecasting_fisher}

We employ standard Fisher information theory to forecast the expected precision on~$\Neff$ and the other cosmological parameters for each survey configuration. The Fisher matrix quantifies the information content of the observed power spectra as a function of the model parameters, and it provides a lower bound on the achievable parameter uncertainties through the Cramér-Rao~inequality. While Fisher forecasts have their limitations and should be used with care, they are a useful and computationally efficient tool to provide guidance for the sensitivity and design of cosmological surveys.\medskip

The Fisher-matrix elements~$F_{\alpha\beta}$ for cosmological parameters~$p_\alpha$ and~$p_\beta$ are computed from the derivatives of the~(theoretical) CMB~power spectra~$C_\ell^{XY}$ with respect to these parameters, and from the covariance~$\mathrm{Cov}_{\ell_1\ell_2}^{XY, WZ}$ between power spectrum measurements,
\begin{equation}
	F_{\alpha\beta} = \sum_{\ell_1, \ell_2} \, \sum_{XY, WZ} \frac{\partial C_{\ell_1}^{XY}}{\partial p_\alpha} \left[ \mathrm{Cov}_{\ell_1\ell_2}^{XY, WZ} \right]^{-1} \frac{\partial C_{\ell_2}^{WZ}}{\partial p_\beta}\, ,
\end{equation}
where the sums over~XY and~WZ run over the CMB~temperature and $E$-mode polarization auto- and cross-spectra~($TT$, $TE$, $EE$) as well as the lensing-deflection power spectrum $C_\ell^{dd} = \ell(\ell+1)\hskip1pt C_\ell^{\phi\phi}$, with the power spectrum of the gravitational potential~$C_\ell^{\phi\phi}$. The fiducial cosmology is taken to be $\Lambda\mathrm{CDM} + \Neff$,\footnote{We fix the sum of neutrino masses to its fiducial value rather than marginalizing over it since~$\Neff$ and~$\sum m_\nu$ are essentially independent at \sfour~precision. We quantify the small effect of additionally varying~$\sum m_\nu$ in~\textsection\ref{sec:results_cvl}.} with the parameter values and step sizes for the numerical derivatives listed in Table~\ref{tab:parameters_fiducial}. %
\begin{table}
	\centering
	\begin{tabular}{l l S[table-format=1.6] S[table-format=1.1e-1]}
			\toprule
		Parameter				& Description																			& {Fiducial Value}	& {Step Size}	\\
			\midrule[0.065em]
		$\omega_\mathrm{b}$		& Physical density of baryons $\omega_\mathrm{b} \equiv \Omega_\mathrm{b} h^2$			& 0.0222			& 8.0e-4		\\
		$\omega_\mathrm{c}$		& Physical density of cold dark matter $\omega_\mathrm{c} \equiv \Omega_\mathrm{c} h^2$	& 0.1197			& 3.0e-3		\\
		$\theta_s$				& Angular size of the sound horizon at decoupling										& 0.010409			& 5.0e-5		\\
		$\num{e9} A_\mathrm{s}$	& Primordial scalar amplitude~(at~$k_\star$)											& 2.196				& 1.0e-1		\\
		$n_\mathrm{s}$			& Primordial scalar spectral index~(at~$k_\star$)										& 0.9655			& 1.0e-2		\\
		$\tau$					& Thomson optical depth due to reionization												& 0.060				& 2.0e-2		\\
		$\Neff$					& Effective number of relativistic species												& 3.044				& 8.0e-2		\\
			\midrule[0.03em]
		$Y_\mathrm{p}$			& Primordial helium abundance															& {`BBN'~~~~}		& 4.8e-3		\\
		$\sum m_\nu$			& Sum of neutrino masses~[\si{eV}]														& 0.06				& 2.0e-2		\\
			\bottomrule
	\end{tabular}
	\caption{Parameter values and step sizes of the numerical derivatives in the Fisher-matrix forecasts for the fiducial $\Lambda\mathrm{CDM} + \Neff$ cosmological model, following~\cite{Allison:2015qca}. The first seven rows indicate the parameters that are varied in all forecasts throughout this work, with the pivot scale for the primordial fluctuations being $k_\star = \SI{0.05}{\per\mega\parsec}$. The primordial helium abundance~$Y_\mathrm{p}$ is set to its BBN-consistent value as a function of~$\omega_\mathrm{b}$ and~$\Neff$. This parameter is not varied independently in our forecasts unless stated otherwise, in which case its fiducial value is $Y_\mathrm{p} = 0.2467$. The sum of neutrino masses is similarly fixed to $\sum m_\nu = \SI{0.06}{eV}$ and only varied when explicitly specified.}
	\label{tab:parameters_fiducial}
\end{table}
The theoretical power spectra~$C_\ell^{XY}$ are computed using~\texttt{CLASS\_delens}~\cite{Hotinli:2021umk, Hotinli:delens}, and their derivatives~$\partial C_\ell^{XY}/\partial p_\alpha$ are calculated via first-order central finite differences. We treat the different surveys and sky patches of the various configurations as independent experiments so that we simply sum their individual Fisher matrices. The marginalized $1\sigma$~uncertainty on any individual parameter~$p_\alpha$ is then obtained from the inverse of the total Fisher matrix as $\sigma(p_\alpha) = \sqrt{(F^{-1})_{\alpha\alpha}}$, satisfying the Cramér-Rao~bound.

We assume a covariance that is diagonal in multipole space and Gaussian for the component-separated CMB~spectra, given
at each multipole~$\ell$ by
\begin{equation}
	\mathrm{Cov}_{\ell_1\ell_2}^{XY, WZ} = \frac{\delta_{\ell_1\ell_2}}{(2\ell_1 + 1) \fsky} \left( \tilde{C}_{\ell_1}^{XY} \tilde{C}_{\ell_1}^{WZ} + \tilde{C}_{\ell_1}^\mathrm{XW} \tilde{C}_{\ell_1}^\mathrm{YZ} \right) ,	\label{eq:covariance}
\end{equation}
where $\tilde{C}_\ell^{XY} = C_\ell^{XY} + C_\ell^{\mathrm{res}, XY}$ are the expected CMB~power spectra as observed after component separation with the effective post-ILC noise power spectra~$C_\ell^{\mathrm{res}, XY}$ from~\textsection\ref{sec:forecasting_ilc}. The sky fraction~$\fsky$ accounts for the reduction in the maximum number of independent modes~$(2\ell_1 + 1)$ due to the finite sky coverage. Weak gravitational lensing of the~CMB by intervening large-scale structure between the last-scattering surface and our detectors generates off-diagonal non-Gaussian covariances between $\ell$-modes that would otherwise be uncoupled on the full sky. At the noise levels of~\sfour\ and other future experiments, these lensing-induced covariance terms have been shown to affect the constraints of parameters that are best constrained through lensing, such as the sum of neutrino masses, at the level of~10\%~\cite{Green:2016cjr, Trendafilova:2023oni}. Since the impact on forecasted~$\Neff$~constraints is negligible for the considered range of experimental configurations, we however do not include these additional terms in our forecasts.

We generally use a multipole range of $2 \leq \ell \leq 5000$ in our forecasts unless otherwise noted. For the realistic configurations, the ground-based~$TT$, $TE$, and~$EE$~spectra span $30 \leq \ell \leq 5000$, while the minimum multipole is~$2$ for the lensing-deflection spectrum~$C_\ell^{dd}$. On large scales, we add primary \Planck~information by including the $TT$~spectrum for $\ell \in [2, 29]$ and the optical-depth prior from polarization, as introduced in~\textsection\ref{sec:configurations_common}. The lower bound of~$\ell = 30$ for the ground-based spectra avoids the largest angular scales that may particularly be affected by systematic effects in ground-based experiments and contribute negligibly to the~$\Neff$~constraint, while these scales are instead captured by~\Planck. The upper bound extends far into the damping tail, from where the sensitivity to~$\Neff$ originates, and well beyond the signal-dominated multipoles of a \sfour-like experiment~(cf.~Fig.~\ref{fig:ilc_residuals}). We adopt the same upper multipole in temperature and polarization: although the temperature foregrounds are more significant, their residuals after component separation are included in the effective post-ILC noise~(\textsection\ref{sec:forecasting_ilc}), so that the small-scale temperature data are appropriately down-weighted rather than discarded, under the assumption that the foreground modeling and component separation remain reliable to this scale. For the variance-limited surveys, we retain the multipole range $2 \leq \ell \leq 5000$ to facilitate a comparison over the same scales.\footnote{Proposed experiments such as CMB-HD~\cite{CMB-HD:2022bsz} are designed to reach significantly higher multipoles, for which additional cosmological information is in principle accessible, but where secondary anisotropies and foregrounds become substantially more challenging to separate from the primary signal.} In this case, we include the~$TT$, $TE$, $EE$, and lensing-deflection spectra down to~$\ell = 2$, while omitting the~$\tau$~prior, since the large-scale polarization is then itself limited by cosmic variance.\medskip

Gravitational lensing of the~CMB not only correlates nearby multipoles, but it also broadens the acoustic peaks and transfers power from large to small angular scales, partially obscuring the primary anisotropy signal, which in particular carries the information about~$\Neff$. The~(partial) removal of this lensing effect, which is referred to as delensing, sharpens the acoustic peaks, makes the damping tail more dominant again, and restores some of the primary signal. This has been shown to noticeably improve parameter constraints for surveys with map depths at the level of~\sfour\ and beyond~\cite{Baumann:2015rya, Green:2016cjr, Ange:2023ygk, Trendafilova:2023xtq, Montefalcone:2025unv, Hotinli:2021umk}. All forecast results presented in this work are inferred from delensed~$TT$, $TE$, and $EE$~spectra computed with~\texttt{CLASS\_delens}~\cite{Hotinli:2021umk, Hotinli:delens}, a modification of the Boltzmann code~\texttt{CLASS}~\cite{Blas:2011rf} that implements delensing on the curved sky to all orders in the lensing potential. It iteratively reconstructs the lensing potential and self-consistently computes the delensed spectra and the lensing-reconstruction noise~$N_\ell^{dd}$ following~\cite{Green:2016cjr}. For this reconstruction, we include all multipoles $\ell \geq 2$, but restrict them in the $TT$~spectrum to $\ell_\mathrm{max}^T = 3000$ to mitigate potential lensing-reconstruction biases from non-Gaussian astrophysical foregrounds~\cite{vanEngelen:2013rla}.\medskip

The Fisher-matrix forecasts are performed using the publicly available \texttt{FisherLens}~code~\cite{Hotinli:2021umk, Ryan:2022qpa, Trendafilova:fisher}, which wraps~\texttt{CLASS\_delens} to compute the power-spectrum derivatives, covariance matrices, and resulting CMB~Fisher matrix for a given experimental configuration. In addition to the baseline forecasting pipeline, \texttt{FisherLens}~also allows us to check the validity of the assumption underlying the Fisher methodology of a Gaussian likelihood using the~DALI~(Derivative Approximation for LIkelihoods) method following~\cite{Sellentin:2014zta, Ryan:2022qpa}. Using this procedure, it was shown in~\cite{Ryan:2022qpa} that the Gaussian Fisher estimates are reliable for our fiducial~$\Lambda\mathrm{CDM} + \Neff$ cosmology. Moreover, the code also includes the capability to estimate biases in the recovered cosmological parameters due to residual foreground contamination~(or other systematic effects), using the extended Fisher formalism of~\cite{Huterer:2004tr, Loverde:2006cj, Amara:2007as}. This provides a quantitative assessment of the systematic risk from imperfect foreground modeling, which we use in~\cite{Raghunathan:inprep} for a detailed investigation of these foreground-induced biases and associated mitigation strategies. The combined functionality of~\texttt{CLASS\_delens} and~\texttt{FisherLens} has been incorporated into the \texttt{DRAFT}~pipeline, which we describe next.

\subsection{DRAFT Tool}
\label{sec:forecasting_draft}

\begin{figure}
	\centering
	\includegraphics{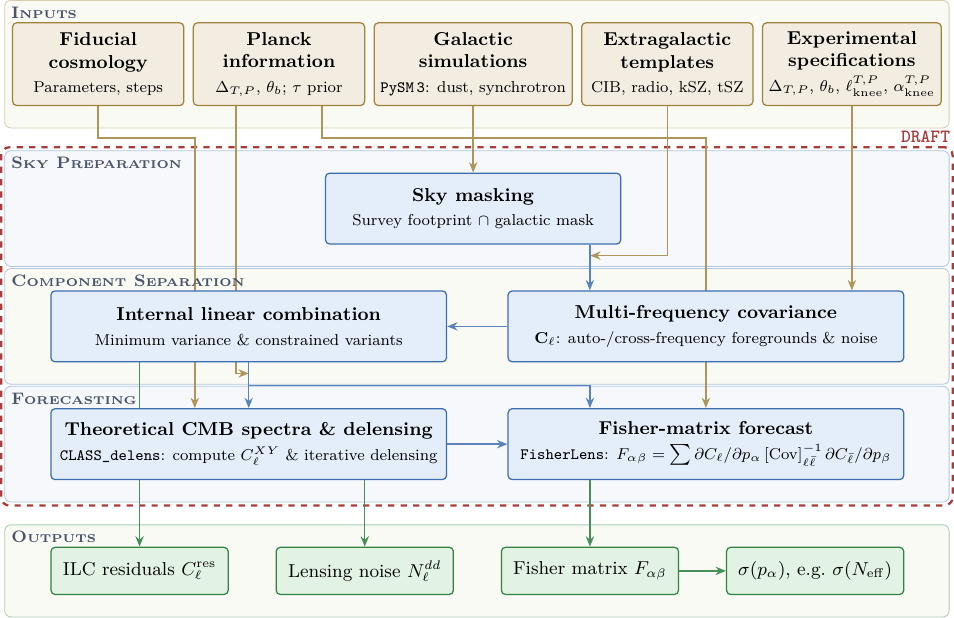}
	\caption{Schematic overview of the \texttt{DRAFT}~tool and the pipeline it implements, from the instrumental and astrophysical inputs to the calculated outputs, especially the projected sensitivity to cosmological parameters. The inputs consist of the fiducial cosmology with the parameter step sizes for the numerical derivatives, the \Planck~data entering through inverse-variance combination of its noise after component separation and the Gaussian $\tau$~prior~(\textsection\ref{sec:configurations_common}), the foreground models from galactic \texttt{PySM\,3}~simulations and extragalactic templates~(\textsection\ref{sec:forecasting_foregrounds}), and the experimental specifications of the considered survey configurations~(Section~\ref{sec:configurations}). The pipeline successively performs sky masking with a galactic mask, such as \Planck~\texttt{GAL090}~(\textsection\ref{sec:forecasting_foregrounds}), the assembly of the multi-frequency covariance~$\mathbf{C}_\ell$ of the noise and foreground spectra, internal-linear-combination component separation~(\textsection\ref{sec:forecasting_ilc}), the iterative delensing performed by~\texttt{CLASS\_delens} on top of its theoretical CMB~spectra, and the Fisher-matrix forecast performed with~\texttt{FisherLens}~(\textsection\ref{sec:forecasting_fisher}). The post-ILC noise spectra are used both in the iterative delensing and in the Fisher covariance, while the delensed spectra and lensing-reconstruction noise are passed onward to the Fisher analysis. The intermediate ILC~residual spectra~$C_\ell^\mathrm{res}$ for a range of survey configurations are publicly released as part of the \texttt{DRAFT}~tool and can therefore also be used as inputs to other forecasting frameworks. The end products are the cosmological Fisher matrix~$F_{\alpha\beta}$ and the projected $1\sigma$~uncertainties~$\sigma(p_\alpha)$ on the cosmological parameters~$p_\alpha$, including~$\sigmaNeff$.}
	\label{fig:draft}
\end{figure}

The forecasting methodology described in~\textsection\ref{sec:forecasting_foregrounds}, \textsection\ref{sec:forecasting_ilc}, and~\textsection\ref{sec:forecasting_fisher} has been integrated into the publicly available \texttt{DRAFT}~tool~\cite{Raghunathan:draft}, which was developed within the Maps-to-Power-Spectra analysis working group of the \sfour~collaboration. The tool provides an end-to-end forecasting pipeline from experimental specifications to projected parameter constraints, with all of the intermediate steps described above, i.e.\ foreground modeling, sky masking, ILC-based component separation, delensing, and Fisher-matrix calculation. We illustrate the pipeline that is implemented by the \texttt{DRAFT}~tool in Fig.~\ref{fig:draft}, %
and summarize its structure and usage in the following.\medskip

The pipeline is organized around a set of modules. The experimental specifications for each survey configuration are stored in a dedicated experiment library that currently includes the \sfour~surveys described in Section~\ref{sec:configurations} as well as various configurations for~\Planck, the South Pole Telescope, the Simons Observatory, and~\mbox{CMB-HD}. For observations spanning a different duration~$t$ than the respective baseline $t_\mathrm{b}$, the white noise levels are rescaled as~$(t/t_\mathrm{b})^{-1/2}$. For combined configurations such as the revised \sfour~design~(\textsection\ref{sec:configurations_revised}), the noise spectra of the contributing surveys are combined via inverse-variance weighting at the frequency-channel level before the~ILC is performed. The extragalactic foreground model is implemented as a library of template power spectra and frequency-scaling functions as described in~\textsection\ref{sec:forecasting_foregrounds}, with all inputs configurable through a parameter file. The galactic foreground power spectra are read from precomputed files derived from the \texttt{PySM\,3}~simulations on different masked sky regions, as also detailed in~\textsection\ref{sec:forecasting_foregrounds}. The multi-frequency covariance matrix~$\mathbf{C}_\ell$ of~\eqref{eq:ilc_weights_mv} is then assembled at each multipole from the noise and foreground auto- and cross-frequency power spectra, and the ILC~weights and residual noise spectra are computed as described in~\textsection\ref{sec:forecasting_ilc}, with the tool supporting the standard minimum-variance~ILC, constrained~ILC, partial~ILC, and cross-ILC, following~\cite{Raghunathan:2023yfe}. For the forecasts presented in the next section, we only employ the \mbox{MV-ILC}~setup and do not explicitly null any foreground components. This choice is somewhat optimistic because it assumes that the shapes of the foreground signals are exactly known, which can lead to biased estimates of cosmological parameters. Since these parameters are however primarily constrained by the~$TE$ and~$EE$~spectra, which are less affected by the largely unpolarized foreground signals, this assumption is acceptable for the purposes of this work~(see Fig.~B3 of~\cite{Raghunathan:2023yfe}). We refer to~\cite{Raghunathan:inprep} for a detailed examination of foreground-induced biases for the survey configurations considered~here.

The resulting ILC~residual spectra are then passed to \texttt{FisherLens}, which uses \texttt{CLASS\_delens} for the calculation of the theoretical CMB~power spectra and their internal, iterative delensing, to calculate the standard Fisher matrix in a given cosmological model as described in~\textsection\ref{sec:forecasting_fisher}. The pipeline outputs the projected $1\sigma$~uncertainties on all varied cosmological parameters, including~$\sigmaNeff$, for a given survey configuration, sky fraction, and observation time. By scanning over these inputs, \texttt{DRAFT}~can therefore be used to explore the dependence of the forecasted constraints on survey-design parameters, cosmology, and other forecasting choices. The computed ILC~residual and lensing-reconstruction noise curves for the \sfour~surveys of Section~\ref{sec:configurations}, as well as various configurations of~\mbox{SPT-3G} and~SO, are distributed as part of the public repository. While~\texttt{DRAFT} provides an end-to-end pipeline, these precomputed noise curves therefore also allow for forecasting with other tools, including other Fisher codes or MCMC-based approaches, without having to rerun~\texttt{DRAFT} for these particular experimental setups.

\section{Forecasted Sensitivity to~\texorpdfstring{$\mathbf{N}_\mathbf{eff}$}{Neff}}
\label{sec:results}

\begin{figure}
	\centering
	\includegraphics{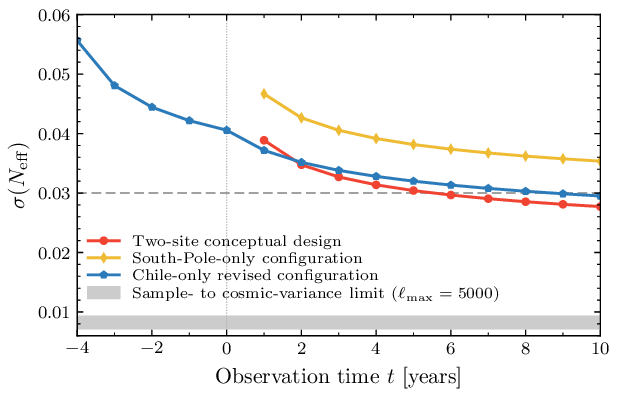}
	\caption{Forecasted $1\sigma$~uncertainty on~$\Neff$ as a function of observation time~$t$ for the \sfour~configurations considered in this work as described in Section~\ref{sec:configurations}. We display the sensitivity of the two-site conceptual design~(red), the South-Pole-only alternative~(orange), and the Chile-only revised configuration~(blue) as obtained using the \texttt{DRAFT}~tool detailed in Section~\ref{sec:forecasting}. The origin $t = 0$ highlighted by the dotted line corresponds to the start of \sfour~operations, with the first forecasting point coming after the first year. For the revised configuration, the curve extends to negative times to cover the preceding observing period of an experiment like the Simons Observatory~(cf.~\textsection\ref{sec:results_revised}; \mbox{\SIrange{-4}{0}{years}} corresponding to calendar years~2029--2033 for a nominal~2033 start of~\sfour). The horizontal dashed line indicates the original \sfour~science target $\sigmaNeff = 0.030$, and the horizontal gray band shows the cosmic-variance-limited floors for multipoles $\ell \leq 5000$ and a sky coverage between $\fsky = 0.62\text{ and }1.0$.}
	\label{fig:results_summary}
\end{figure}

We now present the projected sensitivity, obtained from the forecasting pipeline of the preceding section, of the survey configurations of Section~\ref{sec:configurations} to the effective number~$\Neff$ of relativistic species. The two-site conceptual design reaches $\sigmaNeff = 0.029$ over its seven-year nominal observing period, which therefore meets the original \sfour~science target of $\sigmaNeff = 0.030$~\cite{CMB-S4:2022ght}. The revised configuration with telescopes only located in Chile reaches this threshold after nine years of \sfour~observations through a combined analysis with SO-like~LAT~data, while a cosmic-variance-limited benchmark over the same multipole range sets an ultimate floor of $\sigmaNeff = 0.0073$~[or $\sigmaNeff = 0.0092$ over a sky fraction comparable to the realistic configurations]. We summarize the main forecasts for all configurations in Fig.~\ref{fig:results_summary}%
and Table~\ref{tab:results_summary}, %
\begin{table}
	\centering
	\begin{tabular}{l S[table-format=1.2] S[table-format=2.0] S[table-format=2.4]}
			\toprule
		Configuration 									& {Sky fraction~$\fsky$}	& {Time [\si{yrs}]}	& {$\sigmaNeff$}	\\
			\midrule[0.065em]
		\multicolumn{4}{l}{\textit{Two-site conceptual design~(\textsection\ref{sec:results_conceptual})}}					\\[2pt]
		\quad \sfourwide\ alone							& 0.62						& 7					& 0.031				\\
		\quad \sfourwide\ (without Ultra-deep overlap)	& 0.59						& 7					& 0.032				\\
		\quad \sfourdeep\ alone							& 0.03						& 7					& 0.083				\\
		\quad \sfourwide\ + \sfourdeep					& 0.62						& 7					& 0.029				\\
		\quad \sfourwide\ + \sfourdeep					& 0.62						& 10				& 0.028				\\
			\midrule[0.03em]
		\multicolumn{4}{l}{\textit{South-Pole-only alternative~(\textsection\ref{sec:results_pole})}}						\\[2pt]
		\quad \sfour~South Pole only					& 0.20						& 7					& 0.037				\\
		\quad \sfour~South Pole only					& 0.20						& 10				& 0.035				\\
			\midrule[0.03em]
		\multicolumn{4}{l}{\textit{Chile-only revised configuration~(\textsection\ref{sec:results_revised})}}				\\[2pt]
		\quad \sfour~Hybrid + SO-like LAT				& 0.62						& 7					& 0.031				\\
		\quad \sfour~Hybrid + SO-like LAT				& 0.62						& 9					& 0.030				\\
		\quad \sfour~Hybrid + SO-like LAT				& 0.62						& 10				& 0.030				\\
			\midrule[0.03em]
		\multicolumn{4}{l}{\textit{Cosmic-variance-limited survey~(\textsection\ref{sec:results_cvl})}}						\\[2pt]
		\quad SVL										& 0.62						& {--}				& 0.0092			\\
		\quad CVL										& 1.00						& {--}				& 0.0073			\\
			\bottomrule
	\end{tabular}
	\caption{Forecasted $1\sigma$~uncertainties on~$\Neff$ for the \sfour~configurations considered in this work and the cosmic-variance-limited survey as described in Section~\ref{sec:configurations}. These sensitivities were obtained using the \texttt{DRAFT}~tool with the foreground modeling, component separation, delensing, and Fisher information as explained in Section~\ref{sec:forecasting}, up to a common maximum multipole of $\ell_\mathrm{max} = 5000$. We display the projected constraints for several survey combinations for each configuration, and highlight the respective sky fraction and the observation time of the \sfour~surveys. For the two-site conceptual design, the second row removes the \SI{3}{\percent} of the~\sfourwide\ footprint shared with the ultra-deep survey to avoid double counting and the fourth row uses the non-overlapping~\sfourwide~sky together with the full \sfourdeep~area. For the Chile-only revised configuration, we assume that a large-aperture telescope like the one planned for the Simons Observatory is operating before~\sfour\ starts observations in~2033~(cf.~Fig.~\ref{fig:results_summary}). To highlight the potential future reach of CMB~experiments over the same multipoles, we also forecast the sample-variance-limited~(SVL, $\fsky = 0.62$) and \mbox{cosmic-variance-limited~(CVL, $\fsky = 1.0$) surveys.}}
	\label{tab:results_summary}
\end{table}
and provide additional details and projections in the following.\medskip

We present a survey-design optimization in~\textsection\ref{sec:results_optimization}, which in particular quantifies the competition between sky coverage and noise level. In addition, it motivated the specific configurations proposed and studied for~\sfour. We then discuss the forecasts for the conceptual design with its Chilean and South-Pole sites in~\textsection\ref{sec:results_conceptual}, which is the configuration that meets the mentioned science target in the fewest years of observation. In~\textsection\ref{sec:results_pole}, we turn to the South-Pole-only alternative, which was considered in the analysis of alternatives~\cite{CMB-S4:2023aoa} and requires substantially greater observational effort to reach the same precision on~$\Neff$. The Chile-only revised configuration based on a joint effort of the~\sfour\ and SO-like large-aperture telescopes with observations entirely from the Atacama Desert is discussed in~\textsection\ref{sec:results_revised}. Finally, in~\textsection\ref{sec:results_cvl}, we quantify the future prospects of the CMB~primary anisotropies up to similar angular scales through a cosmic-variance-limited survey and indicate how much room remains for improvement beyond a \sfour-type survey.

\subsection{Survey-Design Optimization}
\label{sec:results_optimization}

Two factors compete in setting the sensitivity of a CMB~survey to~$\Neff$, as introduced in~\textsection\ref{sec:configurations_common}: the effective noise per mode, which controls how far into the damping tail the measurement extends, and the observed sky fraction~$\fsky$, which determines the number of modes available in a survey. In the following, we quantify this competition and establish how to optimize a survey for~$\Neff$~(see also~\cite{CMB-S4:2016ple, SimonsObservatory:2018koc, NASAPICO:2019thw}). We additionally characterize the multipole range over which the~$\Neff$~information is accumulated, which informs how robust the forecasts are to possible small-scale systematics. Together, these considerations motivate the specific configurations described in Section~\ref{sec:configurations} and studied in the remainder of this section.\medskip

\begin{figure}
	\centering
	\includegraphics{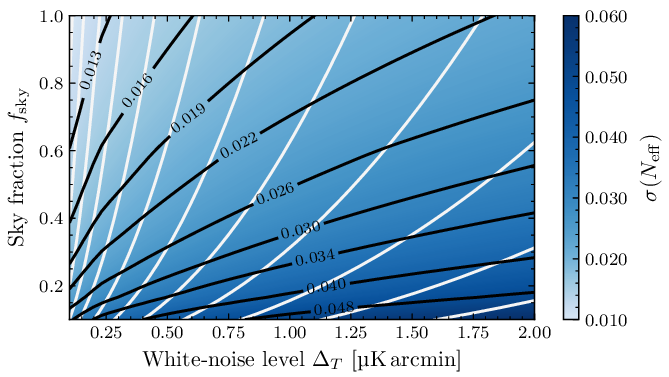}
	\caption{Forecasted $1\sigma$~uncertainty on~$\Neff$ as a function of the white-noise level~$\Delta_T = \Delta_P/\sqrt{2}$ and sky fraction~$\fsky$ for a simplified noise model. Experimental noise is taken as white noise with a \SI{1.4}{arcmin}~beam, and \Planck~data is added on the sky fraction not otherwise covered up to~60\%. Low-frequency noise and foreground modeling are neglected, which means that these values are intended as design-guiding estimates that agree with the more complete treatment used elsewhere in this work at the roughly~20\%~level. The displayed projections are based on delensed CMB~spectra over the multipole range from $\ell_\mathrm{min} = 30$ to $\ell_\mathrm{max}^T = 3000$ and $\ell_\mathrm{max}^P = \ell_\mathrm{max}^d = 5000$, respectively, supplemented by the low-$\ell$ \Planck~temperature power spectrum and $\tau$~prior from polarization as described in~\textsection\ref{sec:configurations_common}. The black lines show contours of constant~$\sigmaNeff$ as labeled. The light-gray lines indicate contours of constant detector-hours, i.e.\ detector count times observation time, which therefore serve as a rough proxy for fixed experimental cost, with each successive line toward the left corresponding to twice the effort.}
	\label{fig:results_sky-noise}
\end{figure}

Figure~\ref{fig:results_sky-noise}%
shows the projected~$\sigmaNeff$ as a function of both the white-noise level~$\Delta_T$ and sky fraction~$\fsky$ of a simple wide-area CMB~survey. It therefore illustrates the relative improvement in $\Neff$~constraints from optimizing the detector noise and/or increasing the sky area observed by the survey. For simplicity, we neglect low-frequency noise~[i.e.\ $\ellKnee^X \to 0$ in the noise spectra~\eqref{eq:noise_model}] and fix the beam width to $\thetaFWHM = \SI{1.4}{arcmin}$. We retain the full multipole range $\ell_\mathrm{min} = 30$ to $\ell_\mathrm{max}^P = 5000$ in polarization, but we cut the temperature spectrum at $\ell_\mathrm{max}^T = 3000$ as an approximate treatment of the neglected foregrounds, since these would otherwise raise the effective noise at small angular scales~(cf.~Fig.~\ref{fig:ilc_residuals}). We additionally include \Planck~data on the portion of the sky that is not covered otherwise up to $\fsky = 0.6$ and on large scales with $\ell < \ell_\mathrm{min}$~(see~\textsection\ref{sec:configurations_common}). Before discussing the figure, we reiterate that the displayed forecasts are only meant to be a guide for how different design choices will tend to impact the projected constraints on~$\Neff$, and we expect that these numerical results are only accurate at roughly the~20\%~level given the mentioned simplifying assumptions.

The iso-$\sigmaNeff$ contours are markedly asymmetric, extending much further along the noise axis than along the sky-fraction axis. For instance, improving the noise level from~\SI{2}{\muKelvin.arcmin} to~\SI{0.25}{\muKelvin.arcmin} at fixed~$\fsky$ only modestly tightens the constraint, whereas increasing~$\fsky$ from~$0.2$ to~$0.6$ at fixed noise yields a substantially larger improvement. In other words, the~$\Neff$~constraints improve more slowly for decreasing noise level at a fixed sky fraction compared to the improvement from increasing the sky fraction at a fixed noise level. Any constraint threshold, including the original \sfour~science target of $\sigmaNeff = 0.030$, is therefore much more efficiently reached by expanding sky coverage than by deepening a narrow field, and reaching $\sigmaNeff \approx 0.02$ requires simultaneously low noise and wide area. This is further exemplified by the light gray lines that highlight contours of constant detector-hours, i.e.\ the number of detectors times the observation time~$t$. These contours can therefore be roughly thought of as lines of constant experimental cost. The underlying reason is the fact that the observable sky fraction for a fixed detector count scales linearly with time, $\fsky \propto t$, while the noise in the absence of foregrounds scales as $\Delta_T \propto t^{-1/2}$~(and analogously for the number of detectors at a fixed observation time).

The observed asymmetry in the $\Neff$~contours reflects the underlying physics of the $\Neff$~information being spread across many peaks of the damping tail over a broad $\ell$~range~(cf.~Fig.~\ref{fig:ilc_residuals}) rather than being concentrated at a single scale~\cite{Hou:2011ec, Baumann:2015rya, Pan:2016zla, Ge:2022qws}. Once the noise is low enough that these peaks are signal-dominated, additional depth yields only marginal returns, while a larger sky area directly multiplies the number of independent modes and proportionally sharpens the constraint. Designing a survey for~$\Neff$ therefore clearly favors wide-field observations within the limits imposed by instrumental and site-related constraints, such as the minimum observable elevation, which directly limits the accessible sky. Such site-related limits do not apply from space, where the entire sky is in principle accessible, making a space-based mission the natural route to large~$\fsky$. On the ground, a new site in the Northern hemisphere could complement those in Chile and at the South Pole.\medskip

\begin{figure}
	\centering
	\includegraphics{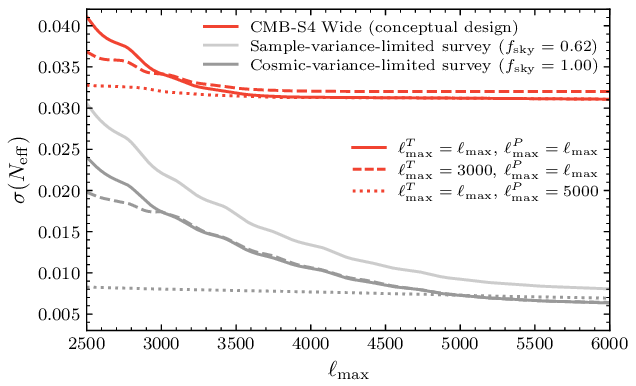}\vspace{-4pt}
	\caption{Forecasted $1\sigma$~uncertainty on~$\Neff$ for the~\sfourwide~survey of the two-site conceptual design and the cosmic-variance-limited survey, as a function of the maximum multipole~$\ell_\mathrm{max}$ included in the analysis. Solid lines use a common cutoff in temperature and polarization, $\ell_\mathrm{max}^T = \ell_\mathrm{max}^P = \ell_\mathrm{max}$, while the dashed and dotted curves instead fix one of the two, $\ell_\mathrm{max}^T = 3000$ and $\ell_\mathrm{max}^P = 5000$, respectively~(with the lensing power spectrum being always included up to~$\ell_\mathrm{max}$ and only the fiducial case shown for the sample-variance-limited survey for simplicity). Comparing these curves isolates the temperature and polarization contributions to the~$\Neff$~sensitivity. For the~\sfourwide~survey, the constraint saturates around $\sigmaNeff = 0.031$ for $\ell_\mathrm{max} \gtrsim 3750$, and extending either temperature or polarization beyond this scale changes it only marginally. (The~\sfourdeep~survey is not included in these forecasts, which does not affect the qualitative result as a function of multipoles, but is the reason why the constraint does not reach $\sigmaNeff = 0.030$ even at $\ell_\mathrm{max} = 5000$.) The variance-limited surveys, by contrast, keep improving well beyond $\ell_\mathrm{max} = 5000$, reflecting $\Neff$~information in the damping tail that remains inaccessible to the~\sfourwide~survey, for which it is obscured by instrumental noise and foreground residuals. The near-coincidence of their dashed curves with the solid ones shows that this additional information is carried almost entirely by polarization. Conversely, once the~$TE$ and~$EE$~spectra are fixed at $\ell_\mathrm{max}^P = 5000$, extending the multipole range for the~$TT$~spectrum adds only marginally to the sensitivity~(dotted curves). We also note that the small oscillations in~$\sigmaNeff$ arise from the varying signal-to-noise ratio of the CMB~peaks and troughs \mbox{as we sweep over them with~$\ell_\mathrm{max}$.}}\vspace{-3pt}
	\label{fig:results_ellmax}
\end{figure}

A complementary question concerns the multipole range from which the $\Neff$~information originates. Small scales $\ell \gtrsim 3000$ are increasingly affected by foreground contamination and other systematics in the temperature power spectrum~(cf.~\textsection\ref{sec:forecasting_foregrounds} and Fig.~\ref{fig:ilc_residuals}). It is therefore useful to know how much of the constraining power would be lost if such a scale cut were imposed as a mitigation strategy~(see also~\cite{Raghunathan:inprep}). Figure~\ref{fig:results_ellmax}%
shows~$\sigmaNeff$ for the~\sfourwide\ survey of the conceptual design as a function of the maximum multipole~$\ell_\mathrm{max}$ in temperature and polarization included in the analysis. (We note that the \sfourdeep~survey is not included in this forecast, which is the reason why the displayed $1\sigma$~constraint asymptotes to~$0.031$, cf.\ Table~\ref{tab:results_summary}.) The constraining power is essentially unaffected for $\ell_\mathrm{max} \gtrsim 3750$, indicating that the $\Neff$~information has saturated by that scale.

To disentangle the roles of temperature and polarization, we additionally vary the two cutoffs separately, holding one fixed at $\ell_\mathrm{max}^T = 3000$ or $\ell_\mathrm{max}^P = 5000$ while the other changes as~$\ell_\mathrm{max}$~(dashed and dotted curves, respectively). We see that the~$TT$~spectrum contributes comparatively little: keeping the polarization range fixed at $\ell_\mathrm{max}^P = 5000$ and extending the temperature spectrum from $\ell_\mathrm{max}^T = 2500$ to~$5000$ tightens the constraint only from $\sigmaNeff = 0.033\text{ to~}0.031$, with the same saturation near $\ell_\mathrm{max} \approx 3750$. Accordingly, cutting only the temperature spectrum to $\ell_\mathrm{max}^T = 3000$ while retaining polarization to~$5000$ reduces the sensitivity by only about~3\%, whereas cutting both spectra to $\ell_\mathrm{max} = 3000$ degrades~$\sigmaNeff$ by approximately~10\%. The majority of this reduction in constraining power therefore originates from the high-$\ell$ polarization modes, where the sharper acoustic peaks of the $E$-mode spectrum carry most of the small-scale $\Neff$~information. Reducing~$\ell_\mathrm{max}^T$ therefore leads to a non-negligible but modest increase in~$\sigmaNeff$, and can serve as a mitigation strategy of last resort if small-scale temperature systematics cannot otherwise be controlled. The complementary case of a cosmic-variance-limited survey, for which the information content extends much further in~$\ell$ and the importance of polarization on small scales is even more pronounced, is discussed in~\textsection\ref{sec:results_cvl}.

\subsection{Two-Site Conceptual Design}
\label{sec:results_conceptual}

Having established the central role of sky coverage in driving the $\Neff$~sensitivity, we now apply the full forecasting pipeline of Section~\ref{sec:forecasting} to the specific \sfour~configurations of Section~\ref{sec:configurations}. We begin with the two-site conceptual design, which combines the wide \sfourwide~survey in Chile with the narrow Pole-based \sfourdeep~observations as detailed in~\textsection\ref{sec:configurations_conceptual}. The projected $\sigmaNeff$~values for this configuration are summarized in the top part of Table~\ref{tab:results_summary}. The~\sfourwide~survey alone reaches $\sigmaNeff = 0.031$.\footnote{We also infer the projected constraints based on larger galactic masks, leaving observed data over~50\% and~56\%, rather than our baseline mask resulting in~62\% of the sky being analyzed, yielding $\sigmaNeff \approx 0.035\text{ and }0.033$, respectively.} This improves to $\sigmaNeff = 0.029$ when combining it with the \sfourdeep~survey~(after conservatively subtracting the entire \sfourdeep~sky fraction of~3\% from~\sfourwide), which therefore meets the \sfour~science target of $\sigmaNeff = 0.030$. As anticipated by the optimization study of~\textsection\ref{sec:results_optimization}, wide sky coverage is the primary driver of the $\Neff$~sensitivity, which is directly illustrated by the substantial difference between the standalone \sfourwide~constraint of~$0.031$ and the value of~$0.083$ obtained from~\sfourdeep\ alone. When combined with the wide survey, the deep observations nonetheless contribute a smaller but non-negligible improvement that reflects their much lower noise in a small area. Beyond~$\Neff$, the same observations also naturally contribute to measuring the sum of neutrino masses~$\sum m_\nu$, providing the most powerful constraints when combined with large-scale-structure information.\footnote{At the sensitivity of the considered \sfour~configurations, the precision of the neutrino-mass inference is limited more significantly by the measurement of the optical depth~$\tau$~(constrained primarily by the low-$\ell$ $EE$~spectrum) and of the expansion history~(constrained primarily by baryon acoustic oscillations) than by the small-scale CMB~precision. Neutrino-mass forecasts are therefore largely insensitive to the precise CMB~configuration. For experiments broadly comparable to the \sfour~conceptual design, combined with DESI-like~measurements of baryon acoustic oscillations, previous work projects $\sigma(\sum m_\nu) \approx \SI{24}{meV}$ for the \Planck~prior $\sigma(\tau) = 0.007$, which tightens to approximately~\SI{14}{meV} for a near-cosmic-variance-limited prior with $\sigma(\tau) = 0.002$~\cite{CMB-S4:2016ple}. We refer for example to~\cite{Kaplinghat:2003bh, Lesgourgues:2012uu, Font-Ribera:2013rwa, Allison:2015qca, CMB-S4:2016ple, Green:2021xzn, Gerbino:2022nvz} for more detailed discussions of neutrino-mass forecasts.}\medskip

We also investigate the dependence of the~$\Neff$~sensitivity on the observation time of the survey. Figure~\ref{fig:results_summary} shows the year-by-year evolution of the combined constraint of~\sfourwide\ and~\sfourdeep, which is the configuration in the fourth line of Table~\ref{tab:results_summary} but with a varying survey duration. For the first few years, there are significant improvements to~$\sigmaNeff$ as expected, with the constraints decreasing by roughly~20\% over the first four years after the initial first-year sensitivity of $\sigmaNeff = 0.039$. The target of $\sigmaNeff = 0.030$ is crossed after approximately 5.5~years of operations. Extending the survey to its nominal seven-year baseline yields the $\sigmaNeff = 0.029$ quoted above, while further extending to ten years provides only an additional~5\% improvement to $\sigmaNeff = 0.028$. These diminishing returns have two different origins in temperature and polarization. On the one hand, the small-scale temperature spectrum is largely limited by foreground residuals by this point and will barely improve with additional observing time. On the other hand, the signal-to-noise ratio in polarization at high multipoles continues to increase incrementally as the instrumental noise slowly falls with the square root of the observation time~(cf.~Fig.~\ref{fig:ilc_residuals}).\footnote{This also implies that further pushing the constraining power of a future survey requires lowering the polarization noise through longer integration times and/or more detectors, while simultaneously improving our knowledge, modeling, and mitigation of foregrounds, in particular in temperature.} Overall, the conceptual design therefore represents a relatively efficient path~(including some margin) to reaching $\sigmaNeff \leq 0.030$.

\subsection{South-Pole-Only Alternative}
\label{sec:results_pole}

We now turn to the South-Pole-only alternative defined in~\textsection\ref{sec:configurations_pole}, in which the entire \sfour~detector count of the conceptual design is deployed to Antarctica to observe $\fsky = 0.25$ over the same nominal seven-year period. After the galactic-latitude cut, we are left with $\fsky = 0.20$, which is less than a third of the analyzed sky of the conceptual design. Applying our \texttt{DRAFT}-based forecasting pipeline to the rescaled noise levels of the \sfourdeep~survey, which constitute a simplified scaling-based estimate rather than the detailed scan-strategy simulations of the other realistic configurations, we project $\sigmaNeff = 0.037$ for an experimental cost comparable to the conceptual design. This is about~22\% above the original \sfour~science target of $\sigmaNeff = 0.030$ and is not substantially improved by a longer survey duration: extending the observation time from seven to ten years only tightens the projected constraint by less than four percent~(see Table~\ref{tab:results_summary}).\footnote{We find that reaching the target $\sigmaNeff = 0.030$ from the South Pole alone requires substantially more detectors than considered here. One option could, for example, be to deploy the equivalent of roughly nine \sfourdeep-like~LATs and observe $\fsky = 0.25$ over a period of 20~years.} We additionally refer to Fig.~\ref{fig:results_summary}, which shows the $\Neff$~sensitivity \mbox{as a function of observation time.}\medskip

This result directly reflects the central role of sky coverage in setting the $\Neff$~sensitivity, as we quantified in~\textsection\ref{sec:results_optimization}. Reaching $\sigmaNeff = 0.030$ at $\fsky = 0.20$ requires driving the effective white noise to $\lesssim \SI{0.30}{\muKelvin.arcmin}$, as can be directly inferred from Fig.~\ref{fig:results_sky-noise}. This depth is not achieved by the \sfourdeep~survey in any frequency channel when observing~3\% of the sky over seven years~(cf.~Table~\ref{tab:noise_conceptual}) and is therefore certainly not reached after rescaling to the larger sky area considered here. At the same time, the projected $\sigmaNeff = 0.037$ remains relatively close to the science target despite the noise levels in every frequency band degrading by approximately~63\% relative to the conceptual ultra-deep survey. The reason is that the temperature ILC~residuals at the frequencies and multipoles most relevant for~$\Neff$ are foreground- rather than instrumental-noise-limited at conceptual \sfour~precision, as can be seen in Fig.~\ref{fig:ilc_residuals}~(see also Fig.~\ref{fig:foregrounds}), so that a uniform increase in instrumental noise translates into a more modest degradation of the constraining power than the rescaling factor alone suggests.\medskip

Among other variants of this single-site option considered in the analysis of alternatives~\cite{CMB-S4:2023aoa}, we also studied the impact of replacing the five-meter three-mirror-anastigmat design of the South-Pole~LAT in the conceptual design with a six-meter crossed-Dragone design comparable to the Chilean~LATs. We find that the $\Neff$~constraints are essentially unaffected by this choice, reinforcing that the accessible sky area, rather than the telescope design, is the dominant factor controlling the damping-tail sensitivity. The South-Pole-only alternative therefore illustrates, in a realistic experimental configuration, the cost of sacrificing sky coverage \mbox{for $\Neff$~science as anticipated in~\textsection\ref{sec:results_optimization}.}

\subsection{Chile-Only Revised Configuration}
\label{sec:results_revised}

In~2024, the \sfour~effort converged on the revised configuration described in~\textsection\ref{sec:configurations_revised}, with one~\sfour~LAT in Chile performing a hybrid scan strategy over both the wide and delensing fields and operating in coordination with the~LAT of an experiment like the Simons Observatory at the same site~\cite{CMB-S4:2025rpp}.\footnote{This was partially inspired by one of the several alternative configurations studied in the analysis of alternatives~\cite{CMB-S4:2023aoa} alongside the Pole-only option of~\textsection\ref{sec:results_pole}. In that configuration, we replaced one of the two \sfour~Chilean~LATs of the conceptual design~(cf.~\textsection\ref{sec:configurations_conceptual}) with an~SO-like~LAT for the entire observation time of~\sfour. In this scenario, the projected constraint on~$\Neff$ at the end of seven years is degraded by about 5\%~relative to the \sfour~conceptual design.} Our forecast for this configuration therefore combines three survey components: the Hybrid Wide and Hybrid~Delensing portions of the \sfour~LAT and the wide-area SO-like~survey at its goal noise levels. Since the latter is planned to begin observations in~2028~(see also footnote~\ref{fn:so}), which is five years before the earliest~\sfour~operational start of~2033, its accumulated data contribute meaningfully to the combined forecast already before~\sfour\ comes online. The combined observational effort and joint analysis span a total sky fraction of approximately~62\% after galactic masking.\medskip

We compare the projected $\Neff$~sensitivity of the revised configuration to the other survey designs in Table~\ref{tab:results_summary}. In addition, Figure~\ref{fig:results_summary} shows the forecasted~$\sigmaNeff$ as a function of observation time, with the origin set to the start of \sfour~operations, which were scheduled to commence in~\num{2033} for the revised configuration. We also include the SO-like observations from~2028 through~2033, which appear at negative times. The target of $\sigmaNeff = 0.030$ is reached by~\num{2042} in this timeline, corresponding to approximately nine years of \sfour~operations and 14~years of cumulative observing time of the~\sfour\ and SO-like~LATs. This is a noticeably longer time than for the conceptual design~(about 5.5~years; \textsection\ref{sec:results_conceptual}) and reflects the reduced number of total detectors compared to the two Chilean~LATs and one Pole~LAT. In comparison, the revised configuration reaches $\sigmaNeff = 0.032$ after the same 5.5~years. We therefore find that this configuration still allows reaching the original \sfour~science target, but on a substantially longer timescale than for the conceptual design. As a final benchmark, we now ask how much scientific room remains for improvement beyond any realized configuration.

\subsection{Cosmic-Variance-Limited Survey}
\label{sec:results_cvl}

We now turn to the cosmic-variance-limited survey introduced in~\textsection\ref{sec:configurations_cvl} to establish the ultimate reach of the CMB~primary anisotropies over the same multipole range $2 \leq \ell \leq 5000$ used for the realistic configurations. This idealized case sets a fundamental floor against which any future experiment can be compared and quantifies how much of the primary-CMB information remains untapped by a survey such as envisioned for~\sfour. Over the reduced sky fraction $\fsky = 0.62$, which is representative of the usable sky for a ground-based survey located in Chile, our forecast yields $\sigmaNeff = 0.0092$. Extending the idealized survey to full sky coverage, $\fsky = 1.0$, tightens the constraint to $\sigmaNeff = 0.0073$. These values are summarized alongside the other configurations in Table~\ref{tab:results_summary} and are a factor of three to four below the $\sigmaNeff = 0.030$~target of the realistic configurations discussed so far. (As we include the large-scale $E$-mode polarization directly, the optical depth is also determined internally, with a cosmic-variance-limited projection of $\sigma(\tau) = 0.0015$.) In this context, it is also worth noting that the achieved delensing efficiency is both very high and important for these results, which can be seen by comparing our delensed forecasts to the two limiting cases of using lensed and unlensed spectra: employing lensed spectra degrades~$\sigmaNeff$ by almost~30\%, while using unlensed spectra improves the constraining power by less than three percent.\footnote{The corresponding gain from perfect delensing relative to the iterative-delensing procedure is somewhat larger at~\sfour~precision, with around six percent for the conceptual design. This is because the iterative lensing reconstruction depends on the small-scale CMB~modes that are noise-limited at~\sfour~precision but cosmic-variance-limited in the CVL~case, albeit over the still restricted range of multipoles $\ell \leq 5000$. Iterative delensing therefore comes closer to saturating its ideal performance for a CVL~survey, leaving less room for improvement from perfect delensing.} The large gain in sensitivity from delensing the CMB~power spectra is due to the fact that this procedure removes the effects of lensing on high-$\ell$ multipoles. In turn, this sharpens the acoustic peaks~(most important for $\ell \lesssim 3000$) and restores the damping tail~(most effective for $\ell \gtrsim 4000$), which are the two features \mbox{in the CMB~spectra that the $\Neff$~constraints mainly originate from.}\medskip

The dependence of the variance-limited sensitivity on the maximum multipole~$\ell_\mathrm{max}$ used in the analysis is shown in Fig.~\ref{fig:results_ellmax}. Unlike the analogous sensitivity curve for the~\sfourwide~survey of the conceptual design displayed in the same figure, for which the constraint saturates already near $\ell_\mathrm{max} \approx 3750$, the CVL~observations reach $\sigmaNeff \approx 0.0073$ by~$\ell_\mathrm{max} = 5000$ and continue to improve well beyond this multipole, e.g.\ $\sigmaNeff(\ell_\mathrm{max} = 6000) = 0.0063$. The relative importance of polarization compared to temperature discussed in~\textsection\ref{sec:results_optimization} for the~\sfourwide~survey is even more pronounced here~(see also~\cite{Galli:2014kla, Scott:2016fad}). Capping the~$TT$~spectrum at $\ell_\mathrm{max}^T = 3000$ leaves the constraint essentially unchanged~[$\sigmaNeff = 0.0064$ at $\ell_\mathrm{max} = 6000$; dashed curve in Fig.~\ref{fig:results_ellmax}], whereas truncating polarization at $\ell_\mathrm{max}^P = 5000$ largely saturates the sensitivity~[$\sigmaNeff = 0.0069$; dotted line]. The dominance of polarization is even clearer at low~$\ell_\mathrm{max}$ since extending polarization alone to $\ell_\mathrm{max}^P = 5000$ for $\ell_\mathrm{max} = 2500$ already reaches $\sigmaNeff = 0.0083$~(dotted), compared to~$0.024$ with all spectra cut off at~$2500$~(solid) and~$0.020$ when only the temperature range is extended to $\ell_\mathrm{max}^T = 3000$~(dashed). (As anticipated, the lensing power spectrum contributes only a modest amount of indirect $\Neff$~constraining power.) In the cosmic-variance limit, the small-scale $\Neff$~information therefore resides overwhelmingly in the~$TE$ and~$EE$~spectra, whose acoustic peaks remain sharp deep into the \mbox{damping tail and with the~$TE$~spectrum carrying most of the temperature information.}

This contrast directly illustrates the interplay of the fundamental and observational limits: in the case of~\sfour, the damping tail becomes dominated by noise and foreground residuals at high~$\ell$ and additional multipoles carry a diminishing amount of information, whereas each additional multipole continues to contribute in the CVL~case. This evolution will be limited only by our ability to extract the primary CMB~signal~(through component separation, delensing, and modeling) from the data dominated by foregrounds, gravitational lensing, and other secondary anisotropies. A survey capable of extracting cosmological information beyond~$\ell = 5000$, such as the proposed CMB-HD~experiment~\cite{CMB-HD:2022bsz}, could therefore access additional sensitivity beyond our nominal CVL~benchmark, \mbox{provided that the associated foregrounds and secondary anisotropies can be mitigated.}\medskip

The forecasts presented so far were obtained within a $\Lambda\mathrm{CDM} + \Neff$~cosmology. Two extensions of this baseline model are particularly well-motivated by the physics of neutrinos and other light relics, and we discuss them in more detail in Section~\ref{sec:implications}: the sum of neutrino masses~$\sum m_\nu$, which directly affects the late-time growth of structure and the lensing of the~CMB, and the primordial helium abundance~$Y_\mathrm{p}$, which is degenerate with~$\Neff$ at the level of the damping tail. We consider both in turn.

We first include the sum of neutrino masses by considering a $\Lambda\mathrm{CDM} + \Neff + \sum m_\nu$~cosmology, treating~$\sum m_\nu$ as an additional free parameter with the Fisher information evaluated around its fiducial value of $\sum m_\nu = \SI{0.06}{eV}$~(see Table~\ref{tab:parameters_fiducial}). At the noise levels of the conceptual design, marginalizing over~$\sum m_\nu$ degrades~$\sigmaNeff$ by only a small fraction of a percent, which indicates that $\Neff$~measurements with an experiment like~\sfour\ as forecasted here are essentially insensitive to the treatment of neutrino mass.\footnote{For comparison, the corresponding effect on~$\sigmaNeff$ of using lensed instead of delensed spectra at the same \sfour~precision is more than two percent, which is an order of magnitude larger.} At the precision of a variance-limited survey, a mild dependence emerges and the inclusion of neutrino masses increases~$\sigmaNeff$ by less than seven percent.\footnote{The effect of varying neutrino masses on the inference of~$\Neff$ remains smaller than the almost 30\%~impact of using lensed instead of delensed spectra noted above, reinforcing that high-fidelity delensing is essential to realize the full sensitivity.} While the parameters~$\Neff$ and~$\sum m_\nu$ are essentially uncorrelated at the percent-level precision of~\sfour, a mild degeneracy emerges in the sample-variance-limited regime where sub-percent precision on~$\Neff$ becomes accessible.

We now consider the primordial helium abundance~$Y_\mathrm{p}$ within a $\Lambda\mathrm{CDM} + \Neff + Y_\mathrm{p}$~cosmology, treating~$Y_\mathrm{p}$ as an additional free parameter with the Fisher information evaluated around its fiducial value of $Y_\mathrm{p} = 0.2467$~(see Table~\ref{tab:parameters_fiducial}). Unlike the case of~$\sum m_\nu$, the inclusion of~$Y_\mathrm{p}$ introduces a well-known degeneracy with~$\Neff$ at the background level: both parameters control the diffusion damping of the small-scale~CMB~anisotropies~($\Neff$ sets the expansion rate at recombination and~$Y_\mathrm{p}$ changes the free-electron density in that epoch), so that an increase in one can be largely compensated by a decrease in the other, leaving the predicted CMB~damping tail nearly unchanged~\cite{Hou:2011ec}. This degeneracy is already substantial at current precision and becomes increasingly limiting at the precision of next-generation experiments like~\sfour\ and in the cosmic-variance-limited limit. The forecasted constraints on both parameters are summarized in Table~\ref{tab:results_extensions}. %
\begin{table}
	\centering
	\setlength{\tabcolsep}{9pt}
	\begin{tabular}{l S[table-format=1.4] S[table-format=1.5] S[table-format=1.3] S[table-format=1.4]}
			\toprule
						& {$\Lambda\mathrm{CDM} + \Neff$}		& {$\Lambda\mathrm{CDM} + Y_\mathrm{p}$}	& \multicolumn{2}{c}{$\Lambda\mathrm{CDM} + \Neff + Y_\mathrm{p}$}	\\
						  \cmidrule(lr){2-2}					  \cmidrule(lr){3-3}						  \cmidrule(lr){4-5}
						& {$\sigmaNeff$}						& {$\sigma(Y_\mathrm{p})$}					& {$\sigmaNeff$}			& {$\sigma(Y_\mathrm{p})$}				\\
			\midrule[0.065em]
		\sfour			& 0.029									& 0.0021									& 0.072						& 0.0043								\\
		SVL				& 0.0092								& 0.00067									& 0.038						& 0.0024								\\
		CVL				& 0.0073								& 0.00053									& 0.030						& 0.0019								\\
			\bottomrule
	\end{tabular}
	\caption{Forecasted $1\sigma$~uncertainties on the effective number of relativistic species~$\Neff$ and the primordial helium abundance~$Y_\mathrm{p}$ from the two-site conceptual design of~\sfour~(cf.~\textsection\ref{sec:configurations_conceptual}), and from the sample-variance-limited~(SVL, $\fsky = 0.62$) and cosmic-variance-limited~(CVL, $\fsky = 1.0$) surveys defined in~\textsection\ref{sec:configurations_cvl}, within three extensions of~$\Lambda\mathrm{CDM}$. The $\Lambda\mathrm{CDM} + \Neff$ cosmology is the model underlying all results presented in this paper except those shown here that include~$Y_\mathrm{p}$. All forecasts use delensed power spectra over $2 \leq \ell \leq 5000$ and the parameters are varied around the fiducial cosmology of Table~\ref{tab:parameters_fiducial}.}
	\label{tab:results_extensions}
\end{table}
At the precision of the conceptual design of~\sfour~(\textsection\ref{sec:configurations_conceptual}), marginalizing over~$Y_\mathrm{p}$ degrades~$\sigmaNeff$ by a factor of~$2.5$ from~$0.029$ to~$0.072$. The relative impact is even larger for variance-limited surveys, with the error bars on~$\Neff$ increasing by a factor of more than four. The corresponding helium-abundance forecasts are $\sigma(Y_\mathrm{p}) = 0.0043$, $0.0024$, and~$0.0019$ for~\sfour, SVL, and~CVL when~$\Neff$ is jointly varied.\footnote{For comparison, current constraints from~\Planck\ combined with~ACT and~SPT yield $\sigmaNeff = 0.23$ and $\sigma(Y_\mathrm{p}) = 0.013$ in a joint $\Lambda\mathrm{CDM} + \Neff + Y_\mathrm{p}$~analysis~\cite{AtacamaCosmologyTelescope:2025nti, SPT-3G:2025bzu}, which tightens to $\sigma(Y_\mathrm{p}) = 0.0083$ when~$\Neff$ is held fixed at its Standard Model~value. The forecasts presented here improve these bounds by a factor of three~(four) when varying~(fixing)~$\Neff$ at \sfour~precision and by a factor of five to seven~(12 to~16) \mbox{at the variance-limited level.}} Holding~$\Neff$ fixed at its Standard Model~value sharpens these bounds by a factor of about~$2$ for~\sfour\ and~$3.5$ for the variance-limited surveys, again reflecting the same underlying degeneracy.\footnote{As in the $\Lambda\mathrm{CDM} + \Neff + \sum m_\nu$~extension, high-fidelity delensing is essential to realize this sensitivity: the use of delensed instead of lensed spectra improves~$\sigma(Y_\mathrm{p})$ by~30\% for a variance-limited survey, since delensing sharpens the acoustic oscillations and restores the shape of the damping tail. Floating~$\sum m_\nu$ in addition to~$Y_\mathrm{p}$ while using delensed spectra degrades the variance-limited forecasts by more than~7\%~(3\%) for~$\Neff$~($Y_\mathrm{p}$), while~$\sigma(\sum m_\nu)$ itself increases by approximately seven percent relative to the $\Lambda\mathrm{CDM} + \Neff + \sum m_\nu$~analysis.}

The relevance of this degradation depends on the physical question asked of the data. In standard~BBN, the primordial helium abundance is set by~$\omega_\mathrm{b}$ and~$\Neff$, with corrections smaller than the precisions discussed here~(see e.g.~\cite{Pitrou:2018cgg, Grohs:2019cae, Fields:2019pfx, Grohs:2023voo, Cooke:2024nqz}; cf.\ also~\textsection\ref{sec:implications_additional}), so that the $\Neff$-only forecasts apply. Departures from this BBN-consistency relation, on the other hand, require non-standard physics that modifies the neutron-to-proton ratio during nucleosynthesis, such as primordial lepton asymmetries, late entropy injection, or neutrino-sector physics altering the freeze-out dynamics~(see~\textsection\ref{sec:implications_additional}). It is in this regime that the degraded constraints represent the genuine projected sensitivity, under which the original \sfour~science target is missed by more than a factor of two. Depending on the particular scenario, external astronomical determinations of~$Y_\mathrm{p}$ could restore some of the lost sensitivity to~$\Neff$. For instance, a Gaussian prior of $\sigma(Y_\mathrm{p}) = 0.0013$ from the recent $Y_\mathrm{p}$~determination based on observations with the Large Binocular Telescope~\cite{Aver:2026dxv}~(or $\sigma(Y_\mathrm{p}) = 0.0040$ from the~EMPRESS~measurement~\cite{Yanagisawa:2025mgx}) tightens the conceptual-design \sfour~forecast to $\sigmaNeff = 0.039~(0.055)$, recovering a substantial fraction of the BBN-consistent sensitivity. The~SVL and CVL~projections correspondingly improve to $\sigmaNeff = 0.021~(0.033)$ and $\sigmaNeff = 0.019~(0.028)$, respectively.\footnote{These joint analyses naturally also sharpen the inference of the helium abundance itself. Combining the CMB~forecasts with the Large-Binocular-Telescope measurement yields $\sigma(Y_\mathrm{p}) = 0.0012$, $0.0011$, and~$0.0011$ for~\sfour, SVL, and~CVL, respectively, while the corresponding combinations with the EMPRESS~measurement give $\sigma(Y_\mathrm{p}) = 0.0029$, $0.0020$, and~$0.0017$.} Nevertheless, a CMB-only measurement of~$Y_\mathrm{p}$ is itself a valuable probe in this context, providing an inference independent of the astronomical observations anchoring the external priors above. The two are therefore highly complementary, and, as we discuss in~\textsection\ref{sec:implications_additional}, the joint analysis remains \mbox{a powerful probe of non-standard scenarios.}

The $\Neff$-$Y_\mathrm{p}$ degeneracy underlying these degraded constraints exists only at the background level and may be partially broken by perturbation-level effects. Free-streaming radiation in particular imprints a characteristic phase shift on the acoustic peaks~(see Section~\ref{sec:introduction} and~\textsection\ref{sec:implications_radiation}). Because~$\Neff$ conventionally parameterizes free-streaming radiation, as is the case for the cosmic neutrino background, the shifts in the acoustic peaks depend on~$\Neff$ but not on~$Y_\mathrm{p}$, which breaks the damping-tail degeneracy~\cite{Bashinsky:2003tk, Baumann:2015rya}. While the phase shift is a small effect, its quantitative importance both in general and especially within a $\Lambda\mathrm{CDM} + \Neff + Y_\mathrm{p}$~cosmology has recently been illustrated and quantified in~\cite{Montefalcone:2025unv}. Without this perturbation-level information~(for instance, if the radiation propagates as a perfect fluid rather than free-streaming; see~\textsection\ref{sec:implications_radiation}), the degradation of constraining power with varying~$Y_\mathrm{p}$ is even more severe. The degeneracy-breaking power of the phase shift nonetheless weakens toward small scales since the acoustic oscillations themselves are exponentially damped, so that the peaks are resolved with diminishing signal-to-noise ratio even in the cosmic-variance limit. This is also why marginalizing over~$Y_\mathrm{p}$ inflates~$\sigmaNeff$ by a larger relative factor in the variance-limited regime~(more than four) than at \sfour~precision~(about~$2.5$). A deeper survey accrues most of its additional constraining power in the damping tail along the degenerate direction, while the orthogonal phase-shift information saturates, so that the absolute~$\sigmaNeff$ keeps improving even as the relative penalty for varying~$Y_\mathrm{p}$ grows. The near-optimal delensing achieved in our forecasts, which sharpens these peaks, is therefore essential to extract all the information contained in the~CMB~data.\medskip

In summary, a sample- or cosmic-variance-limited survey sets a $\sigmaNeff$~floor that is a factor of three to four below the realistic \sfour~configurations considered above, leaving a substantial fraction of the primary-CMB information untapped by current and near-term~CMB~surveys. We also considered two parameter extensions of our fiducial $\Lambda\mathrm{CDM} + \Neff$~cosmology, which further sharpen this picture. Marginalizing over the sum of the neutrino masses leaves the variance-limited~$\sigmaNeff$ essentially unchanged, whereas marginalizing over the primordial helium abundance~$Y_\mathrm{p}$ inflates the uncertainty on~$\Neff$ by a factor of about four due to the well-known background-level degeneracy between~$\Neff$ and~$Y_\mathrm{p}$. The perturbation-level phase shift from free-streaming radiation partially mitigates this degeneracy, which underscores both that these perturbation effects are highly detectable and crucial at \sfour~precision~(cf.~also~\cite{Montefalcone:2025unv}), and that the practical relevance of the degeneracy depends on the physical scenario being tested~(see~\textsection\ref{sec:implications_radiation}). The significant range of improvement that remains in observations of the primary~CMB~anisotropies at large multipoles provides a concrete incentive for future ground- and space-based~CMB~experiments beyond the next decade. The associated physics implications \mbox{could be transformative and we discuss them in the following section.}

\section{Implications for Neutrinos and New Physics}
\label{sec:implications}

The forecasted constraints on~$\Neff$ presented in the previous section have broad implications for a wide range of well-motivated physics, both within and beyond the Standard Model of particle physics~(see e.g.~\cite{Lesgourgues:2013sjj, Abazajian:2013bxd, Abazajian:2016hbv, Alexander:2016aln, CMB-S4:2016ple, Lattanzi:2017ubx, Wallisch:2018rzj, Green:2019glg, Asadi:2022njl, Gerbino:2022nvz, Dvorkin:2022jyg, Green:2022bre, Antel:2023hkf} for reviews). Among many other interesting scenarios, a \sfour-type measurement with $\sigmaNeff \approx 0.030$ would for the first time probe the energy density contributed by any thermal relic with spin~$\geq 1/2$ that was ever in equilibrium with the Standard~Model. A CVL~measurement with $\sigmaNeff \lesssim 0.01$ would extend this reach to the entire landscape of light thermal relics, including the lightest possible scalar. While the majority of BSM~scenarios predict additional contributions with $\Delta\Neff > 0$, new physics can also reduce~$\Neff$ below the SM~prediction, making precise and accurate measurements valuable in both directions.\footnote{Measuring~$\Neff$ constrains or excludes many departures from the Standard Model in a largely model-independent way since it probes the total radiation energy density irrespective of its origin~(the spin-dependent thresholds of~\textsection\ref{sec:implications_relics} allow entire classes of thermal relics to be excluded, for example). In contrast, pinpointing the source of a potential deviation $\Neff \neq 3.044$ will generally require complementary cosmological information, such as the free-streaming phase shift~(see~\textsection\ref{sec:implications_radiation}), the comparison of the radiation content at recombination and during~BBN~(cf.~\textsection\ref{sec:implications_additional}) or CMB~spectral distortions~(see e.g.~the reviews~\cite{Chluba:2019kpb, Chluba:2019nxa, Chluba:2025wxp}), as well as independent astrophysical and laboratory probes. The latter are especially valuable in this context: a robust cosmological detection of $\Neff \neq 3.044$ would point to a comparably smaller set of viable scenarios, which could be further narrowed down through the mentioned model-agnostic cosmological probes before being targeted \mbox{by dedicated terrestrial experiments designed to identify the underlying physics.}} In this section, we discuss these and other physics implications, quoting detection significances throughout based on reference sensitivities of $\sigmaNeff = 0.030$ for a~\sfour-like survey and $\sigmaNeff = 0.01$ for a variance-limited CMB~survey~(see Table~\ref{tab:results_summary} for our detailed forecasts). We first review the contribution of light thermal relics to~$\Neff$ and specific BSM~particle candidates in~\textsection\ref{sec:implications_relics}. We then describe in~\textsection\ref{sec:implications_radiation} how the same data probe not only the energy density but also the physical properties of the radiation sector, including the free-streaming nature of neutrinos and potential interactions in the dark sector. Finally, we survey additional implications for early-universe physics in~\textsection\ref{sec:implications_additional}, ranging from the stochastic gravitational-wave background and light dark matter to big bang nucleosynthesis and tests of the Standard Model prediction itself.

\subsection{Light Thermal Relics}
\label{sec:implications_relics}

Any particle with a mass $m \lesssim \SI{1}{eV}$ contributes to the radiation energy density measured in the~CMB if it was in thermal equilibrium with the Standard-Model~plasma at any point in the early universe and subsequently decoupled while still being relativistic.\footnote{While the cosmic microwave background is released at a temperature of about $T_{\gamma, \mathrm{dec}} \approx \SI{0.3}{eV}$, the imprints in the damping tail are generated at higher temperatures. This means that thermal particles with non-zero masses $m \lesssim \SI{1}{eV}$ generally contribute to~$\Neff$ as inferred from the~CMB. These light-but-massive relics behave like SM~neutrinos in that they observationally contribute to the radiation energy density at early times when they are relativistic, parameterized by~$\Neff$, and to the matter energy density at late times when they are non-relativistic, parameterized by~$\sum m_\nu$. The latter makes them additionally accessible to large-scale-structure observations~(see~\cite{DePorzio:2020wcz, Xu:2021rwg, Banerjee:2025gwe, Kumar:2025gkw, DePorzio:2026zba} for additional details), which otherwise probe essentially the same~$\Neff$ as the~CMB. The relevant mass threshold is, however, epoch-dependent: relics with masses of order~\si{MeV} are still relativistic at the higher temperatures probed by big bang nucleosynthesis~($T \sim \SI{1}{MeV}$) and contribute to the radiation density inferred there~(cf.~\textsection\ref{sec:implications_additional}).} The size of this contribution depends on two quantities: the effective number of internal degrees of freedom of the particle, which encodes both its internal degrees of freedom and its spin statistics, and the temperature at which it decoupled from the thermal bath. If a species~$X$ decouples at a temperature~$T_\mathrm{F}$, when the number of entropic degrees of freedom in the plasma is~$g_{*S}(T_\mathrm{F})$, its contribution to~$\Neff$ is illustrated in Fig.~\ref{fig:neff_freezeout} and given by
\begin{equation}
	\Delta\Neff = \frac{4}{7} n_X \left( \frac{g_{*S}(T_{\nu, \mathrm{dec}})}{g_{*S}(T_\mathrm{F})} \right)^{\!4/3}\, ,
\end{equation}
where $T_{\nu, \mathrm{dec}} \approx \SI{1}{MeV}$ is the neutrino decoupling temperature and $g_{*S}(T_{\nu, \mathrm{dec}}) = 10.75$. The effective number of internal degrees of freedom is captured by~$n_X$, with $n_X = 1$ for a real scalar, $n_X = 7/4$ for a Weyl fermion, and $n_X = 2$ for a massless vector boson. The dependence on~$g_{*S}(T_\mathrm{F})$ means that particles that decoupled at higher temperatures are diluted by the subsequent annihilation of heavier SM~particles, leading to smaller contributions to~$\Neff$. Conversely, particles that remained longer in thermal equilibrium and decoupled at lower temperatures, when fewer SM~degrees of freedom were present, contribute more. This also means that the minimum value of~$\Delta\Neff$ arises for a particle that decoupled at a temperature above the electroweak symmetry breaking scale of $T_\mathrm{EWSB} \approx \SI{160}{GeV}$ when $g_{*S}(T_\mathrm{F} > T_\mathrm{EWSB}) = 106.75$ in the Standard Model. In this limit, the contributions from a single species of each spin are
\begin{equation}
	\Delta\Neff \approx
		\begin{cases}
			0.027	& \text{real scalar (spin-0)}\, ,			\\
			0.047	& \text{Weyl fermion (spin-1/2)}\, ,		\\
			0.054	& \text{massless vector boson (spin-1)}\, .
		\end{cases}
\end{equation}
To reiterate, these values represent the smallest possible contribution to~$\Neff$ for any particle that was ever in thermal equilibrium with the Standard Model in a standard cosmological history.\footnote{These thresholds assume a standard thermal history after the one relic particle decouples. If a heavy, long-lived particle decays predominantly into SM~radiation after the light relic has frozen out, the resulting entropy injection dilutes its energy density relative to the photon bath and effectively reduces the observable~$\Delta\Neff$. This is however also degenerate with the uncertainty on the reheating temperature and the question of whether the relic has been in thermal equilibrium. In addition, a significant dilution of the stated thresholds requires a large increase in the degrees of freedom in the very early universe. While this is possible in extensions of the Standard Model, these models generically come with many additional light particles that would in turn increase~$\Delta\Neff$, e.g.\ the large increase in particle content in the Minimal Supersymmetric Standard Model can be compensated by just three of these particles being light. We refer to~\cite{Wallisch:2018rzj} for an expanded discussion on these aspects and additional context. Conversely, light species that were never in thermal equilibrium with the Standard Model, such as the relativistic states of a completely secluded dark sector, are not subject to these thresholds, but still contribute to~$\Neff$ with an abundance suppressed by the fourth power of the dark-to-photon temperature ratio~(cf.~\textsection\ref{sec:implications_additional}).\label{fn:thresholds}} A measurement of~$\Neff$ consistent with the SM~prediction at a precision below these thresholds would therefore exclude the existence of any such thermal relic, which would provide a powerful, model-independent probe of the particle content of the early universe and BSM~physics. In the context of our forecasts of Section~\ref{sec:results}, this means that a \sfour-like experiment will start probing all light relics and be sensitive to Weyl fermions and vector bosons at the level of~$1.6\sigma$ and~$1.8\sigma$, respectively. Achieving~$\sigmaNeff \approx 0.01$ will provide nearly~$3\sigma$ sensitivity to light scalars, almost~$5\sigma$ sensitivity to Weyl fermions, and more than~$5\sigma$ sensitivity to vector bosons, as summarized in Table~\ref{tab:relics_significance},%
\begin{table}
	\centering
	\begin{tabular}{l c S[table-format=1.3] S[table-format=1.1] S[table-format=1.1] c c}
			\toprule
		\multirow{2}{*}{Relic species}	& \multirow{2}{*}{Spin}	& \multicolumn{1}{c}{\multirow{2}{*}{Minimum $\Delta\Neff$}}	& \multicolumn{2}{c}{Minimum $N_\sigma$}	& \multicolumn{2}{c}{$T_\mathrm{F}$ excluded at $2\sigma$}	\\
																																  \cmidrule(lr){4-5}						  \cmidrule(lr){6-7}
										& 						& 																& {\sfour}		& {SVL}						& {\sfour}		& {SVL}										\\
			\midrule[0.065em]
		Real scalar particle			& $0$					& 0.027															& 0.9			& 2.7						& \SI{470}{MeV}	& {all}										\\
		Weyl fermion					& $1/2$					& 0.047															& 1.6			& 4.7						& \SI{41}{GeV}	& {all}										\\
		Massless vector boson			& $1$					& 0.054															& 1.8			& 5.4						& \SI{75}{GeV}	& {all}										\\
		Gravitino (low-scale SUSY)		& $3/2$					& 0.059															& 2.0			& 5.9						& {--}			& {--}										\\
			\bottomrule
	\end{tabular}
	\caption{Minimal contribution to~$\Neff$ from a single thermal relic of a given spin that decoupled above all Standard Model mass thresholds~(cf.~Fig.~\ref{fig:neff_freezeout}), the corresponding detection significances $N_\sigma = \Delta\Neff/\sigmaNeff$, and the approximate decoupling temperature~$T_\mathrm{F}$ below which the contribution reaches the $2\sigma$~threshold $\Delta\Neff = 2\sigmaNeff$. Prominent examples are axions and other pseudo-Nambu-Goldstone bosons for the spin~$0$ case, sterile neutrinos and other dark fermions are typical spin-$1/2$ particles, and dark photons are common vector bosons. The gravitino contributes at least as a Weyl fermion~($\Delta\Neff = 0.047$), with the larger value shown being predicted by concrete low-scale supersymmetry-breaking models~(see the main text). We adopt the reference sensitivities $\sigmaNeff = 0.03$ for a~\sfour-type experiment and $\sigmaNeff = 0.01$ for a sample-variance-limited survey~(the specific forecasts of Table~\ref{tab:results_summary} are tighter, so that the actual significances are larger by up to about a third). At $\sigmaNeff = 0.01$, even the minimal contributions exceed~$2\sigma$ so that a relic of any spin is detectable at more than~95\% confidence for any decoupling temperature in the standard thermal history~(``all'').}
	\label{tab:relics_significance}
\end{table}
with implications for particle physics that cannot be overstated.

If instead the species decoupled at lower temperatures, the contribution to~$\Neff$ is correspondingly larger, as illustrated in Fig.~\ref{fig:neff_freezeout}. For example, a scalar that decoupled just after the QCD~phase transition at a freeze-out temperature of $T_\mathrm{F} \approx \SI{120}{MeV}$ contributes $\Delta\Neff \approx 0.26$, while a Weyl fermion freezing out at the same temperature contributes $\Delta\Neff \approx 0.46$. Particles decoupling between the QCD~and electroweak transitions generically contribute $\Delta\Neff \in [0.027, 0.13]$, depending on their spin and the precise decoupling temperature~(see also Table~\ref{tab:relics_significance} for the decoupling temperatures~$T_\mathrm{F}$ below which the respective $\Neff$~contribution becomes detectable at~$2\sigma$), which places them squarely within the range of sensitivity of~SO, \sfour, and future experiments.\medskip

Such new light particles, which would thermally freeze out in the early universe and contribute to~$\Neff$ as described, are ubiquitous in many well-motivated extensions of the Standard Model~\cite{Jaeckel:2010ni, Brust:2013ova, Essig:2013lka, Alexander:2016aln, Wallisch:2018rzj, Green:2019glg, Asadi:2022njl, Dvorkin:2022jyg, Green:2022bre, Antel:2023hkf}. In the following, we briefly discuss the most prominent candidates and summarize the implications of the forecasted $\Neff$~measurements.

\paragraph{Axions and other pseudo-Nambu-Goldstone bosons.} The QCD~axion, which was introduced to solve the strong~CP problem, is the most prominent spin-0 pseudo-Nambu-Goldstone boson~\cite{Peccei:1977hh, Weinberg:1977ma, Wilczek:1977pj, Hook:2018dlk}. More general axion-like particles arise in BSM~scenarios from spontaneous breaking of a global shift symmetry~\cite{Graham:2015cka, Arkani-Hamed:2016rle, Chacko:2016hvu, DiLuzio:2020wdo, Adams:2022pbo}. Further well-known examples are familons~\cite{Davidson:1981zd, Wilczek:1982rv, Reiss:1982sq, Feng:1997tn} and majorons~\cite{Chikashige:1980ui, Chacko:2003dt}, which arise from the spontaneously broken global symmetries of family and lepton number, respectively, as well as light moduli appearing in string compactifications~\cite{Svrcek:2006yi, Arvanitaki:2009fg, Cicoli:2012aq, Acharya:2012tw, Marsh:2015xka}. If such a particle was thermally produced\hskip1pt\footnote{These particles do not have to be produced thermally. Several non-thermal mechanisms can efficiently produce a non-relativistic population, which may also constitute~(part of) the dark matter. For instance, the QCD~axion is commonly assumed to originate as a cold condensate through the misalignment mechanism~\cite{Preskill:1982cy, Abbott:1982af, Dine:1982ah, Marsh:2015xka}, in which case it acts as a dark-matter component and does not contribute to~$\Neff$. Nevertheless, even when such a non-thermal population dominates the present-day density, an additional thermal population generally arises if the couplings to the Standard Model and the reheating temperature are sufficiently large, with its contribution to~$\Neff$ being independent of the non-thermal abundance.} before the electroweak phase transition, it contributes $\Delta\Neff = 0.027$. For decoupling temperatures between the electroweak and QCD~transitions, the contribution grows to values within the range $[0.027, 0.064]$. Since the time of decoupling depends on the coupling strength to the Standard Model, measurements of~$\Neff$ can place limits on these interactions~\cite{Cadamuro:2011fd, Brust:2013ova, Chacko:2015noa, Baumann:2016wac, Ferreira:2018vjj, DEramo:2018vss, Arias-Aragon:2020shv, Ferreira:2020bpb, Dror:2021nyr, Caloni:2022uya}. For standard decoupling scenarios, these constraints depend on the reheating temperature of the universe, since these species must have been thermalized for the bounds to apply.\footnote{This dependence on the beginning of the hot big bang becomes an additional feature to constrain the reheating temperature if axion couplings are detected through other measurements.} On the other hand, if the axions couple to matter, they may also recouple to Standard Model particles below their mass thresholds, which allows for less stringent, but reheating-independent limits on their interaction strengths~\cite{Baumann:2016wac, Ghosh:2020vti, Ferreira:2020bpb, Green:2021hjh, DEramo:2021usm, Badziak:2024qjg, DEramo:2024jhn, Badziak:2025mkt, Barbieri:2026ewj}. In either case, the $\Neff$~measurements projected in Section~\ref{sec:results} probe axion couplings in regimes that are complementary to, and in some cases more constraining than, laboratory and astrophysical searches.

\paragraph{Dark fermions and sterile neutrinos.} Any light fermion that was in thermal equilibrium with the Standard Model contributes $\Delta\Neff \geq 0.047$ per Weyl degree of freedom. This includes a sterile neutrino, a right-handed neutrino in Dirac mass models, and any other dark-sector fermions. Measurements of~$\Neff$ at the level of the forecasted \sfour~(variance-limited)~survey will be sensitive to one of these species at the level of~$1.6\sigma$~(around~$5\sigma$). Some phenomenological scenarios include coupling these particles to the Standard Model via dipole, anapole, and four-fermion interactions~\cite{Abazajian:2012ys, Brust:2013ova, Magill:2018jla, Luo:2020sho, Brdar:2020quo}. For instance, sterile neutrinos that mix with active neutrinos and thermalize at temperatures around or below the QCD~phase transition can produce much larger~$\Delta\Neff$ depending on the mixing parameters and the thermal history, and are already highly constrained with current measurements~\cite{Abazajian:2012ys, Abazajian:2017tcc, Gariazzo:2019gyi, Hagstotz:2020ukm, Aloni:2023tff, GarciaEscudero:2025orc}. In another example, the three right-handed neutrinos in models where neutrinos are Dirac particles can be thermalized through their Yukawa couplings at high temperatures, which results in $\Delta\Neff \gtrsim 0.14$~\cite{Abazajian:2019oqj, Adshead:2020ekg, Biswas:2021kio}. We forecasted in Section~\ref{sec:results} that this scenario will be testable close to the~$5\sigma$ level with a \sfour-type experiment and at even higher significance for a potential successor survey.

\paragraph{Dark photons.} A massless, thermal spin-1 particle contributes like two real scalars due to its two polarizations, i.e.\ $\Delta\Neff = 0.054$ if it decoupled above the electroweak scale and detection significances of almost~$2\sigma$ at the level of~\sfour~(more than~$5\sigma$ for~CVL). The most prominent example is a massless~$U(1)$ gauge boson, known as a dark photon, which kinetically mixes with the SM~hypercharge boson, with another option being dipole interactions with the SM~fermions~\cite{Holdom:1985ag, Galison:1983pa, Ackerman:2008kmp, Pospelov:2008zw, Vogel:2013raa, Pierce:2014spa, Adshead:2022ovo, Caputo:2026pdw}. It is also worth noting that even massive dark photons with masses below the \si{MeV}~scale can contribute significantly if they were produced thermally~\cite{Redondo:2008ec, Berger:2016vxi, Ibe:2019gpv}. In this regime, $\Neff$~measurements provide constraints on the dark photon mass that are complementary to laboratory searches and astrophysical bounds~\cite{Caputo:2021eaa}.

\paragraph{Gravitino.} The gravitino is the unique elementary particle with spin-3/2 and is a universal prediction of supergravity. Since its mass is set by the supersymmetry~(SUSY) breaking scale, the gravitino can be light enough to be relativistic at recombination in low-scale SUSY-breaking scenarios~\cite{Weinberg:1982zq, Moroi:1993mb, Bolz:2000fu, Feng:2010ij}. As a spin-3/2 particle with two on-shell helicity states, which are equivalent to a Weyl fermion at high temperatures, a light gravitino contributes $\Delta\Neff = 0.047$ if it decoupled well before electroweak symmetry breaking. Since it necessarily exists in scenarios with additional degrees of freedom compared to the Standard Model, concrete models of supersymmetry breaking actually predict that a light gravitino effectively contributes $\Delta\Neff = 0.059$~\cite{Pierpaoli:1997im, Ichikawa:2009ir, Osato:2016ixc}, corresponding to a sensitivity of about~$2\sigma$ for a \sfour-like experiment and clearly exceeding the~$5\sigma$ statistical threshold for a CVL~survey.

\subsection{Properties of the Radiation Sector}
\label{sec:implications_radiation}

Measurements of the CMB~damping tail and acoustic peak structure are sensitive not only to the total energy density of non-photon radiation as parameterized by~$\Neff$, but also to the physical properties of that radiation. These measurements can in particular shed light on whether the radiation propagates as free-streaming species or behaves like a self-interacting fluid. These distinct signatures provide powerful additional insights into the nature of the radiation sector that are directly accessible with the data constraining~$\Neff$. We highlight a few of these testable properties in the following.

\paragraph{Free-Streaming and Fluid-Like Radiation.} Free-streaming relativistic particles, such as Standard Model neutrinos after their decoupling, travel faster than the sound speed of the photon-baryon fluid and generate a gravitational drag that slightly shifts the acoustic oscillations. In addition to an amplitude shift, this produces a characteristic, multipole-dependent phase shift in the CMB~temperature and polarization power spectra that was first identified analytically in~\cite{Bashinsky:2003tk} and has since been detected in~\Planck, ACT, and SPT~data~\cite{Follin:2015hya, Baumann:2015rya, Montefalcone:2025unv}. It has also been measured in large-scale-structure observations~\cite{Baumann:2019keh, Whitford:2024ecj}, which will contribute additional constraining power in the future~\cite{Baumann:2017gkg, Baumann:2019keh, Montefalcone:2025mbg}. This phase-shift imprint, which approaches a constant at small scales, is a particularly clean signature since it is only produced by free-streaming radiation under the assumption of adiabatic initial conditions~\cite{Baumann:2015rya}.\footnote{While isocurvature perturbations also produce a phase shift in the acoustic oscillations, they are generally expected to be scale dependent and, therefore, easily distinguishable from the free-streaming-induced shifts~\cite{Baumann:2015rya}. The much better resolution of the acoustic peaks in the CMB~power spectra in \sfour-like and future surveys will however naturally also enable better constraints on isocurvature modes in this way.}

The detection of this phase shift in current cosmological data therefore directly demonstrates that the cosmic neutrino background propagated to a very high degree as free-streaming radiation before recombination rather than as a tightly coupled fluid. Radiation that instead propagates as a perfect fluid, whether due to self-interactions or due to interactions with the Standard Model or another dark sector, imprints a much smaller phase shift~(but due to the presence of baryons remains non-zero)~\cite{Bashinsky:2003tk, Baumann:2015rya, Montefalcone:2025ibh}. This makes it possible to separately constrain the energy density carried by free-streaming radiation~($\Neff$) and by fluid-like radiation~(often parameterized by~$N_\mathrm{fluid}$, or equivalently by the free-streaming fraction of the total non-photon radiation density), as developed in~\cite{Bell:2005dr, Friedland:2007vv, Baumann:2015rya, Brust:2017nmv, Blinov:2020hmc, Brinckmann:2020bcn, Saravanan:2025cyi, Mescia:2026rxx}.\footnote{Free-streaming and fluid-like radiation have identical effects at the background level, but free-streaming particles additionally imprint a shift in the amplitude and phase of the acoustic oscillations at the perturbation level. Constraints on~$N_\mathrm{fluid}$ are therefore mainly driven by the damping tail and the time of matter-radiation equality, while~$\Neff$ is additionally sensitive to the perturbation-level information. In current data, this asymmetry translates into constraints on~$N_\mathrm{fluid}$ that are substantially weaker than those on~$\Neff$~\cite{Baumann:2015rya, Brust:2017nmv, Blinov:2020hmc, Brinckmann:2020bcn, Saravanan:2025cyi} and makes phase-shift measurements essentially independent of the helium abundance~$Y_\mathrm{p}$~\cite{Montefalcone:2025unv}.} Being able to make this distinction is physically important because many BSM~scenarios predict dark radiation that interacts strongly enough to not be free-streaming prior to recombination. Non-Abelian dark gauge bosons and light species with strong self-couplings naturally behave as a fluid before recombination~\cite{Jeong:2013eza, Buen-Abad:2015ova, Lesgourgues:2015wza}. In addition, there are various scenarios in which the radiation transitions between fluid-like and free-streaming behavior over cosmic history. Neutrinos with non-standard self-interactions or couplings to a light scalar can decouple later than in the standard cosmology and propagate as a fluid until some time before or even after recombination~\cite{Cyr-Racine:2013jua, Lancaster:2017ksf, Choi:2018gho, Kreisch:2019yzn, Ghosh:2019tab, Taule:2022jrz, Brinckmann:2022ajr, Montefalcone:2025ibh}. In dark-sector scenarios with mass thresholds, the dark radiation can conversely transition from free-streaming to a fluid-like state through a phase transition or recoupling event~\cite{Buen-Abad:2017gxg, Aloni:2021eaq, Brinckmann:2022ajr, Cvetko:2025kda, Sharma:2026ngx}. Future CMB~experiments will significantly tighten the constraints on both~$\Neff$ and~$N_\mathrm{fluid}$, with, for example, the lower bound on the free-streaming fraction projected to rise from about~82\% today to more than~96\% at \sfour~precision~\cite{Saravanan:2025cyi}~(see also~\cite{Baumann:2015rya, Brinckmann:2020bcn}). Expressed in terms of the amplitude of the phase shift itself relative to its expected SM~value, the current constraints at the~12\%~level are projected to improve to about~2.6\% with a \sfour-like experiment, while the CVL~floor lies around~1.3\%~\cite{Montefalcone:2025unv}. These projections illustrate that future surveys will allow for a significantly \mbox{more precise characterization of the radiation content.}

\paragraph{Neutrino Self-Interactions.} The neutrino-induced phase shift can be employed as a direct probe of neutrino interactions in the early universe. If neutrinos have non-standard self-interactions beyond their weak interactions, for instance, they would remain coupled longer and the onset of free-streaming would be delayed. This reduces the size of the phase shift and modifies the~CMB and matter power spectra in a way that is distinct from simply changing~$\Neff$~(as just discussed). This in turn enables the inference of bounds on the interaction strength of a potential universal Fermi interaction or flavor-specific couplings, for example~\cite{Cyr-Racine:2013jua, Archidiacono:2013dua, Oldengott:2017fhy, Forastieri:2015paa, Forastieri:2019cuf, Das:2020xke, RoyChoudhury:2020dmd, Venzor:2022hql, Das:2023npl, He:2023oke, Camarena:2023cku, Camarena:2024daj, Poudou:2025qcx, Whitford:2025dmq}~(see also~\cite{Berryman:2022hds} for a recent review). The mentioned phase shift can also be used to directly constrain the temperature-dependent neutrino scattering rates, with the same functional form as in the free-streaming case and only its asymptotic amplitude rescaled~\cite{Montefalcone:2025ibh}. Current data from~\Planck, ACT, and~SPT already indicate that neutrinos have been free-streaming since deep in the radiation-dominated era~\cite{Montefalcone:2025unv, Montefalcone:2025ibh}. This means that measurements at the level of~\sfour\ and beyond, with correspondingly improved sensitivity to the acoustic peak structure, will naturally further tighten these bounds~\cite{Montefalcone:2025unv}.

\paragraph{Dark Matter-Neutrino/Dark Radiation Interactions.} Interactions between dark matter and neutrinos or dark radiation leave additional imprints in the~CMB beyond those from the radiation sector alone~\cite{Mangano:2006mp, Ackerman:2008kmp, Cyr-Racine:2012tfp, Wilkinson:2014ksa, Cyr-Racine:2015ihg, Buen-Abad:2017gxg, Olivares-DelCampo:2017feq, Gluscevic:2019yal, Mosbech:2020ahp, Boddy:2022knd}. If dark matter scatters off neutrinos, the resulting momentum exchange slows the effective sound speed of the neutrino perturbations through a mechanism known as dark-matter loading, which further modifies the phase shift in a way that is proportional to the interacting dark matter abundance and largely independent of the radiation energy density~\cite{Ghosh:2024wva}. This provides a distinctive signature of dark matter-neutrino or dark matter-dark radiation scattering that is directly accessible through the acoustic peak structure of the~CMB. More broadly, models with such dark sectors, which may also be secluded~(see~\textsection\ref{sec:implications_additional}), moreover predict correlated effects on the CMB~damping tail and the matter power spectrum~\cite{Blennow:2012de, Cyr-Racine:2012tfp, Buen-Abad:2015ova, Chacko:2015noa, Chacko:2016kgg, Buen-Abad:2017gxg}, so that precise $\Neff$~measurements \mbox{can simultaneously constrain the internal dynamics of the dark sector.}

\subsection{Further Probes of Early-Universe Physics}
\label{sec:implications_additional}

Beyond light thermal relics and the properties of the radiation sector, precise $\Neff$~measurements probe a broad range of interesting early-universe physics. Their unique reach originates from the fact that the radiation energy density is preserved by the cosmological expansion unless there are entropy-injection events. This implies that any process that altered the radiation content between reheating and recombination leaves an imprint in~$\Neff$, regardless of the details of the underlying mechanism. The same measurement therefore is simultaneously sensitive to non-equilibrium production that goes beyond the discussed thermal relics, neutrino-sector physics not captured by the already mentioned radiation-perturbation effects, and the consistency of the radiation content across cosmic epochs. Each of the following scenarios would be sharpened by a \sfour-like measurement and could be transformed by a cosmic-variance-limited survey, including in the limit where the inferred~$\Neff$ is consistent with the SM~prediction itself.

\paragraph{Stochastic Gravitational Wave Background.} Any stochastic gravitational-wave background~(SGWB) of cosmological origin contributes to the total radiation energy density and is therefore constrained by~$\Neff$. Sources include inflationary gravitational waves, first-order phase transitions, cosmic strings, and preheating dynamics~(see~\cite{Amin:2014eta, Caprini:2018mtu, Allahverdi:2020bys, Hindmarsh:2020hop, Dvorkin:2022jyg, Caldwell:2022qsj, Achucarro:2022qrl, Athron:2023xlk} for reviews). Importantly, the $\Neff$-derived bound provides an integrated constraint over a very broad frequency range, unlike the relatively narrow bands of direct-detection experiments~\cite{Maggiore:1999vm, Smith:2006nka}. Current \Planck~constraints on~$\Neff$ already provide competitive limits on the integrated gravitational-wave energy density at frequencies far below those accessible to ground-based interferometers, such as~LIGO, Virgo, and~KAGRA, or the planned LISA~satellites. A measurement of $\sigmaNeff \approx 0.030$ around the SM~value would correspondingly improve these bounds, with a CVL~survey pushing the constraints by more than an order of magnitude. The integrated~$\Neff$~limit is moreover directly relevant for assessing cosmological interpretations of the stochastic background recently detected at nano-Hertz frequencies by pulsar-timing arrays~\cite{NANOGrav:2023gor, EPTA:2023fyk, Reardon:2023gzh, Xu:2023wog}. While the signal is consistent with an astrophysical population of inspiraling supermassive-black-hole binaries~\cite{NANOGrav:2023hfp}, a number of proposed early-universe sources for the signal would generate gravitational-wave energy densities at frequencies above the nano-Hertz band that already approach or saturate current $\Neff$~bounds~(see e.g.~\cite{Bringmann:2023opz, NANOGrav:2023hvm, Ellis:2023oxs, Ben-Dayan:2025bqd}), with future $\Neff$~measurements therefore providing decisive discriminating tests. In general, future cosmological inferences of~$\Neff$ will therefore probe the~SGWB across a broad frequency range that is complementary to other ways of detecting gravitational waves or inferring their effects~\cite{Meerburg:2015zua, Lasky:2015lej, Burke-Spolaor:2018bvk, KAGRA:2021kbb, Renzini:2022alw, LISACosmologyWorkingGroup:2022jok}.

\paragraph{MeV-Scale Thermal Dark Matter.} Light dark matter particles with masses in the \si{MeV}~range may remain in thermal equilibrium with the Standard Model plasma through neutrino decoupling and then become non-relativistic and annihilate into either the electron-photon bath or the neutrino bath. In these cases, they modify the entropy transfer between the electromagnetic and neutrino sectors, and therefore shift~$\Neff$ in a coupling- and spin-dependent way~\cite{Serpico:2004nm, Ho:2012ug, Boehm:2013jpa, Nollett:2013pwa, Nollett:2014lwa, Green:2017ybv, Knapen:2017xzo}. Electrophilic species heat the photon bath after neutrino decoupling and reduce~$\Neff$ below its SM~value~(see also below), whereas neutrinophilic species heat the neutrino bath and increase~$\Neff$. In both cases the size of the shift depends on the internal degrees of freedom of the particle~(cf.\ \textsection\ref{sec:implications_relics}) and on whether this proceeds in $s$- or $p$-wave annihilation~\cite{Nollett:2013pwa, Nollett:2014lwa, Escudero:2018mvt}. Current~\Planck\ and BBN~data already exclude thermal relics that couple exclusively to the electromagnetic or to the neutrino sector and are lighter than approximately \SIrange[range-units=single, range-phrase = --]{3}{10}{MeV}, depending on their spin and coupling structure~\cite{Sabti:2019mhn, Depta:2019lbe, Giovanetti:2021izc, Chu:2022xuh, An:2022sva, An:2024nsw}. Beyond the relics themselves, $\Neff$~measurements separately constrain the mediator that couples the dark sector to the Standard Model or its decay products. These measurements therefore access the much broader class of dark matter-baryon interactions, providing a complementary probe to more direct tests in terrestrial experiments, astrophysical observations, and cosmology~\cite{Green:2017ybv, Knapen:2017xzo, Gluscevic:2019yal, Boddy:2022knd, An:2024nsw}. Future \sfour-like measurements, as forecasted in this work, will therefore significantly extend the bounds on \si{MeV}-scale thermal relics and further constrain their viability, especially in combination with BBN~constraints.

\paragraph{Freeze-In Production.} Beyond thermal freeze-out, light relics can also be populated via freeze-in, in which feeble interactions slowly populate the abundance of the species without ever bringing it into thermal equilibrium~\cite{Hall:2009bx}. As an example, in ultraviolet freeze-in scenarios, the production rate grows with temperature so that the abundance is set near the reheating temperature, which implies that~$\Delta\Neff$ depends on both the coupling strength and the reheating scale~\cite{Elahi:2014fsa, Adshead:2016xxj, Bernal:2019mhf}. This mechanism is particularly relevant for axion-like particles produced via the Primakoff process, massless dark photons, and light right-handed neutrinos~\cite{Dessert:2018khu, Luo:2020fdt, Biswas:2022vkq, Caloni:2024olo, Jain:2024dtw}.

\paragraph{Dark Radiation from Heavy Particle Decays.} In many SM~extensions, heavy and long-lived particles can decay partially into light dark-sector species, injecting radiation that contributes to~$\Neff$. This mechanism is generic in supersymmetric and string-motivated cosmologies, where moduli, saxions, or heavy gravitinos can decay into light hidden-sector states such as axinos and axions~\cite{Banks:1993en, deCarlos:1993wie, Choi:1996vz, Choi:2013lwa}. The contribution to~$\Neff$ depends on the mass, lifetime, and branching ratios of the decaying particle, with typical predictions in the range of $\Delta\Neff \approx 0.01$ to $\Delta\Neff \approx 1$ depending on the model~\cite{Cicoli:2012aq, Higaki:2012ar, Hasenkamp:2012ii, Hebecker:2014gka, Cicoli:2015bpq, Bleau:2023fsj}. In addition to a more precise determination of~$\Neff$ in future CMB~experiments, the comparison between the inferred values from the~CMB and~BBN~(see below) will further bound many of these scenarios.

\paragraph{Secluded Dark Sectors.} The mechanisms discussed so far generate dark radiation from the visible sector, whether through thermal freeze-out, decays, or freeze-in. Even a purely gravitationally coupled hidden sector, which was never in thermal equilibrium with the Standard Model, can contribute to~$\Neff$ if it contains relativistic species. The size of this contribution is suppressed by the fourth power of the ratio of the dark-sector temperature to the visible-sector~(photon) temperature, which is set by how entropy was shared between the two sectors at their last common contact or by asymmetric reheating~\cite{Berezhiani:1995am, Berezhiani:2000gw, Feng:2008mu, Ackerman:2008kmp, Blennow:2012de, Cyr-Racine:2012tfp, Boddy:2014yra, Chacko:2015noa, Adshead:2016xxj, Arkani-Hamed:2016rle, Craig:2016lyx, Chacko:2018vss, Cyr-Racine:2021oal, Ireland:2022quc, Batell:2025hmx}. A colder hidden sector therefore yields a smaller~$\Delta\Neff$, which can fall below the contribution of even a single thermalized real scalar~(see~\textsection\ref{sec:implications_relics}). On the other hand, a secluded sector whose temperature is comparable to that of the photons can produce a sizable signal, in particular if the sector contains many relativistic degrees of freedom. Concrete realizations include mirror and twin dark sectors~\cite{Berezhiani:1995am, Berezhiani:2000gw, Feng:2008mu, Boddy:2014yra, Chacko:2015noa, Craig:2016lyx, Chacko:2018vss, Cyr-Racine:2021oal}. Because such a sector can itself contain coupled dark matter and free-streaming or fluid-like dark radiation, it imprints correlated signatures in the CMB~damping tail, the acoustic peak structure, and the matter power spectrum that go beyond its contribution to the total radiation density alone~(cf.~\textsection\ref{sec:implications_radiation}). Precise $\Neff$~measurements therefore probe the existence and thermal history of dark sectors even when they are entirely decoupled from the visible universe.

\paragraph{Scenarios with $\mathbf{N_{eff} < 3}$.} While the discussion has focused primarily on positive~$\Delta\Neff$, precise and accurate measurements are equally sensitive to scenarios in which new physics reduces~$\Neff$ below the SM~value of~$3.044$, i.e.\ $\Delta\Neff < 0$. This can occur if either the neutrino density is smaller than in the Standard Model or the photon density is larger. For instance, particles that transfer their entropy to the electromagnetic sector after neutrino decoupling~(such as the electrophilic \si{MeV}-scale relics discussed above) heat the photon bath relative to the neutrinos, diluting the neutrino energy density relative to that of the photons. Concrete realizations include \si{MeV}-scale dark photons or scalars that annihilate or decay into electron-positron pairs at temperatures below about~\SI{1}{MeV}~\cite{Cadamuro:2010cz, Escudero:2018mvt, Ibe:2019gpv, Escudero:2026mgw}. Other scenarios resulting in negative~$\Delta\Neff$, and being constrained to varying degrees, include those with low reheating temperatures that prevent the neutrino bath from fully thermalizing~\cite{Kawasaki:2000en, deSalas:2015glj, Hasegawa:2019jsa, Abazajian:2023reo, Barbieri:2025moq}, first-order phase transitions that reheat the photon bath after neutrino decoupling~\cite{Bai:2021ibt, Deng:2023twb, Xu:2025zsv}, and neutrinophilic mediators that decouple the neutrino sector from the photon plasma at non-standard temperatures~\cite{Berlin:2018ztp, Escudero:2019gzq}. Interestingly, recent ACT~and SPT~data in combination with~\Planck\ yield central values below the SM~expectation, $\Neff \approx 2.8$~\cite{AtacamaCosmologyTelescope:2025nti, SPT-3G:2025bzu}, which has further motivated the systematic exploration of such scenarios~\cite{Escudero:2026mgw}. While these measurements remain consistent with $\Neff = 3.044$ at the~$<2\sigma$~level, \sfour-like and certainly CVL~experiments will be able to conclusively test whether this trend persists or is a statistical fluctuation.

\paragraph{Neutrino Lifetime.} If neutrinos are unstable, their decays modify the radiation content and perturbation structure of the universe in ways that are directly constrained by CMB~measurements. In the case of invisible decays, $\nu_\mathrm{H} \to \nu_\mathrm{l} + \phi$, where~$\nu_\mathrm{H}$ and~$\nu_\mathrm{l}$ are the heavier and lighter neutrino, respectively, and~$\phi$ is a light scalar such as a majoron, the decay products redistribute the neutrino energy and suppress the free-streaming anisotropic stress that leads to the characteristic phase shift in the CMB~acoustic peaks~\cite{Hannestad:2005ex, Archidiacono:2013dua, Funcke:2019grs, Escudero:2020ped, Gerbino:2022nvz}. This provides a sensitive probe of the neutrino lifetime, resulting in bounds that are many orders of magnitude more stringent than current laboratory and oscillation constraints~\cite{Escudero:2019gfk, Chacko:2019nej, Chacko:2020hmh, Barenboim:2020vrr, FrancoAbellan:2021hdb, Chen:2022idm}. Future CMB~experiments will further tighten these bounds, extending the reach to longer lifetimes and providing complementary constraints to next-generation neutrino telescopes and experiments.

\paragraph{Neutrino Chemical Potential.} A large chemical potential in the neutrino sector, which may arise from a primordial lepton asymmetry, for instance, increases the neutrino energy density and shifts~$\Neff$ above the Standard Model value~\cite{Lesgourgues:1999wu, Abazajian:2002qx, Mangano:2010ei, Castorina:2012md, Grohs:2016cuu, Gerbino:2022nvz}. While current constraints on the neutrino chemical potential are relatively weak from the~CMB alone, they will significantly improve with future CMB~inferences of~$\Neff$~(and~$Y_\mathrm{p}$)~\cite{Oldengott:2017tzj, Escudero:2022okz, Froustey:2024mgf}. These bounds will sharpen further and provide complementary sensitivity when additionally combined with BBN~observations of the primordial helium abundance, since the helium yield is independently affected by the electron neutrino chemical potential through the neutron-to-proton ratio~\cite{Escudero:2022okz, Yanagisawa:2025mgx}.

\paragraph{Primordial Helium Abundance.} The primordial helium mass fraction~$Y_\mathrm{p}$ can be jointly constrained with~$\Neff$ from the CMB~damping tail, where the two are anti-correlated through their common effect on the diffusion-damping scale~\cite{Hou:2011ec}, a degeneracy that is broken by the free-streaming phase shift~(see~\textsection\ref{sec:results_cvl} and~\textsection\ref{sec:implications_radiation}). In addition to the inference of~$Y_\mathrm{p}$ from CMB~anisotropies and astronomical observations of metal-poor extragalactic regions and the intergalactic medium~\cite{Izotov:2014fga, Aver:2015iza, Cooke:2018qzw, Cooke:2024nqz, Yanagisawa:2025mgx, Aver:2026dxv}, the primordial helium abundance could in principle also be probed through the \SI{21}{cm}~signal from the dark ages~\cite{Mondal:2023xjx} and the cosmological recombination radiation via CMB~spectral distortions. The helium-line features of these distortions encode the unreprocessed value of~$Y_\mathrm{p}$ and could ultimately be measured by a future spectrometer~\cite{Rubino-Martin:2007tua, Chluba:2019kpb, Chluba:2019nxa, Chluba:2015gta, Chluba:2016bvg, Hart:2020voa, Chluba:2025wxp}. A~\sfour-like constraint on~$\Neff$ from the~CMB alone will simultaneously provide a precise CMB-derived measurement of~$Y_\mathrm{p}$ that can be compared to astronomical determinations together with BBN~predictions. In fact, inferring the value of~$Y_\mathrm{p}$ from the~CMB should become more precise at that level~(see our forecasts in~\textsection\ref{sec:results_cvl}) than the value derived from~BBN and light-element-abundance measurements~(cf.~e.g.~\cite{Pitrou:2018cgg, Grohs:2019cae, Lague:2019yvs, Fields:2019pfx, Pisanti:2020efz, Yeh:2020mgl, Matsumoto:2022tlr, Yeh:2022heq, Grohs:2023voo, Cooke:2024nqz, Yanagisawa:2025mgx, Aver:2026dxv}). This therefore offers a non-trivial consistency test of the standard model of cosmology.

\paragraph{Comparison Across Cosmic Epochs.} Relatedly, big bang nucleosynthesis and the cosmic microwave background independently probe the radiation content of the universe at vastly different epochs: $T \sim \SI{1}{MeV}$ or about one second after the big bang and $T \sim \SI{0.3}{eV}$ or about \num{400000}~years later. Comparing the two and/or inferring the value of~$\Neff$ during~BBN from CMB~measurements of~$Y_\mathrm{p}$ provides powerful tests of physics at and between these epochs. Any discrepancy will signal new physics that altered the radiation content, such as the decay of a massive particle into dark radiation, a cosmological phase transition producing relativistic species, or late changes to the neutrino sector~\cite{Fischler:2010xz, Menestrina:2011mz, Baumann:2015rya, Berlin:2019pbq, Escudero:2019gfk, Bai:2021ibt, Aloni:2021eaq, Yeh:2022heq, Sobotka:2023bzr, Buckley:2024nen, Garny:2024ums, Joseph:2026gws}. Current data are consistent with a common value, with joint~BBN and~CMB analyses yielding $\Neff \lesssim 3.1$ at~$2\sigma$~\cite{Giovanetti:2024eff, Yeh:2026pil, Goldstein:2026iuu}. \sfour-like and future measurements will sharpen this comparison in combination with ongoing improvements in astronomical measurements of the primordial helium abundance~$Y_\mathrm{p}$ and the deuterium-to-hydrogen ratio~(see the previous paragraph). With the inference of~$\Neff$ at~BBN from the~CMB becoming competitive with or even surpassing direct inferences from~BBN and light-element abundances as just discussed, the comparison will be possible at the level where even small inter-epoch variations could be detected or excluded~\cite{Baumann:2015rya}. The forecasted constraints of Section~\ref{sec:results} would therefore open a direct window into the thermal history of the universe between the first second and recombination.

\paragraph{Confirmation of the Standard Model prediction.} Finally, we emphasize that inferring a value of~$\Neff$ from CMB~data that is consistent with the SM~value of~$3.044$ is itself a powerful result. It would be another important test of the standard models of both cosmology and particle physics, and place strong bounds on many of their alternatives and extensions. This applies to SM~neutrinos as well as the corrections away from the number of three neutrinos from non-instantaneous neutrino decoupling and finite-temperature QED~effects at the level of~$0.044$~\cite{Akita:2020szl, Froustey:2020mcq, Bennett:2020zkv, Cielo:2023bqp, Drewes:2024wbw, Binder:2024vmy, Ihnatenko:2025kew, Escudero:2025kej}. A measurement of~$\sigmaNeff \approx 0.030$ would confirm these corrections at the level of about~$1.5\sigma$, while a measurement with $\sigmaNeff \approx 0.01$ would detect them at close to~$5\sigma$, providing a direct test of the physics of neutrino decoupling and the Standard Model of particle physics.

\section{Conclusions and Outlook}
\label{sec:conclusions}

The early universe is the most extreme laboratory available to physicists, reaching temperatures, densities, and timescales far beyond any scale that can be reproduced on Earth. Neutrinos and other light species leave distinct, characteristic, and measurable imprints in cosmological observations, in particular of the cosmic microwave background. This allows for precise measurements of the effective number of relativistic species~$\Neff$ through which we can probe the particle content and the thermal history of our universe, among other interesting physics.\medskip

The \sfour~effort was started and matured with the two design-driving science goals\hskip1pt\footnote{The full \sfour~science program extended substantially further and was organized along two additional themes beyond the fundamental-physics goals of inflation and the dark universe~\cite{CMB-S4:2016ple, Abazajian:2019eic, CMB-S4:2022ght, CMB-S4:2023cdr}. On the one hand, the survey would have provided a precise map of the matter distribution through CMB~lensing with direct implications for the sum of neutrino masses, dark energy, and tests of gravity on cosmological scales as well as constraints on dark-matter properties and possible non-gravitational interactions. It would have also characterized the formation history and baryonic feedback of galaxy clusters through their thermal and kinematic Sunyaev-Zel'dovich signatures. On the other hand, beyond these cosmological probes, the planned wide and high-cadence survey of the millimeter-wave sky would have opened a new window onto the transient and time-variable universe, ranging from gamma-ray-burst afterglows and multi-messenger follow-up of neutrino and gravitational-wave events to outer-solar-system bodies.} of detecting primordial gravitational waves, and measuring~$\Neff$ at the one-percent level to test the radiation content of the universe and gain qualitatively new insights into physics within and beyond the Standard Model of particle physics. Dedicated analysis techniques and forecasts for inferring the primordial tensor-to-scalar ratio~$r$ from \sfour~observations were presented in~\cite{CMB-S4:2020lpa}. In this paper, we present Fisher-matrix forecasts of the projected sensitivity to~$\Neff$ of a few of the proposed \sfour~survey configurations, which were considered during its extensive design phase before the decision by the main funding agencies to not carry the project forward. To exemplify the range of considered configurations, we studied the two-site conceptual design, a South-Pole-only alternative, and the Chile-only revised configuration combining one \sfour~large-aperture telescope in the Atacama Desert with the corresponding telescope of a survey such as the Simons Observatory. The two options with Chilean components allow reaching the original science target of $\sigmaNeff = 0.030$ over time scales ranging from less than six to nine years, while a configuration deployed entirely at the South Pole at a cost roughly equivalent to that of the conceptual design misses the target sensitivity by more than~20\% over the nominal seven-year observation time. In addition, we reiterated and highlighted through a simplified optimization study that the number of independent observed modes and, therefore, the maximum observable sky area should drive the survey design together with optimal component separation and delensing. We further complement these insights from simplified and realistic configurations by considering a cosmic-variance-limited survey that sets the ultimate reach $\sigmaNeff = 0.0073$ of the CMB~primary anisotropies over the same multipole range $\ell \leq 5000$, also highlighting the additional room that could be explored with further experimental, computational, and theoretical work.

To enable these forecasts, we built an end-to-end pipeline that integrates multi-frequency modeling of galactic and extragalactic foregrounds, sky masking, component separation through internal linear combination, iterative delensing, and a Fisher analysis. This pipeline and intermediary data products, such as the residual internal-linear-combination~(noise) spectra, are publicly released as the~\texttt{DRAFT}~tool~\cite{Raghunathan:draft}, which can also be directly applied to future survey-design exercises and studies of systematics.\footnote{The present work focuses on the statistical sensitivity of a survey like~\sfour\ with one realistic realization of foregrounds. A complementary study of foreground-induced biases and associated mitigation strategies is based on the same \texttt{DRAFT}~pipeline~\cite{Raghunathan:inprep}. It will inform future choices, strategies, and requirements to better understand and model foregrounds for any actual analysis at the observational precision of~\sfour\ and beyond.}\medskip

The precision levels forecasted in this work translate into a rich landscape of physical implications, which we discussed in some detail in Section~\ref{sec:implications}. A measurement at the level of $\sigmaNeff \approx 0.030$ will place the first meaningful constraints on any thermal relic with spin~$\geq 1/2$ that was ever in equilibrium with the Standard Model, while a measurement approaching the cosmic-variance-limited floor would probe the entire landscape of light thermal relics down to the $\Delta\Neff = 0.027$ contribution of a single real scalar that decoupled before the electroweak phase transition~(see also footnote~\ref{fn:thresholds} in this context). The same data carry information beyond the total radiation energy density. This includes the free-streaming nature of the cosmic neutrino background and dark radiation through the associated phase shift of the acoustic oscillations, the entropy balance between the neutrino and photon sectors, thermal dark matter, and an integrated constraint on any cosmological stochastic gravitational-wave background far below the frequencies accessible to direct-detection experiments, for instance. Moreover, inferences of~$\Neff$ at this level would not only test physics beyond the Standard Model of particle physics, but also provide consistency tests of the standard models through joint inferences with the primordial helium abundance~$Y_\mathrm{p}$ and exquisite insights into the physics of neutrino decoupling.\medskip

Looking beyond the survey configurations considered here, the physics insights, design optimization, and forecasting tools developed within the \sfour~collaboration directly inform the planning of current- and next-generation~CMB surveys. The Simons Observatory and the South Pole Observatory will continue to sharpen our view of the damping tail over the coming decade, and experiments such as CCAT-prime will be instrumental in better characterizing and therefore mitigating foreground contamination~\cite{CCAT-Prime:2021lly}. Beyond lowering the instrumental noise, such progress in observationally characterizing and theoretically understanding the foregrounds will determine how much of the $\Neff$~information imprinted at small angular scales can ultimately be extracted. Continued work on foreground modeling and component separation on small scales, in addition to the large scales relevant for $B$-mode science, is therefore of direct cosmological relevance. On the instrumental side, SO~and other proposed ground- and space-based experiments will achieve greater depth, and subsequently approach and eclipse the \sfour~precision on~$\Neff$ forecasted in this work. In particular, the central role of sky coverage established in~\textsection\ref{sec:results_optimization} makes a full-sky space mission, such as~the Probe of Inflation and Cosmic Origins~(PICO)~\cite{NASAPICO:2019thw} or a similar satellite with higher angular resolution, an attractive avenue for~$\Neff$~science. In parallel, upcoming and proposed spectroscopic and photometric galaxy surveys will provide complementary measurements of~$\Neff$, both independently and through joint analyses with the~CMB~\cite{Baumann:2017gkg, Sprenger:2018tdb, CosmicVisions21cm:2018rfq, Baumann:2019keh, PUMA:2019jwd, Sailer:2021yzm, MoradinezhadDizgah:2021upg, Karkare:2022bai, Shi:2022drq, Lee:2023uxu, Euclid:2024imf, MoradinezhadDizgah:2026hrg}, offering cross-checks across cosmological observables and further improvements in the overall constraining power. The trajectory of~$\Neff$~measurements in the next decade will therefore continue to sharpen, enabled by a complementary suite of experiments that will eventually reach the highlighted precision targets over longer timescales than originally envisaged.\medskip

The work performed within the \sfour~collaboration~(and project) and its impact extend far beyond the forecasts and tools presented in this and other works. Over the past one-and-a-half decades, it also assembled a compelling physics case for~CMB and other surveys, convened the cosmological community, and laid the technical groundwork. With the effective number of relativistic species~$\Neff$ and its broad physics implications being one example, this legacy will continue to shape the design and operation of current CMB~experiments and to motivate the ones that will follow. Decisive steps in instrumentation, modeling, analysis, phenomenology, and theory, together with the sustained collaboration of the observational and theoretical physics communities, offer a highly promising path to uncover qualitatively new physics. Driven by these efforts, observations of the cosmic microwave background will continue to push our understanding of the universe as a laboratory for physics from the smallest to the largest scales.

\vskip20pt
\paragraph{Acknowledgments}
C.\,T.~is supported by the Richard~S.\ Morrison Fellowship at Case Western Reserve University and was previously supported by the US~Department of Energy~(DOE) under Grant~\mbox{DE-SC0010129}. S.\,R.~acknowledges the support of~Michael and Ester~Vaida, and the National Science Foundation~(NSF) via Award~\mbox{OPP-1852617}. S.\,R.~and C.\,T.~were supported by the Center for AstroPhysical Surveys~(CAPS) at the National Center for Supercomputing Applications~(NCSA), University of Illinois Urbana-Champaign. B.\,W.~acknowledges support by the Bezos Membership at the Institute for Advanced Study, the~EDUCATE~Excellence Centre funded by the Swedish Research Council through Grant~\mbox{2022-06627}, NSF~under Grant~\mbox{PHY-1820775}, the Simons Foundation Modern Inflationary Cosmology Initiative under Grant~\mbox{SFARI~560536}, the Swedish Research Council under Contract No.~\mbox{638-2013-8993}, and~DOE under Grants~\mbox{DE-SC0009919} and~\mbox{DE-SC0019035} during the course of this project, with Nordita being supported in part by~NordForsk. J.\,M.~was supported by~DOE under Grant~\mbox{DE-SC0010129}, by~NASA through ADAP~Grant~\mbox{80NSSC24K0665}, and by~NSF through Grant~\mbox{AST-2510926}. K.\,N.\,A.~was supported by NSF~Theoretical Physics Program Grant~\mbox{PHY-2210283}. C.\,B.~acknowledges partial support by the Italian Space Agency LiteBIRD~Project~(ASI~Grants~\mbox{2020-9-HH.0} and~\mbox{2016-24-H.1-2018}), the~InDark and LiteBIRD~Initiatives of~INFN, Project~\mbox{SPACE-IT-UP} by the Italian Space Agency and Ministry of University and Research, Contract No.~\mbox{2024-5-E.0}, the RadioForegroundsPlus Project~\mbox{HORIZON-CL4-2023-SPACE-01}, \mbox{GA~101135036}, and the \mbox{CMB-Inflate}~project funded by the European Union's Horizon~2020 Research and Innovation Staff Exchange under the Marie Skłodowska-Curie Grant Agreement No.~101007633. K.\,K.\,B.~acknowledges support from the~NSF under Grant~\mbox{PHY-2413016}. G.\,C.~acknowledges funding from the European Union~(ERC, POLOCALC, Project No.~101096035) and from the Italian Ministry of Research~(Young Researcher Grant~\mbox{MSCA2024\_0000016}). F.-Y.\,C.-R.~acknowledges the support of the~NSF through CAREER~Grant~2440096. K.\,F.~is grateful for support from the \mbox{Jeff \& Gail Kodosky}~Endowed Chair in Physics at the University of Texas, and acknowledges support from the US~Department of Energy, Office of Science, Office of High Energy Physics program under Award~\mbox{DESC-0022021} and the Swedish Research Council~(Contract No.~\mbox{638-2013-8993}). H.\,G.\,E.~is supported by Grant~63667 from the John Templeton Foundation. The opinions expressed in this publication are those of the authors and do not necessarily reflect the views of the John Templeton Foundation. M.\,G.~is funded by the European Union~(ERC, RELiCS, Project No.~101116027). Views and opinions expressed are however those of the authors only and do not necessarily reflect those of the European Union or the European Research Council Executive Agency. Neither the European Union nor the granting authority can be held responsible for them. D.\,G.~is supported by~DOE under Grant~\mbox{DE-SC0009919}. K.\,M.\,H.~acknowledges DOE~Award~\mbox{DE-SC0024462}. L.\,K.~acknowledges support from DOE~Office of Science Award~\mbox{DE-SC0009999}. M.\,L.~and M.\,M.\,S~are supported by DOE~Grants~\mbox{DE-SC0023183} and~\mbox{DE-SC0011637}. P.\,L.~acknowledges funding from NASA~Space Grant, and the Emmett and Gladys~W.~Technology Fund. G.\,M.~acknowledges support from the US~Department of Energy, Office of Science, Office of High Energy Physics program under Award~\mbox{DESC-0022021}. M.\,R.~is partially supported by the InDark~Initiative of~INFN and by~INAF. C.\,L.\,R.~acknowledges support from the Australian Research Council's Discovery Project scheme~(No.~DP260100705). This work received support from~DOE under Contract No.~\mbox{DE-AC02-76SF00515} to SLAC~National Accelerator Laboratory~(E.\,S.). This document was prepared by~\sfour\ using the resources of the Fermi National Accelerator Laboratory~(Fermilab), a US~Department of Energy, Office of Science, Office of High Energy Physics~HEP~User Facility~(S.\,M.\,S.). Fermilab~is managed by Fermi Forward Discovery Group,~LLC, acting under Contract No.~89243024CSC000002. S.\,W.~was supported in part by DOE~Grant~\mbox{DE-FG02-85ER40237}.

This work made use of computational resources at the Illinois Campus Cluster~(S.\,R.\ and C.\,T.), the Institute for Advanced Study~(B.\,W.), \mbox{ManeFrame~II}~(J.\,M.~and C.\,T.), \mbox{ManeFrame~III}~(J.\,M.), the National Energy Research Scientific Computing Center~(NERSC), and the National Supercomputer Centre~(NSC) Tetralith~(B.\,W.). The Illinois Campus Cluster is operated by the Illinois Campus Cluster Program in conjunction with~NCSA and is supported by funds from the University of Illinois Urbana-Champaign. Resources on~\mbox{ManeFrame~II} and~\mbox{ManeFrame~III} were provided by Southern Methodist University's O'Donnell Data Science and Research Computing Institute. NERSC~is a US~Department of Energy User Facility, with work performed using NERSC~Award~\mbox{HEP-ERCAP0036526}. NSC~Tetralith is provided by the National Academic Infrastructure for Supercomputing in Sweden~(NAISS) under Projects~\mbox{2025/5-729} and~\mbox{2025/6-464}, which is partially funded by the Swedish Research Council through Grant~\mbox{2022-06725}.

We acknowledge the use of~\texttt{CAMB}~\cite{Lewis:1999bs}, \texttt{CLASS}~\cite{Blas:2011rf}, \texttt{CLASS\_delens}~\cite{Hotinli:2021umk}, \texttt{FisherLens}~\cite{Hotinli:2021umk}, \texttt{HEALPix}~\cite{Gorski:2004by}, and \texttt{IPython}~\cite{Perez:2007ipy}, and the Python packages~\texttt{healpy}~\cite{Zonca:2019vzt}, \texttt{Matplotlib}~\cite{Hunter:2007mat}, \texttt{NumPy}~\cite{Harris:2020xlr}, and~\texttt{SciPy}~\cite{Virtanen:2019joe}.

\clearpage
\phantomsection
\addcontentsline{toc}{section}{References}
\bibliographystyle{utphys}
\bibliography{references}

\end{document}